%% file: model.tex
\documentclass{aa}

\usepackage[varg]{txfonts}

\usepackage{natbib}
\bibpunct{(}{)}{;}{a}{}{,}

\defcitealias{2013A&AAlibert}{A13}

\defcitealias{NGPPS2}{Paper II}
\def\papertwo{\citetalias{NGPPS2}}

\defcitealias{NGPPS3}{Paper III}
\def\paperthree{\citetalias{NGPPS3}}

\usepackage{amssymb, amsmath}
\usepackage{wasysym}

\usepackage{siunitx}
\DeclareSIUnit\erg{erg}
\DeclareSIUnit\year{yr}
\DeclareSIUnit\au{au}

\input{defs.tex}

\usepackage{tikz}
\usetikzlibrary{arrows,shapes,decorations.markings}

\usepackage{hyperref}
\hypersetup{pdfauthor=A. Emsenhuber et al., pdftitle=The New Generation Planetary Population Synthesis (NGPPS). I., colorlinks=true, urlcolor=blue, linkcolor=blue, citecolor=blue}

\title{The New Generation Planetary Population Synthesis (NGPPS)}
\subtitle{I. Bern global model of planet formation and evolution, model tests, \\ and emerging planetary systems}
\titlerunning{The New Generation Planetary Population Synthesis (NGPPS). I.}

\author{Alexandre Emsenhuber\inst{\ref{uofa},\ref{unibe},\ref{lmu}} \and Christoph Mordasini\inst{\ref{unibe}} \and Remo Burn\inst{\ref{unibe},\ref{mpia}} \and Yann Alibert\inst{\ref{unibe}} \and Willy Benz\inst{\ref{unibe}} \and Erik Asphaug\inst{\ref{uofa}}}
\authorrunning{A.~Emsenhuber et al.}

\institute{
	Lunar and Planetary Laboratory, University of Arizona, 1629 E. University Blvd., Tucson, AZ 85721, USA \label{uofa}
	\and
	Physikalisches Institut, Universit\"at Bern, Gesellschaftsstrasse 6, 3012 Bern, Switzerland\label{unibe}
	\and
	Universit\"ats-Sternwarte M\"unchen, Ludwig-Maximilians-Universit\"at M\"unchen, Scheinerstra\ss{}e 1, 81679 M\"unchen, Germany \\
	\email{emsenhuber@usm.lmu.edu} \label{lmu}
	\and
	Max-Planck-Institut f\"ur Astronomie, K\"onigstuhl 17, 69117 Heidelberg, Germany\label{mpia}
}

\date{Received 1 June 2020 / Accepted 6 July 2021}

\abstract
{The explosion of observational data on exoplanets gives many constraints on theoretical models of planet formation and evolution. Observational data probe very large areas of the parameter space and many different planet properties.}
{Comparing theoretical models with observations allows one to take a key step forward towards understanding planetary systems. It however requires a model able to (i) predict all the necessary observable quantities (not only masses and orbits, but also radii, luminosities, magnitudes, or evaporation rates) and (ii) address the large range in relevant planetary masses (from Mars mass to super-Jupiters) and distances (from stellar-grazing to wide orbits).}
{We have developed a combined global end-to-end planetary formation and evolution model, the Generation III \textit{Bern} model, based on the core accretion paradigm. This model solves as directly as possible the underlying differential equations for the structure and evolution of the gas disc, the dynamical state of the planetesimals, the internal structure of the planets yielding their planetesimal and gas accretion rates, disc-driven orbital migration, and the gravitational interaction of concurrently forming planets via a full \textit{N}-body calculation. Importantly, the model also follows the long-term evolution of the planets on gigayear timescales after formation including the effects of cooling and contraction, atmospheric escape, bloating, and stellar tides.}
{To test the model, we compared it with classical scenarios of Solar System formation. For the terrestrial planets, we find that we obtain a giant impact phase of protoplanet-protoplanet collisions provided enough embryos ($\sim\num{100}$) are initially emplaced in the disc. For the giant planets, we find that Jupiter-mass planets must accrete their core shortly before the dispersal of the gas disc to prevent strong inward migration that would bring them to the inner edge of the disc. Regarding the emergence of entire planetary systems, many aspects can be understood with the comparison of the timescales of growth and migration, the capture into resonances, and the consequences of large-scale dynamical instabilities caused by the gravitational interactions of protoplanets, including the situation when a second core starts runaway gas accretion.}
{The Generation III \textit{Bern} model provides one of the most comprehensive global end-to-end models of planetary system formation and evolution developed so far, linking a multitude of crucial physical processes self-consistently. The model can form planetary systems with a wide range of properties. We find that systems with only terrestrial planets are often well-ordered (in period, mass, and radius), while giant-planet bearing systems show no such similarity. In a series of papers, the model will be used to perform extensive planetary population syntheses, putting the current theoretical understanding of planet formation and evolution to the observational test.}


\begin{document}
\maketitle

\section{Introduction}

Since the discovery of the first exoplanet detected around a main sequence star \citep{1995NatureMayorQueloz}, the number of known exoplanets has greatly increased. These planets span a wide range of masses and sizes, and they were detected using various techniques, such as radial velocity, transit, direct imaging, and microlensing. Despite all these observational constraints, the exact formation pathways are not yet certain. To highlight this, we first discuss possible formation mechanisms for different planet kinds.

Giant planets have been found orbiting their host star over a wide range of periods. Some are of the order of days or tens of days, which is well within the orbit of Mercury \citep{2011MayorArxiv,2014ApJFabrycky}; others were detected at large separations using the direct imaging technique \citep{2008ScienceMarois,2010ScienceLagrange,2013ApJRameau,2015ScienceMacintosh,2017A&AChauvin,2018AAKeppler}.

Most giant planets are thought to form via the core accretion mechanism as gravitational instability \citep{1997ScienceBoss,2003ApJBoss} is found to work only at large separation (several tens of astronomical units, \citealp{2005ApJRafikov,2021AASchib}), though the clumps could migrate after formation \citep{2010MNRASNayakshin}, and for bodies above about \SI{5}{\mj} \citep{2018ApJSchlaufman} or even the deuterium-burning limit \citep{2010ApJKratter}. On the other extreme, for very close-in giant planets, core accretion \citep{1974IcarusPerriCameron,1980PThPhMizuno} requires for in situ formation a very strong pile-up of solids. While this has been proposed \citep{2016ApJBoley,2016ApJBatygin,2018ApJLBaileyBatygin}, the possibility remains heavily debated. A scenario where these planets formed further out and were subsequently moved to their final location \citep{1996NatureLin} is usually considered more likely.

In the standard view, giant planets form from embryos located beyond the ice line, where solids are abundant owing to the volatiles being present in the solid form. This allows the embryo to form rapidly enough before the dispersal of the gas disc, which occurs in a time frame of several million years \citep{2001ApJHaisch,2010AAFedele,2018MNRASRichert}. Embryos initially accrete solids and a small quantity of gas. The further growth results in the accretion of gas, which is governed by the ability of the planet to radiate away the accretion energy \citep{1996IcarusPollack,2015ApJLeeChiang}. The cooling process becomes more efficient as the mass increase, so that when the planet reaches a mass of several times that of the Earth, the amount of solids and gas are equal (the critical mass, \citealp{1982PSSStevenson}). Once the accretion rate becomes greater than what the disc is able to supply, the envelope can no longer remain in equilibrium with the surrounding nebula and it contracts.

This process is further complicated by the implications of planetary migration \citep{2014PPVIBaruteau,2016SSRvBaruteau}. The final mass and location of the planet depends thus on the interplay between growth and migration, not to mention the interactions with the other planets forming in the same system.

Observations show that the giant planets are divided into two sub-groups depending on the host-star metallicity \citep{2013ApJDawson,2018ApJBuchhave}. Hot-Jupiters around metal-poor stars exhibit lower stellar obliquity and eccentricity than the ones around metal-rich stars. The usual concept of inward migration due to interaction with the gas disc \citep{1979ApJGoldreichTremaine,1997IcarusWard,2002ApJTanaka} cannot account for the obliquity of these planets, which more likely were brought there by few-body interactions combined with tidal circularisation \citep{2018ARAADawson}.

For the distant giant planets, core accretion is still favoured \citep{2019ApJWagner}. A possible formation pathway for some of these distant planets is accretion in the inner region of the disc followed by close encounters and scattering \citep{2019AAMarleau}. This pathway is supported by evidence that it is able to reproduce the distribution of eccentricities of giant planets \citep{2008ApJChatterjee,2008ApJJuric,2008ApJFordRasio,2010ApJRaymond,2017AASotiriadis}, and that most giant planet-harbouring system are multiple \citep{2014ApJKnutson,2016ApJBryan,2019ApJWagner}.

Exoplanets include planets unknown in the solar system, those between the Earth and Neptune \citep{2011MayorArxiv,2011ApJYoudinB,2012ApJSHoward}. The density of these planets vary more than one order of magnitude \citep{2015ApJHatzesRauer,2020AAOtegi}.

Sub-Neptunes exhibit a low bulk density, indicating the presence of a gaseous envelope \citep{2014ApJWeissMarcy,2015ApJRogers}. This implies that they mostly formed in a time scale comparable with the lifetime of the protoplanetary disc. However, whether they formed early (in the same way as the core of giant planets) or towards the end of the disc \citep{2016ApJLeeChiang} is not yet settled.

Super-Earths on the other hand are compatible with being gas-free. They are not constrained by the lifetime of the protoplanetary disc and can form over longer periods of time \citep{2019AALambrechts,2018AAOgihara}. These could also have had a gaseous envelope in the past that was removed by, for instance, atmospheric escape \citep[e.g.][]{2014ApJJin} or giant impacts \citep{2018SSRvSchlichting}.

Multi-planetary systems provide additional information. Many super-Earth systems have similar mass and spacing \citep{2017ApJMillholland,2018AJWeiss}, though this is debated \citep{2020AJZhu,2020ApJWeissPetigura}. However, most of the planet pairs are out of mean-motion resonances \citep{2014ApJFabrycky}. The low number of planets in mean-motion resonances (MMR) may be surprising, as gas-driven migration is efficient at capturing the planets in MMRs. But it is possible for the resonances to be broken during the retreat of the gas disc \citep{2017AALiu} or after the dispersal by dynamical instabilities \citep{2016ApJInamdar,2017MNRASIzidoro,2021AAIzidoro}.
The mutual inclinations remain relatively low \citep{2011ApJSLissauer,2012ApJFangMargot} and they exhibit low-to-moderate eccentricities \citep{2016PNASXie,2019AJMills}.

Many models have been developed to capture the above-mentioned effects. We may cite \citet{1996IcarusPollack}, \citet{2004ApJIda1,2004ApJIda2}, \citet{2004A&AAlibert,2005A&AAlibert}, \citet{2011MNRASMiguel}, \citet{2014MNRASColemanNelson}, \citet{2016MNRASCridland,2017MNRASCridland}, \citet{2017AALiu}, \citet{2017AAOrmel}, \citet{2017MNRASRonco}, \citet{2018MNRASNdugu}, \citet{2018ApJChambers}, \citet{2018MNRASAlessi}, \citet{2019AABitschA}, \citet{2019AAJohansen}, \citet{2019MNRASBooth}, or \citet{2020MNRASAlessiA,2020MNRASAlessiB}, to mention only a few.
To capture all of the above effects, models of planetary formation must include many physical processes occurring during the formation of the systems, which lead to feedbacks and non-linearities. Then, comparing the outcomes of global end-to-end models with observations is a key step for the understanding of the origins and evolution of planets systems.

As the model has many parameters, a large number of planets with different properties are required to constrain their possible values. The model must then be able to predict all the necessary observable quantities for the different observational techniques, not only masses and distances, but also radii (for transits), luminosities, magnitudes (for direct imaging), and evaporation rates. To leverage the enormous amount of statistical observational data on exoplanets, the models should also be able to make quantitative predictions which can be compared statistically with the actual planetary population. For this, planetary population synthesis \citep{2004ApJIda1,2009A&AMordasinia} is a frequently used approach.

In this work, we introduce a strongly improved and extended version the \textit{Bern} global end-to-end model of planetary formation and evolution for multi-planetary systems. The model combines the work of \citet{2013A&AAlibert}, hereafter \citetalias{2013A&AAlibert} (inclusion of \textit{N}-body interactions) and the internal structure calculations and long-term thermodynamical evolution model of \citet{2012A&AMordasiniB,2012A&AMordasiniC}. Here, we track the planets with full \textit{N}-body interactions, in contrast to \citet{2010ApJIdaLin} for instance, who introduced a semi-analytical approach, an improvement over previous works, such as \citet{2004ApJIda1}, to follow planet-planet interactions. The model follows the formation of many embryos, as it usually obtained from the end stage of the runaway growth of solids \citep{1998IcarusKokubo}, so that both terrestrial as well as giant planetary systems can be obtained.

The structure of this work is as follows: in the first part, we describe our global model. In Sect.~\ref{sec:model}, we introduce the new version of our model with a general overview of its conception, along with its relationship to previous work. Detailed description are provided in Sect.~\ref{sec:disc} for the stellar and nebular components, in Sect.~\ref{sec:planet} for the planets, and in Sect.~\ref{sec:dyn} for the migration and dynamical evolution. In a second part we perform some tests for different kind of planets and show possible resulting systems. In Sect.~\ref{sec:terrestrial}, we aim at reproducing formation of terrestrial planets with our improved model to determine whether it is applicable to kind of planets. The formation of giant planets and the implications for Jupiter are discussed in Sect.~\ref{sec:giants}. Finally, in Sect.~\ref{sec:results}, we apply the presented model to specific systems to assess the interaction between the different mechanisms occurring during the formation and evolution of planetary systems.

This work is the first of a series of several. In a companion paper, \citet[][hereafter referred to as \papertwo]{NGPPS2}, we will use this model to compute synthetic populations of planetary systems and perform statistical analysis. In subsequent articles, we will perform more detailed comparison with observations, and analyse various parameters that we have in the present model.

\section{The \textit{Bern} model}
\label{sec:model}

\subsection{History}
\label{sec:model-hist}

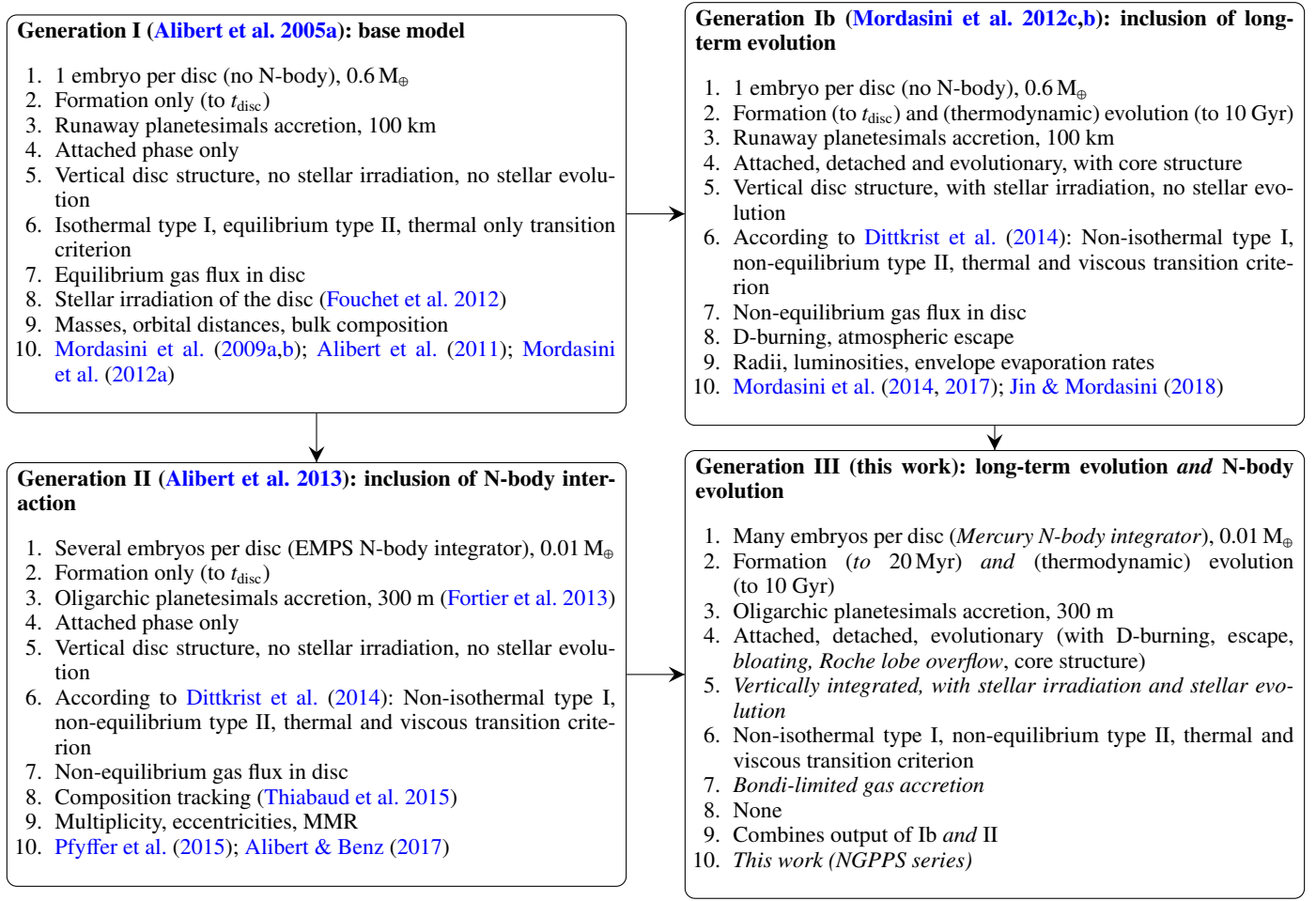
\begin{figure*}
	{\small
	\underline{Physical mechanisms and base assumptions included in all model generations}
	\begin{itemize}
		\setlength\itemsep{0em}
		\setlength\parskip{0pt}
		\item Formation paradigm: core accretion
		\item Protoplanetary disc model: solution of 1D evolution equation for gas surface density in an axissymetric constant $\alpha$-disc with photoevaporation
		\item Solid accretion: rate equation (Safronov-type) from planetesimals of a single size; planetesimals are represented by a solid surface density with dynamical state
		\item Gas accretion and planet interior structure: from solving 1D radially symmetric hydrostatic planet interior structure equations
		\item Orbital migration: gas disc-driven, types~I and~II
	\end{itemize}
	{\centering
		\underline{Evolution of physical mechanisms considered in various model generations}
	}

	\begin{tikzpicture}
	\node [rectangle, draw=black, align=flush center, rounded corners, text width=0.46\textwidth] (genone) at (0,0) {	\parbox[t]{0.99\textwidth}{
			\textbf{Generation I \citep{2005A&AAlibert}: base model}
			\begin{enumerate}
			\setlength\itemsep{0em}
			\setlength\parskip{0pt}
			\item 1 embryo per disc (no N-body), $\SI{0.6}{\mearth}$
			\item Formation only (to $t_\mathrm{disc}$)
			\item Runaway planetesimals accretion, 100 km
			\item Attached phase only
			\item Vertical disc structure, no stellar irradiation, no stellar evolution
			\item Isothermal type I, equilibrium type II, thermal only transition criterion
			\item Equilibrium gas flux in disc
			\item Stellar irradiation of the disc \citep{2012A&AFouchet}
			\item Masses, orbital distances, bulk composition
			\item \citet{2009A&AMordasinia,2009A&AMordasinib,2011A&AAlibert,2012A&AMordasiniA}
			\end{enumerate}
	}};
	\node [rectangle, draw=black, align=flush center, rounded corners, text width=0.46\textwidth] (genoneb) at (9.5,0) {\parbox[t]{0.99\textwidth}{
			\textbf{Generation Ib \citep{2012A&AMordasiniB,2012A&AMordasiniC}: inclusion of long-term evolution}
			\begin{enumerate}
			\setlength\itemsep{0em}
			\setlength\parskip{0pt}
			\item 1 embryo per disc (no N-body), $\SI{0.6}{\mearth}$
			\item Formation (to $t_\mathrm{disc}$) and (thermodynamic) evolution (to~10~Gyr)
			\item Runaway planetesimals accretion, 100 km
			\item Attached, detached and evolutionary, with core structure
			\item Vertical disc structure, with stellar irradiation, no stellar evolution
			\item According to \citet{2014A&ADittkrist}: Non-isothermal type I, non-equilibrium type II, thermal and viscous transition criterion
			\item Non-equilibrium gas flux in disc
			\item D-burning, atmospheric escape
			\item Radii, luminosities, envelope evaporation rates
			\item \citet{2014AAMordasiniA,2017A&AMordasini,2018ApJJin}
			\end{enumerate}
			}};
	\node [rectangle, draw=black, align=flush center, rounded corners, text width=0.46\textwidth] (gentwo) at (0,-6.5) {\parbox[t]{0.99\textwidth}{
		\textbf{Generation II \citep{2013A&AAlibert}: inclusion of N-body interaction}
		\begin{enumerate}
		\setlength\itemsep{0em}
		\setlength\parskip{0pt}
		\item Several embryos per disc (EMPS N-body integrator), $\SI{0.01}{\mearth}$
		\item Formation only (to $t_\mathrm{disc}$)
		\item Oligarchic planetesimals accretion, 300 m \citep{2013A&AFortier}
		\item Attached phase only
		\item Vertical disc structure, no stellar irradiation, no stellar evolution
		\item According to \citet{2014A&ADittkrist}: Non-isothermal type I, non-equilibrium type II, thermal and viscous transition criterion
		\item Non-equilibrium gas flux in disc
		\item Composition tracking \citep{2015A&AThiabaud}
		\item Multiplicity, eccentricities, MMR
		\item \citet{2015A&APfyffer,2017A&AAlibert}
		\end{enumerate}
		}};
	\node [rectangle, draw=black, align=flush center, rounded corners, text width=0.46\textwidth] (gentwob) at (9.5,-6.5) {\parbox[t]{0.99\textwidth}{
		\textbf{Generation III (this work): long-term evolution \textit{and} N-body evolution}
		\begin{enumerate}
		\setlength\itemsep{0em}
		\setlength\parskip{0pt}
		\item Many embryos per disc (\textit{Mercury N-body integrator}), $\SI{0.01}{\mearth}$
		\item Formation (\textit{to \SI{20}{\mega\year}}) \textit{and} (thermodynamic) evolution (to~10~Gyr)
		\item Oligarchic planetesimals accretion, 300 m
		\item Attached, detached, evolutionary (with D-burning, escape, \textit{bloating, Roche lobe overflow}, core structure)
		\item \textit{Vertically integrated, with stellar irradiation and stellar evolution}
		\item Non-isothermal type I, non-equilibrium type II, thermal and viscous transition criterion
		\item \textit{Bondi-limited gas accretion}
		\item None
		\item Combines output of Ib \textit{and} II
		\item \textit{This work (NGPPS series)}
		\end{enumerate}
	}};

	\draw[decoration={markings,mark=at position 1 with {\arrow[scale=2,>=stealth]{>}}},postaction={decorate}] (genone) -- (genoneb);
	\draw[decoration={markings,mark=at position 1 with {\arrow[scale=2,>=stealth]{>}}},postaction={decorate}] (genone) -- (gentwo);
	\draw[decoration={markings,mark=at position 1 with {\arrow[scale=2,>=stealth]{>}}},postaction={decorate}] (genoneb) -- (gentwob);
	\draw[decoration={markings,mark=at position 1 with {\arrow[scale=2,>=stealth]{>}}},postaction={decorate}] (gentwo) -- (gentwob);

	\end{tikzpicture}
	}

	\caption{Overview of the physical mechanisms included in the \textit{Bern} model. At the top, the processes and base assumptions made in all model generations are given. The four boxes below show the four model generations with the main paper introducing each generation. The further points are: (1) number of initial embryos per disc, N-body integrator type, initial embryo mass, (2) phases simulated, (3) planetesimal accretion mode and planetesimal size, (4) phases with calculation of the planets’ internal structure, (5) disc model characteristics, (6) orbital migration: type I, type II, transition criterion from type I to II (here thermal refers to a criterion only with the ratio between the Hill radius and the scale height of the disc, while `thermal and viscous' refers to the full criterion of \citealt{2006IcarusCrida}), (7) disc-limited gas accretion rate, (8) later additions and improvements, (9) additional output relative to older generation, (10) population synthesis publications using this generation. In the bottom right panel, text in italic indicates new elements.}
	\label{fig:ModelGens}
\end{figure*}

The model presented in this work, the Generation~III \textit{Bern} model, combines the formation and evolution stage of planetary system. It is based on many contributions in the field that aim to study different aspects of the physics of planetary formation and evolution. We thus start by a short history of the series of model, and its different branches that we couple together in this work. A graphical sketch of the different generations of the \textit{Bern} model is provided in Fig.~\ref{fig:ModelGens}.

The original model was introduced in \citet{2004A&AAlibert,2005A&AAlibert} for individual planets, then used in \citet{2009A&AMordasinia,2009A&AMordasinib} for entire planetary populations. We refer to it as Generation I, which computed the formation on a single planet until the gas disc disperses. The model subsequently diverged into two different branches: one with the aim to follow the long-term evolution of the formed planet \citep[Generation Ib;][]{2012A&AMordasiniB,2012A&AMordasiniC} while the other obtained the ability to form multi-planetary systems with an improved description of the planetesimals disc \citep[Generation II;][]{2013A&AAlibert,2013A&AFortier}. In this work we bring these two variants of the model back together so that we can follow the formation and the long-term evolution of multi-planetary systems. At the same time, we extend the model with new elements, which are shown in italic on Fig.~\ref{fig:ModelGens}.

Previous versions of the model have been extensively described in referenced papers (see also \citealp{2014PPVIBenz,2015IJAsBMordasini,2018BookMordasini} for reviews and the interactions between the different mechanisms involved in planet population syntheses). We nevertheless describe this new version in the remainder of this section.

\subsection{General description}
\label{sec:model-desc}

\begin{figure*}
\centering
\includegraphics{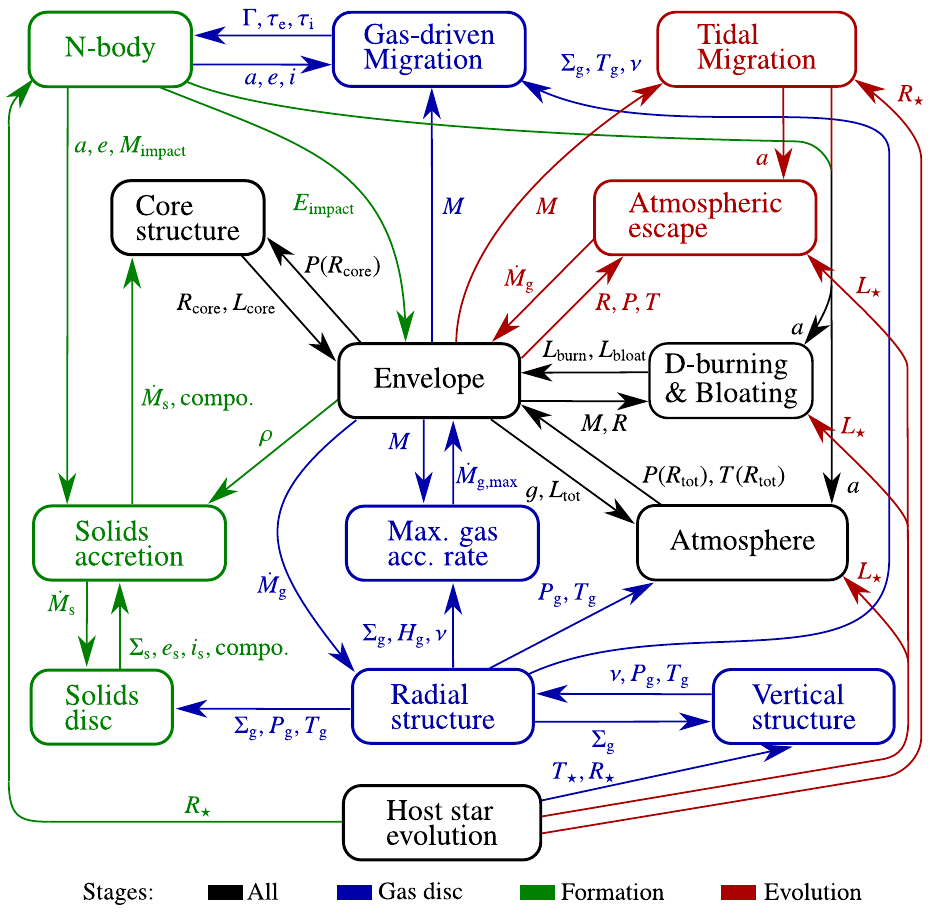}
\caption{Sub-modules and most important exchanged quantities of the Generation III \textit{Bern} model. The colours denote the stages at which processes are considered. Blue indicates processes active in the formation stage, but only before the dispersal of the gas disc. Green processes are considered during the entire formation stage, even after the dispersal of the gas disc. Processes in red are only considered during the evolution stage. The processes in black are included in all stages.}
\label{fig:modules}
\end{figure*}

We base our study on the \textit{Bern} model of planetary formation and evolution. This global model self-consistently computes the evolution of the gas and planetesimals discs, the accretion of gas and solids by the protoplanets, their internal and atmospheric structure, as well as interactions between the protoplanets and between the gas disc and the protoplanets. We provide a diagram of the main components of the overall model as well as the most important exchanged quantities in Fig.~\ref{fig:modules}.

In our coupled formation and evolution model, we first model the planets' main formation phase for a fixed time interval (set to \SI{20}{\mega\year}, see the related discussion in Section~\ref{sec:terrestrial} regarding the impact of this specific choice). Afterwards, in the evolutionary phase, we follow the thermodynamical evolution of each planet individually to \SI{10}{\giga\year}.

\subsubsection{Formation phase}
\label{sec:model-form}

During the formation stage (0--\SI{20}{\mega\year}), the model follows the evolution of a gaseous protoplanetary disc and the dynamical state of planetesimals (Section~\ref{sec:disc}). These serve as sources for the accretion of the protoplanets (Section~\ref{sec:planet}). The lifetime of the gas disc is shorter than the simulated formation stage, so that solids accretion in a gas-free environment can also take place. The gas disc also leads to planetary migration, and interactions (scattering, collisions) between the concurrently growing proptoplanets are tracked via a \textit{N}-body integrator (Section~\ref{sec:dyn}).

The formation of planets is based on the core accretion paradigm \citep{1980PThPhMizuno,1996IcarusPollack}: First, a solid core is formed, and once it becomes massive enough, it starts to bind a significant H/He envelope. Core growth results from the accretion of planetesimals. Gas accretion is initially governed by the ability of the planet to radiate away the energy released by the accretion of both solids and gas. Once the gas accretion rate of the envelope exceeds the limit from the disc, the envelope can no longer maintain equilibrium with the disc; hence it subsequently contracts and passes into the detached phase. \citep{2000IcarusBodenheimer}.

\subsubsection{Evolution phase}
\label{sec:model-evo}
The long-term evolution of the planets (\SI{20}{\mega\year} - \SI{10}{\giga\year}) is calculated by solving, like already in the formation phase, the standard spherically symmetric internal structure equations, but with different boundary conditions, and taking into account different physical effects like atmospheric escape, or radius inflation. In this phase, the planets evolve individually; \textit{N}-body interactions and the accretion of planetesimals are no more considered. The orbits and masses of the planets may however still evolve because of effects like tides and atmospheric escape.

As described in \citet{2012A&AMordasiniB}, the coupling between the formation and evolution phases is made self-consistently, that is both the compositional information as well as the gravothermal heat content given by the formation model are given to the evolution model as initial conditions.

Regarding the temporal evolution, we now also take the thermal energy content of the planet's core into account for a planet's luminosity, as described in \citet{2019AALinder}. This is important for core-dominated low-mass planets \citep[e.g.][]{2014ApJLopezFortney}. As in previous calculations, the other gravothermal energy sources are the cooling and contraction of the H/He envelope, the contraction of the core, and radiogenic heating due to the presence of long-lived radionuclides in the core.

\subsubsection{Envelope structure}
\label{sec:model-struct}

The calculation of the internal structure of all planets (Section~\ref{sec:planet}) during their entire formation and evolution is a crucial aspect of the Bern Model, as visible from its central position in Figure \ref{fig:modules}. It not only yields the planetary gas accretion rate in the attached phase but is also key for the accretion of planetesimals via the drag enhanced capture radius. It also yields the radii and luminosities that on one hand enter multiple other sub-modules, and on the other hand are key observable quantities. The internal structure model assumes that planets have an onion-like spherically symmetric structure with an iron core, a silicate mantle, and depending of a planet's accretion history, a water ice layer and a gaseous envelope made of pure H/He. In contrast to earlier syntheses predicting planetary radii \citep{2012A&AMordasiniB,2014AAMordasiniA}, we now use self-consistently the iron mass fraction as given by the disc compositional model (according to \citealp{2014AAThiabaud}, Section~\ref{sec:disc-compo}), instead of assuming a fixed 2:1 silicate:iron mass ratio.

Physical effects that are included in the model besides the usual cooling and contraction are XUV-driven atmospheric escape \citep{2014ApJJin}, D-burning \citep{2012A&AMolliere}, Roche-lobe overflow, and bloating of the close-in planets \citep{2021AASarkis}.

Compared to some other 1D internal structures models in the literature \citep[e.g.][]{2013MNRASVazan,2016AAVenturini,2020ApJVallettaHelled}, the model is simplified first by assuming that the gaseous envelope consists of pure H/He, while accreted solids sink to the core. In this sense, the model is similar to the original \citet{1996IcarusPollack} model. We neglect thus the consequences of heavy element enrichment and compositional gradients in the envelope. These effects will be added in future work. One should note that also other modern models make use of the simplification of pure H/He envelopes \citep{2021IcarusDAngelo}. Including enrichment would generally speed up gas giant formation \citep{2016AAVenturini}. Second, the effects of hydrodynamic flows affecting the (upper) envelope structure and cooling behaviour are currently also neglected \citep{2020MNRASali-dib,2021AAMoldenhauer}.

On the other hand, our internal structure model allows to model the entire `life' of planets from $t=0$ to \SI{10}{\giga\year}, modelling and coupling self-consistently all phases (attached, detached, evolutionary), for both the gaseous envelope and the solid core. Importantly, the model is capable of calculating the internal structure and temporal evolution of planets ranging in mass from \SI{e-2}{\mearth} to the lithium-burning limit (about 63 Jovian masses, \citealp{2001RMPBurrows}). It includes besides the standard aspects (accretion, cooling, contraction) also atmospheric escape, bloating, Roche-lobe overflow, and deuterium burning. In particular, this makes it possible to model planets that reside very close to their host star. This quite unique general applicability to very different planet types reflects the needs arising from a population synthesis calculation.

As shown in Figure \ref{fig:modules}, atmospheric escape is only included in the evolution phase starting at \SI{20}{\mega\year}. In reality, it would start immediately once the gas disc has dissipated and the planets start to `see' the stellar XUV irradiation. This could lead to a certain under-estimation of atmospheric escape. The consequences should, however, be small, since escape continues to be important for at least the first \SI{100}{\mega\year} when stars are in the saturated phase of high XUV emission, and not only for the first \SI{20}{\mega\year}. The effect that atmospheric escape can destabilise resonant chains for sufficiently high mass loss \citep{2020ApJMatsumotoOgihara} is thus not included. On the other hand, we include during the entire formation phase (also after gas disc dissipation) the accretion of planetesimals, which also changes planet masses.

In the following sections, we describe in detail all the sub-modules visible in Figure \ref{fig:modules}.

\section{Star and protoplanetary disc}
\label{sec:disc}

\subsection{Stellar model}

Instead of assuming a fixed \SI{1}{\lsun} stellar luminosity for a \SI{1}{\msun} star as in previous model generations, stellar evolution is now considered by incorporating the stellar evolution tracks from \citet{2015AABaraffe}. These provide the radius $\rstar$, luminosity $\lstar$ and temperature $\tstar$ for a given stellar mass $\mstar$ at any moment. Stellar temperature and radius are used for the outer boundary conditions of the gas disc; stellar radius is also used in the \textit{N}-body integrator to detect collisions with the star and to calculate the stellar tidal migration. Finally, the stellar irradiation enters into the calculation of the outer (atmospheric) temperature (at $\tau=2/3$) of the planets' interior structure as described in \citet{2012A&AMordasiniB} and radius bloating (Sect.~\ref{sec:bloat}).

\subsection{Gas disc}

The protoplanetary gas disc is modelled with a 1D radial axisymetric structure. The evolution is given by solving the viscous diffusion equation as function of the time $t$ and orbital distance $r$ \citep{1952ZNatALust,1974NMRASLyndenBellPringle},
\begin{equation}
	\frac{\partial\sigmag}{\partial t}=\frac{1}{r}\frac{\partial}{\partial r}\left[3r^{1/2}\frac{\partial}{\partial r}\left(r^{1/2}\nu\sigmag\right)\right]-\sigmadotgphoto-\sigmadotgplan,
\end{equation}
where $\sigmag=\int_{-\infty}^\infty\rho\mathrm{d}z$ is the surface density of gas, and $\sigmadotgphoto$ and $\sigmadotgplan$ are the sink terms related to photo-evaporation (Section~\ref{sec:photo}) and accretion by the planets respectively. The viscosity is parametrised, following \citet{1973A&AShakuraSunyaev}, with
\begin{equation}
\nu=\alpha c_s H
\label{eq:nu}
\end{equation}

This equation is solved on a grid spaced regularly in $\log$ with 3400 points that extends from the inner location of the disc $\rin$ (an initial condition) to $\rmax=\SI{1000}{\au}$. At these two locations, the surface density is fixed to zero.

\subsubsection{Vertical structure}

The disc's vertical structure is computed at each step of the evolution following the approach of \citet{1994ApJNakamoto}. This change is necessary to accommodate the new stellar model with variable quantities. With this approach, the link between the outer and midplane temperatures is given by
\begin{equation}
	\sigmaSB\tmid^4=\frac{1}{2}\left(\frac{3}{8}\tau_R+\frac{1}{2\tau_P}\right)\dot{E}+\sigmaSB\tsurf^4
\end{equation}
with $\tmid$ the disc mid-plane temperature, $\tsurf$ the temperature due to irradiation (see below), $\sigmaSB$ the Stefan-Boltzmann constant, $\tau_R$ and $\tau_P$ are the Rosseland and Planck mean optical depths respectively, and $\dot{E}$ is the viscous dissipation rate. This formula yields the mid-plane temperature both in the optically-thick (the term with $\tau_R$) and optically-thin (the term with $\tau_P$) regimes.
The Rosseland optical depth is given by $\tau_R=\kappa_\mathrm{disc}(\rhomid,\tmid)\Sigma$ where $\rhomid=\Sigma/(\sqrt{2\pi}H)$ is the central density, $H=c_s/\Omega$ the disc's vertical scale height, $c_s=\sqrt{\kB\tmid/(\mu m_\mathrm{H})}$ the isothermal sound speed, $\mu=2.24$ the mean molecular weight of the gas, and $m_\mathrm{H}$ the mass of an hydrogen atom. The opacity $\kappa_\mathrm{disc}$ is given by the maximum of the opacities computed according to \citet{1994ApJBell} (which accounts for micrometre size with a fixed interstellar dust-to-gas ratio of \SI{1}{\percent}, independently of the dust-to-gas ratio chosen for the solids disc) and \citet{2014ApJSFreedman} (which gives molecular opacities for a grain-free gas). For the Planck optical depth, we follow further \citet{1994ApJNakamoto} and set $\tau_P=2.4\tau_R$.

It is clear that this treatment of the opacities is simplified: in reality, the evolution of the dust via coagulation, fragmentation, and drift influences via the resulting grain opacity the thermal and density structure of the disc. This structure in turn feeds back onto the dust evolution, meaning that the processes must be treated together in a self-consistently coupled way \citep{2015ApJGorti,2020AASavvidou}.

Such a more realistic coupled model affects for example the disc lifetime, the local dust-to-gas ratio \citep{2015ApJGorti}, or -- in the context of planet formation -- the locations of the outward migration zones \citep[][see Sect. \ref{sec:dyn-mig-map}]{2020AASavvidou}. They also show that the ratio of Planck and Rosseland opacity is in reality not simply a constant as currently assumed. In \citet{2020AAVoelkel} we have recently coupled the \citet{2012A&ABirnstiel} dust-pebble evolution model to the Bern Model. Based on this, future version of the Bern Model will include also a more physically realistic grain opacity and therefore disc structure model. This will in particular also include the dependency of the disc opacity on the stellar metallicity, which is currently not taken into account.

In equilibrium, the radiative flux is identical as the viscous dissipation rate, which is given by
\begin{equation}
	\dot{E}=\Sigma\nu\left(r\frac{\partial\Omega}{\partial r}\right)^2=\frac{9}{4}\Sigma\nu\Omega^2
\end{equation}
with $\Omega$ being the Keplerian angular frequency at distance $r$ from the star. The second equality holds only if purely the mass of the central star is accounted for in the Keplerian frequency, that is $\Omega=\sqrt{\ggrav\mstar/r^3}$, $\ggrav$ being the gravitational constant. The self-gravity of the disc has been neglected.

The disc's outer temperature due to irradiation is given by
\begin{equation}
	\tsurf^4=\tstar^4\left[\frac{2}{3\pi}\left(\frac{\rstar}{r}\right)^3+\frac{1}{2}\left(\frac{\rstar}{r}\right)^2\frac{H}{r}\left(\frac{\partial\ln{H}}{\partial\ln{r}}-1\right)\right]+\tirr^4+\tcd^4
\end{equation}
following \citet{2005A&AHueso}, but also accounting for the direct irradiation through the disc's mid plane. The first term inside the bracket is the irradiation of the star onto a flat disc. The second term in the square brackets accounts for the flaring of the disc at large separation. In our case, we do not compute this factor explicitly and instead adopt $\partial\ln{H}/\partial\ln{r}=9/7$ \citep{1997ApJChiang}.

The $\tirr$ term accounts for the direct irradiation through the disc midplane. It is computed as
\begin{equation}
\tirr^4 = \frac{\lstar}{16 \pi r^2 \sigmaSB} e^{-\tau_\mathrm{mid}},
\label{eq:tirr}
\end{equation}
which is the black-body equilibrium temperature accounting for the optical depth through the midplane of the disc $\tau_\mathrm{mid}=\int\rhomid\kappa(\rhomid,\tmid)\mathrm{d}r$. This contribution is usually important only at the very end of the disc lifetime while it clears; otherwise, the optical depth confines the contribution to the very innermost region. However, taking this contribution in account is necessary to provide a smooth transition of the temperature at the surface of the planets (see Sec.~\ref{sec:struct}) from the time when they are embedded in the nebula to the time when they are exposed to the direct stellar irradiation.

The last term accounts for the heating by the surrounding environment (molecular cloud), which we set constant to $\tcd=\SI{10}{\kelvin}$. We thus neglect possible variations of this background temperature depending on the stellar cluster environment in which a star and its planetary system are born \citep{2006ApJKrumholz,2018MNRASNdugu}. On the other hand, different cluster environments and thus different levels of the interstellar FUV field \citep{2008ApJFatuzzoAdams} are taken into account by varying in the population syntheses (see \papertwo) the magnitude of the external photoevaporation rate $\mwind$. External photoevaporation is likely the most important environment-related factor for discs \citep{2020MNRASWinter}.

\subsubsection{Disc photoevaporation}
\label{sec:photo}

\begin{figure*}
\centering
\includegraphics{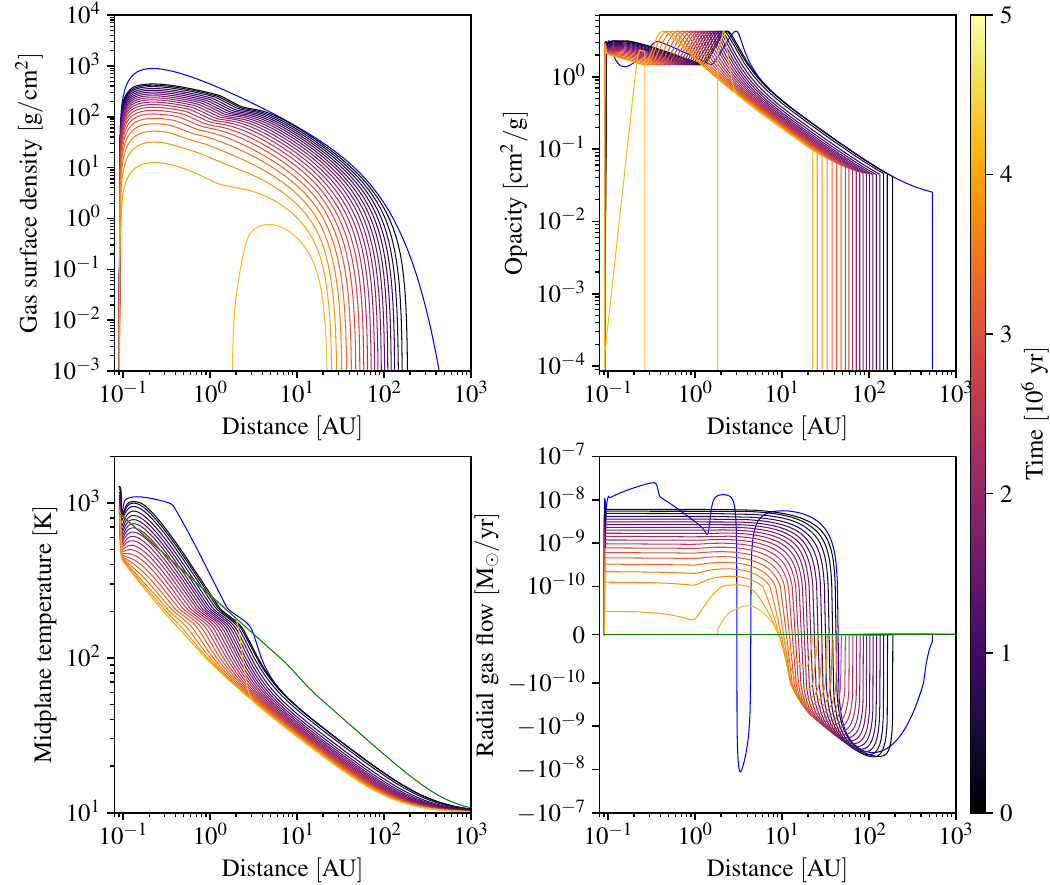}
\caption{Time evolution of the surface density (\textit{top left}), opacity $\kappa$ (\textit{top right}), midplane temperature (\textit{bottom left}), and radial flow rate (\textit{bottom right}) of a protoplanetary disc. The lines represent each one snapshot the state, and are spaced by about \SI{2e5}{\year}. The blue line in both panels shows the initial profile, which has not yet been evolved at all, and is therefore not in equilibrium. The green line in the temperature profile shows the profile at disc's dispersal, which is given by the equilibrium temperature with the host star's luminosity.}
\label{fig:gas_disc_str}
\end{figure*}

\begin{table*}
    \centering
	\caption{Initial conditions and parameters for the example system. The upper part contains the gas disc properties, the middle part the planetesimals disc properties, and the bottom part show planetary embryos properties.}
	\label{tab:params-ex}
	\begin{tabular}{ll}
			\hline
			Quantity & Value \\
			\hline
			Stellar mass $\mstar$ & \SI{1}{M_\odot} \\
			Reference surface density $\sigmanorm$ at \SI{5.2}{\au} & \SI{145}{\gram\per\square\centi\meter} \\
			Initial gas disc mass $\mgas$ & \SI{3.90E-2}{\msun} \\
			Inner edge of the gas disc $\rin$ & \SI{0.091}{\au} (\SI{10}{\day}) \\
			Characteristic radius of the gas disc $\rcutg$ & \SI{66.5}{\au} \\
			Disc viscosity parameter $\alpha$ & \num{2e-3} \\
			External photoevaporation rate $\mwind$ & \SI{6.42E-7}{\msun\per\year} \\
			Power law index of the gas disc $\betag$ & 0.9 \\
			\hline
			Dust-to-gas ratio & \SI{3.4}{\percent} \\
			Planetesimal disc mass & \SI{348}{\mearth} \\
			Power law index of the solids disc $\betas$ & 1.5 \\
			Characteristic radius of the solids disc $\rcuts$ & $\rcutg/2$ \\
			Planetesimal radius & \SI{300}{\meter} \\
			Planetesimal density (rocky) & \SI{3.2}{\gram\per\cubic\centi\meter} \\
			Planetesimal density (icy) & \SI{1}{\gram\per\cubic\centi\meter} \\
			\hline
			Embryo mass $\mstart$ & \SI{1e-2}{\mearth} \\
			Opacity reduction factor $\fopa$ & \num{3e-3} \\
			\hline
	\end{tabular}
\end{table*}

Photoevaporation in the protoplanetary discs is the principal means of controlling their lifetimes. For the prescription, we follow \citet{2012A&AMordasiniC}. In this scheme, we include contributions from both internal (due the host start itself) and external (due to nearby massive stars in the birthplace of the system) sources.

For the external photo-evaporation, we use the far-ultraviolet (FUV) description of \citet{2003ApJMatsuyama}. FUV radiation (6--\SI{13.6}{eV}) creates a neutral layer of dissociated hydrogen whose temperature is $T_\mathrm{I}\approx\SI{e3}{\kelvin}$. The corresponding sound speed is then
\begin{equation}
c^2_\mathrm{s,I} = \frac{\kB T_\mathrm{I}}{\mu_\mathrm{I}m_\mathrm{H}},
\label{eq:csphev}
\end{equation}
where the mean molecular weight $\mu_\mathrm{I}=1.35$ for the dissociated gas. It corresponds to the gravitational radius (where the sound speed equals the escape velocity) of
\begin{equation}
r_\mathrm{g,I} = \frac{\ggrav\mstar}{c^2_\mathrm{s,I}}.
\label{eq:rgrav}
\end{equation}
We assume that mass is removed uniformly outside of $\beta_\mathrm{I}r_\mathrm{g,I}$ with $\beta_\mathrm{I}=0.14$ \citep[similar to][]{2012MNRASAlexanderPascucci}, so that the rate is given by
\begin{equation}
\sigmadotgphotoext(r)=\left\{
\begin{array}{ll}
    0 & \text{for}~r<\beta_\mathrm{I}r_\mathrm{g,I} \\
    \frac{\mwind}{\pi(\rmax^2-\beta^2_\mathrm{I}r^2_\mathrm{g,I})} & \text{otherwise}
\end{array}
\right.
\end{equation}
with $\mwind$ a parameter that provides the total mass loss rate if the disc would extend to $\rmax=\SI{1000}{\au}$. In practice however, the actual mass loss rate due to external photoevaporation is clearly smaller than that parameter, as the disc does not extend up to $\rmax$, but to a dynamically obtained radius which results from the interplay of viscous spreading (increasing the outer radius) and external photoevaporation (decreasing the outer radius).

For the internal photoevaporation, we follow \citet{2001MNRASClarke}, which in turn is based on `weak stellar wind' case of \citep{1994ApJHollenbach}. Here, extreme-ultraviolet (EUV; $>\SI{13.6}{eV}$) creates a layer of ionised hydrogen whose temperature is $T_\mathrm{II}\approx\SI{e4}{\kelvin}$ and with a mean molecular weight $\mu_\mathrm{II}=0.68$. The sound speed and gravitational radius are computed in analogy with Eqs.~(\ref{eq:csphev}) and~(\ref{eq:rgrav}). The scaling radius $r_{14}=\beta_\mathrm{II}r_\mathrm{g,II}/\SI{e14}{\centi\meter}$ follows \citet{2001MNRASClarke} while we select again $\beta_\mathrm{II}=0.14$ following \citet{2012MNRASAlexanderPascucci}. With this, we can estimate the base density with
\begin{equation}
n_0(r_{14})=k_\mathrm{Hol}\Phi_{41}^{1/2}r_{14}^{-3/2},
\end{equation}
where we set $k_\mathrm{Hol}=\num{5.7e4}$ following the hydrodynamical simulations of \citet{1994ApJHollenbach} and $\Phi_{41}=0.1 \sqrt{M_\star/M_\odot}$, which is the ionising photon luminosity in the units of $\SI{e41}{\per\second}$. The distance-dependent base density can then be calculated as
\begin{equation}
n_0(r)=n_0(r_{14})\left(\frac{r}{r_\mathrm{g,II}}\right)^{-\frac{5}{2}} .
\end{equation}
We further follow \citet{2001MNRASClarke} to get $\sigmadotgphotoint = 2 c_\mathrm{s,II} n_0 m_\mathrm{H}$ outside of $\beta_\mathrm{II} r_\mathrm{g,II}$.

The final photoevaporation rate is given by the sum of the effects of host star + nearby massive stars with
\begin{equation}
\sigmadotgphoto = \sigmadotgphotoext + \sigmadotgphotoint.
\end{equation}

\subsubsection{Initial gas surface density profile and example}

We initialise the gas surface density profile with \citep{2004MNRASVerasArmitage}
\begin{equation}
\sigmag(t=0) = \sigmanorm\left(\frac{r}{\rnorm}\right)^{-\betag}\exp{\left(-\left(\frac{r}{\rcutg}\right)^{2-\betag}\right)}\left(1-\sqrt{\frac{\rin}{r}}\right)
\end{equation}
where $\rnorm=\SI{5.2}{\au}$ is the reference distance, $\betag=0.9$ the power-law index \citep{2010ApJAndrews}, $\rcutg$ the characteristic radius for the exponential decay and $\rin$ the inner edge of the disc.

The conversion between the total mass and the normalisation surface density $\sigmanorm$ at $\rnorm$ is obtained with
\begin{equation}
\mgas = \frac{2\pi\sigmanorm}{2-\betag}\left(\rnorm\right)^{\betag}\left(\rcutg\right)^{2-\betag}.
\end{equation}
It should be noted that this formula neglects the lack of gas within $\rin$, but since the total mass is dominated by the outer disc as we have a shallow power-law, there has in practice very limited effect.

An example of evolution of such as disc, without any planets (i.e. $\sigmadotgplan=0$), is provided in Fig.~\ref{fig:gas_disc_str}. The initial conditions and parameter are provided in Table~\ref{tab:params-ex} (note that the table also contains planetesimals disc properties that are not used here). The lifetime of that disc is nearly \SI{5.3e6}{yr}.

The temporal evolution shows overall a decrease in the surface density. A hole forms inside roughly \SI{2}{\au} by about \SI{4.7}{\mega\year}. The change in the temperature profile initially between 1.5 and \SI{3}{\au} and that moves inwards is due to a maximum in the opacity (top right panel). This different behaviour is reflected in the surface density as the temperature affects the sound speed, hence the viscosity.

For the midplane temperature, the direct irradiation term is only important in the innermost region (within about \SI{0.2}{\au}) until a few hundred thousand years before the dispersal of the gas disc. The last profile before dispersal shows an increase of temperature within \SI{0.2}{\au} due to this contribution; otherwise the midplane temperature remains below the equilibrium temperature, apart from the inner region (<\SI{<3}{\au}) at early times.

We compared the results obtained here with prescriptions from other works, as for instance \citet{2015AABitschA}. We find that in general, for a given stellar accretion rate (which is the parametrisation of the \citealp{2015AABitschA} prescription) we obtain lower surface density profiles by \SI{30}{\percent} coupled with larger temperature by \num{20}--\SI{40}{\percent}. The two models cannot be very well compared directly due to the different underlying assumptions, like constant radial flow rate in \citet{2015AABitschA}. Our models accounts for the full evolutionary equation for the surface density including photoevaporation and gas accretion by the protoplanets, which means we have the radial flow rate varying with distance (bottom right panel of Fig.~\ref{fig:gas_disc_str}, where the radially constant inflow in the inner disc, and the viscous spreading (outflow) in the outer disc can be seen). There are other model assumptions that result in the differences between the surface density and temperature in the two models: 1) the stellar luminosity, which in our case it starts with roughly \SI{3}{\lsun} as predicted by the \citet{2015AABaraffe} tracks whereas \citet{2015AABitschA} begins with \SI{1.5}{\lsun} following \citet{1998AABaraffe}, 2) the opacity which affects the relation between midplane and disc photospheric temperature, and 3) the different approach of including stellar irradiation (vertically integrated assuming an equilibrium for the flaring angle versus an explicit 1D vertical structure with radiative transfer).

\subsection{Planetesimal disc}
\label{sec:solids disc}

Planetesimals are represented by a fluid-like description, that is they are modelled not as individual particles but on a grid as a surface density ($\sigmasol$) with eccentricity ($\eplan$) and inclination ($\iplan$) as dynamical state.

\subsubsection{Dynamical state}

For the time evolution of the dynamical state, we use the approach of \citet{2013A&AFortier} and explicitly solve the differential equations describing the change of eccentricity and inclination. In this framework, these are stirred by both the protoplanets, and to a lesser extent the other planetesimals, and damped by drag from the gas disc. The equations for the root mean square (RMS) of the planetesimals' eccentricity $e_\mathrm{plan}$ and inclination $i_\mathrm{plan}$ read as
\begin{eqnarray}
\esqdotplan & = & \left.\esqdotplan\right|_\mathrm{drag} + \left.\esqdotplan\right|_\mathrm{VS,M} + \left.\esqdotplan\right|_\mathrm{VS,plan} \\
\isqdotplan & = & \left.\isqdotplan\right|_\mathrm{drag} + \left.\isqdotplan\right|_\mathrm{VS,M} + \left.\isqdotplan\right|_\mathrm{VS,plan}.
\end{eqnarray}
The contributions from the aerodynamical drag, stirring by the protoplanets and the planetesimals are denoted by `drag', `VS,M' and `VS,plan' respectively. The dynamical state is followed during the entire formation stage. The drag term is only evaluated while the gas disc is still present. After the dissipation of the gas disc, the term is set to 0.

The form of the drag term depends on the regime: Epstein, Stokes (laminar) or quadratic (turbulent). The distinction between those regimes is made using the criterion proposed by \citet{2004AJRafikov} using the molecular Reynolds number $\mathrm{Re}_\mathrm{mol}=\vrel\rplan/\nu_\mathrm{mol}$, where $\nu_\mathrm{mol}=\lambda c_\mathrm{s}/3$ is the molecular viscosity, $\lambda=(n_\mathrm{H_2}\sigma_\mathrm{H_2})^{-1}$ the gas molecules' mean free path, $n_\mathrm{H_2}$ the number density assuming all of the gaseous molecules having hydrogen mass, $\sigma_\mathrm{H_2}$ their collisional cross-section, $\rplan$ the planetesimals' radius,
\begin{equation}
\vrel=\vkep\sqrt{\eta^2+5/8\eplan^2+1/2\iplan^2}
\end{equation}
their relative velocity,
\begin{equation}
\eta=-\frac{1}{2\Omega r \rhomid}\frac{\partial p}{\partial r}
\end{equation}
the deviation between the gas and Keplerian velocities due the support of the gas by the radial pressure gradient, $\rhomid$ the midplane gas density, and $\vkep=\Omega r$ the Keplerian velocity. When $\rplan<\lambda$, the gas drag is assumed to be in the Epstein regime. Otherwise, if $\mathrm{Re}_\mathrm{mol}>20$, the gas drag is taken to be in the quadratic (or turbulent) regime and in the Stokes regime if not.

The expressions for the drag in the quadratic regimes are \citep{1976PThPhAdachi,2006IcarusChambers},
\begin{eqnarray}
\left.\esqdotplan\right|_\mathrm{drag} & = & -\frac{\eplan^2}{\tau_\mathrm{drag}}\sqrt{\eta^2+\frac{5}{8}\eplan^2+\frac{1}{2}\iplan^2} \\
\left.\isqdotplan\right|_\mathrm{drag} & = & -\frac{\iplan^2}{2\tau_\mathrm{drag}}\sqrt{\eta^2+\frac{5}{8}\eplan^2+\frac{1}{2}\iplan^2},
\end{eqnarray}
where
\begin{equation}
\tau_\mathrm{drag}=\frac{8\rhoplan\rplan}{3C_\mathrm{D}\rhomid\vkep}
\end{equation}
is the gas drag time scale and $C_\mathrm{D}=1$.

In the Stokes regimes the drag expressions are
\begin{eqnarray}
\left.\esqdotplan\right|_\mathrm{drag} & = & -\frac{3}{2}\frac{\lambda\rhomid\eplan^2}{\rhoplan\rplan^2} \\
\left.\isqdotplan\right|_\mathrm{drag} & = & -\frac{3}{4}\frac{\lambda\rhomid\iplan^2}{\rhoplan\rplan^2},
\end{eqnarray}
while in the Epstein regime they read as
\begin{eqnarray}
\left.\esqdotplan\right|_\mathrm{drag} & = & -\eplan^2\frac{c_\mathrm{s}\rhomid}{\rhoplan\rplan} \\
\left.\isqdotplan\right|_\mathrm{drag} & = & -\frac{\iplan^2}{2}\frac{c_\mathrm{s}\rhomid}{\rhoplan\rplan}
\end{eqnarray}
\citep{1976PThPhAdachi,2004AJRafikov,2013A&AFortier}. We also want to point out that we do not model the formation of gap in the gas disc by giant planets. This means that drag in the vicinity of such planets might be overestimated, resulting in lower eccentricities and inclination. As consequence, the accretion rate of planetesimals would be overestimated in this stage, which affects the heavy element contents of the planets.

As in \citet{2013A&AFortier}, the stirring by the protoplanets follows the approach of \citet{2010AAGuilera}, where the stirring of \citet{2002IcarusOhtsuki} is modulated with the separation from the protoplanets. The contribution reads as
\begin{eqnarray}
\left.\esqdotplan\right|_\mathrm{VS,M} & = & \sum_{\jmath=1}^{n}f_{\Delta,\jmath}\left(\frac{\Omega\mplanetj}{6\pi b\mstar}\right)P_\mathrm{VS} \\
\left.\isqdotplan\right|_\mathrm{VS,M} & = & \sum_{\jmath=1}^{n}f_{\Delta,\jmath}\left(\frac{\Omega\mplanetj}{6\pi b\mstar}\right)Q_\mathrm{VS}
\end{eqnarray}
where the sum is over all the protoplanets present in the system,
\begin{equation}
f_{\Delta,\jmath}^{-1} = 1+\left(\frac{|r-\aplanetj|}{b\rhill}\right)^5
\end{equation}
is the modulation due to separation so that the perturbation is effectively restricted to the planet's feeding zone,
\begin{equation}
\rhill=\aplanetj\sqrt[3]{\frac{\mplanetj}{3\mstar}}
\label{eq:rhill}
\end{equation}
the planet's Hills radius, and $b = 5$ is the half-width of the feeding zone (see Sect.~\ref{sec:fz}). The terms $P_\mathrm{VS}$ and $Q_\mathrm{VS}$ are given by \citep{2013A&AFortier},
\begin{eqnarray}
P_\mathrm{VS} & = & \left[\frac{73\eplanh^2}{10\Lambda^2}\right]\ln{\left(1+10\Lambda^2/\eplanh^2\right)} \nonumber\\
& & +\left[\frac{72 I_\mathrm{PVS}(\beta)}{\pi\eplanh\iplanh}\right]\ln{\left(1+\Lambda^2\right)} \\
Q_\mathrm{VS} & = & \left[\frac{4\iplanh^2+0.2\iplanh\eplanh^3}{10\Lambda^2\eplanh}\right]\ln{\left(1+10\Lambda^2\eplanh^2\right)} \nonumber\\
& & +\left[\frac{72 I_\mathrm{QVS}(\beta)}{\pi\eplanh\iplanh}\right]\ln{\left(1+\Lambda^2\right)}. \\
\end{eqnarray}
Here, $\eplanh=r\eplan/\rhill$ and $\iplanh=r\iplan/\rhill$ are respectively the reduced planetesimals' eccentricity and inclination, $\Lambda=\iplanh(\eplanh^2+\iplanh^2)/12$, $\beta=\iplan/\eplan$, while for $I_\mathrm{PVS}$ and $I_\mathrm{QVS}$ we use the approximations obtained by \citet{2006IcarusChambers}:
\begin{eqnarray}
I_\mathrm{PVS}(\beta) & \simeq & \frac{\beta-0.36251}{0.061547+0.16112\beta+0.054473\beta^2}, \\
I_\mathrm{QVS}(\beta) & \simeq & \frac{0.71946-\beta}{0.21239+0.49764\beta+0.14369\beta^2}.
\end{eqnarray}

The stirring by the other planetesimals is given by, following \citet{2002IcarusOhtsuki},
\begin{eqnarray}
\left.\esqdotplan\right|_\mathrm{VS,plan} & = & \frac{1}{6}\sqrt{\frac{\ggrav r}{\mstar}}\sigmasol\hplan P_\mathrm{VS} \\
\left.\isqdotplan\right|_\mathrm{VS,plan} & = & \frac{1}{6}\sqrt{\frac{\ggrav r}{\mstar}}\sigmasol\hplan Q_\mathrm{VS}
\end{eqnarray}
with
\begin{equation}
\hplan=\sqrt[3]{\frac{2\mplan}{3\mstar}},
\end{equation}
and $\mplan=4/3\pi\rplan^3\rhoplan$, the mass of a planetesimal.

To set the initial dynamical state, we assume that the disc is initially in a cold state, that is only the self-stirring of the planetesimals contributes to their eccentricities and inclinations. In other words, this assumes that the embryos appear instantly at the beginning of the simulation. The equilibrium values can be derived by equating the contributions of self-stirring and damping \citep{2003IcarusThommes,2006IcarusChambers}, which results in
\begin{equation}
\eplan = 2.31\frac{\mplan^{4/15}r^{1/5}\rhoplan^{2/15}\sigmag^{1/5}}{C_D^{1/5}\rho^{1/5}\mstar^{2/5}}
\end{equation}
and
\begin{equation}
\iplan=\frac{1}{2}\eplan.
\end{equation}

We also compared our prescription for the dynamical state with gamma-stirring from, for instance, \citet{2008ApJIda} and \citet{2013ApJOkuzumiOrmel}. Although this is not straightforward due to the differences in the sources, we find that, generally, the eccentricities resulting from $\gamma$-stirring are larger than the self-stirring from the planetesimals, but lower than the stirring by the forming protoplanets. Thus, accounting for the stirring of planetesimals by turbulent diffusion in the disc would increase their eccentricities at locations far away from growing protoplanets. Close to the growing protoplanets however, where the planetesimals' eccentricities are important for the solids accretion rate, neglecting this effect does not significantly affect planetesimals' eccentricities.

\subsubsection{Size, initial surface density profile, and evolution}

To roughly take into account the observational \citep[e.g.][]{2018ApJAnsdell} and theoretical \citep[e.g.][]{2014ApJBirnstiel} finding that solids have a more concentrated distribution than the gas, the initial surface density profile of planetesimals now follows a steeper slope than the one of the gas disc \citep{2019ApJLenz,2020AAVoelkel}. This leads to a higher concentration of solids in the inner part of the disc.

As already in the Generation II Model \citep{2013A&AAlibert}, we assume a constant planetesimals radius of \SI{300}{\meter} throughout the disc, which is a strong assumption and simplification. There is an ongoing discussion about the characteristic primordial planetesimal size in the literature. Observations of extrasolar debris belts \citep{2021MNRASKrivov}, the presence of hypervolatile ices in comets that can only be preserved in impacts involving small bodies \citep{2021IcarusGolabekJutzi}, direct size determinations by stellar occultations \citep{2019NatAsArimatsu} and some theoretical studies \citep{2009ApJFraser,2013AJSchlichting} suggest small (\SI{\sim1}{\kilo\meter}) characteristic planetesimals sizes. On the other hand, the absence of small craters on Pluto \citep{2019SciSinger}, the size distribution in the asteroid belt \citep{2009IcarusMorbidelli}, and the theoretical predictions of planetesimal formation models (e.g. \citealt{2020ApJKlahr}) rather point at \SI{\sim100}{\kilo\meter} planetesimals. The first two points can, however, also be explained with other effects \citep[][although the former work makes no determination about the initial size frequency distribution of planetesimals]{2017ApJZheng,2018ApJWei}.

In the more specific context of the simulations presented here, this choice was made for the following reasons: 1) small planetesimals undergo sufficient eccentricity and inclination damping by the disc gas to sustain a planetesimal accretion rate in the oligarchic growth regime that is high enough to build giant planet cores during typical disc lifetimes \citep{2013A&AFortier}. We note that the Generation I and Ib Bern Models assumed in contrast runaway planetesimal accretion as \citet{1996IcarusPollack}. In the runaway regime, the eccentricities and inclinations of the planetesimals are assumed to remain low even without damping by the disc gas. Therefore, fast core growth occurred in these models also with 100 km planetesimals, which was the assumed size in these early model generations. 2) their drift time scales are longer than typical lifetimes of gas discs \citep{2019AABurn} and 3) this size was shown to be able to reproduce several of the known exoplanet properties across a wide range of masses \citep{2013A&AFortier}. In any case, the constant planetesimal size is an important limitation of the model. Including in the Bern Model an explicit model for the evolution of the solid building blocks across the entire size range (dust-pebble-planetesimals) is thus subject of ongoing research. A first important step was recently made in \citet{2020AAVoelkel} where we have coupled the dust-and-pebble model of \citet{2012A&ABirnstiel} and the planetesimal formation model of \citet{2019ApJLenz} to our global model. These effects are, however, not yet included in the Generation III Model presented here.

To set the initial surface density profile of planetesimals, we thus use a slightly different description than for the gas, that is,
\begin{equation}
	\Sigma_s(t=0)=\sigmas0f_\mathrm{s}(r)\left(\frac{r}{r_0}\right)^{-\betas}\exp{\left(-\left(\frac{r}{\rcuts}\right)^{2}\right)}
	\label{eq:sigmas0}
\end{equation}
with the power-law exponent is set to $\betas=1.5$, as in the MMSN, and $\rcuts=\rcutg/2$ is the exponential cutoff radius of the solids, set half the value of the gas disc following \citet{2018ApJAnsdell}. This formula also enables us to model relatively sharp outer edges of the solids disc \citep{2014ApJBirnstiel}.

The reference surface density value $\sigmas0$ is adjusted so that the bulk solids-to-gas ratio remains to the prescribed value (e.g. \SI{1}{\percent}).

The surface density of planetesimals is reduced by accretion onto and ejection by the protoplanets to ensure mass conservation (see Sect.~\ref{sec:model-core}), or removed entirely if $\eplan^2>0.95$. Our model only includes ejection (Sect.~\ref{sec:model-pla-ejec}) and not scattering by the forming planets. Thus, we do not have redistribution of material to other regions of the disc by planets, as obtained by \citet{2017IcarusRaymondIzidoro} for instance. Finally, the planetesimals disc remains after the dispersal of the gas disc; the only difference is that the damping terms for eccentricity $\left.\esqdotplan\right|_\mathrm{drag}$ and inclination $\left.\isqdotplan\right|_\mathrm{drag}$ vanish.

\subsubsection{Compositional model}
\label{sec:disc-compo}

\begin{figure*}
\centering
\includegraphics{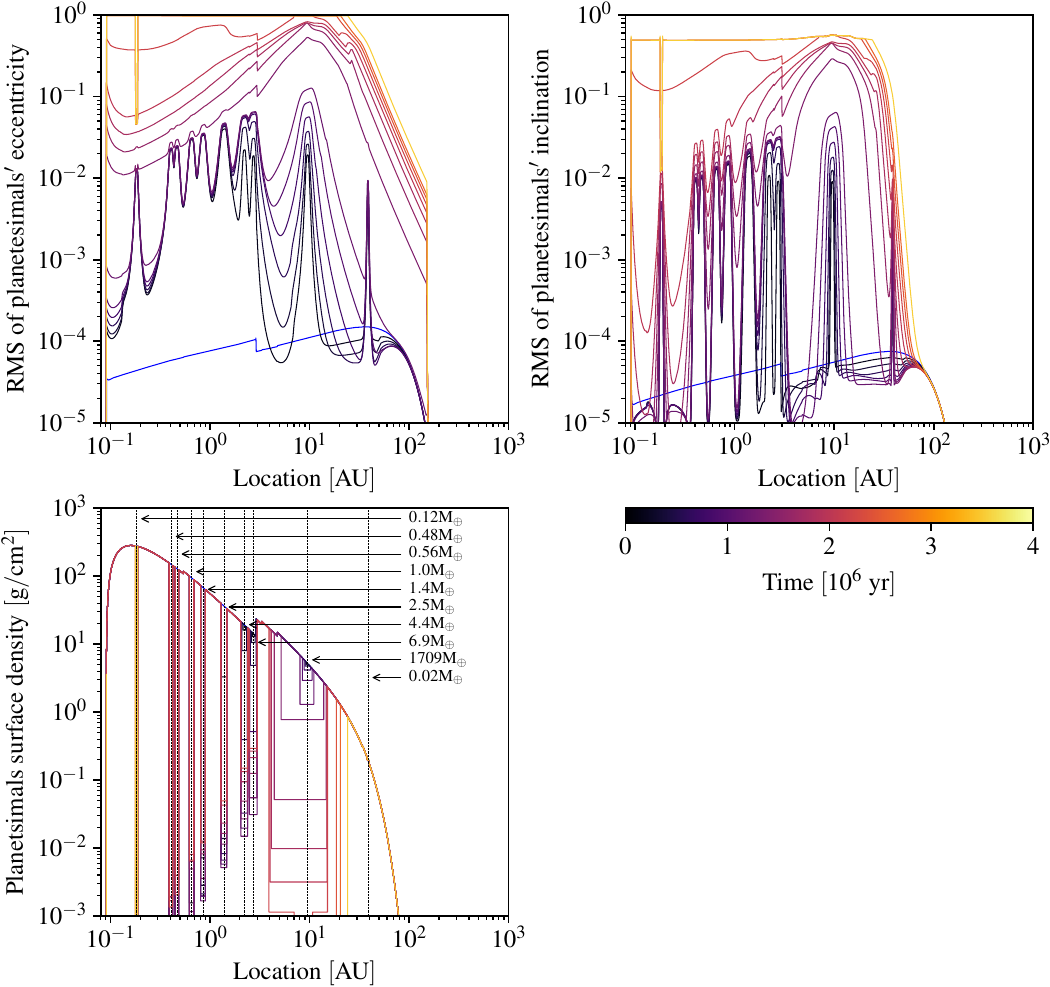}
\caption{Time evolution of the RMS of planetesimals' eccentricity (\textit{top left}), inclination (\textit{top right}), and surface density (\textit{bottom left}) of a circumstellar disc that also contains 10 embryos. The lines represent temporal snapshots of the three quantities, and are spaced by about $\SI{2e5}{yr}$. The blue line in both panels denote the initial profile. The dashed vertical lines represent the location of the embryos, which is fixed in this case. \textit{N}-body interactions were also disabled. The lifetime of the gas disc is shorter than the case presented in Fig.~\ref{fig:gas_disc_str} due to the accretion by the protoplanets.}
\label{fig:plan_disc_str}
\end{figure*}

The Bern model includes the simple condensation model of \citet{2014AAThiabaud} and \citet{2014AAMarboeufB}. The initial abundance of volatile and refractory species is identical to the one given in \citet{2014AAMarboeufA}. Volatile species are composed of H, O, C, and S atoms whose abundance reflect solar composition \citep{2003ApJLodders}. The relative abundances of the molecules are set according to interstellar medium. Then at each location in the disc at $t=0$, we check whether each molecule is the solid or gas phase assuming local thermodynamical equilibrium. This yields the fraction of heavy elements that is locally condensed and thus contributes to the solid surface density (the ice line locations), and the chemical composition. This composition is tracked into the protoplanets when a propotoplanet accretes planetesimals, and in giant impacts between protoplanets. This yields in particular the final iron to silicate ratio and the volatile mass fraction of all the planets.

The factor $f_\mathrm{s}(r)$ in Eq.~(\ref{eq:sigmas0}) for the initial planetesimal surface density accounts for the mass fraction of all elements that are in the solid phase at a given location. To compute its value, we use the aforementioned condensation model.
Only the contribution of molecules in the solid phase are accounted for the resulting solid surface density.
Thus, the value of $f_\mathrm{s}$ in the inner locations is the mass fraction of condensed to total solids and this value increases by small jumps each time an ice line is crossed until it becomes unity at large separation.

For the density of planetesimals, we assume that in the region where only refractory materials contributes to the solid phase $\rho_\mathrm{plan}=\SI{3.2}{\gram\per\cubic\centi\meter}$ while when volatiles are in the solid phase we take $\rho_\mathrm{plan}=\SI{1}{\gram\per\cubic\centi\meter}$. This transition corresponds to the H\textsubscript{2}O-ice line in all discs, which induces the largest surface density jump because H\textsubscript{2}O makes up $\sim$\SI{60}{\percent} of all ices in mass \citep{2014AAMarboeufB}.

\subsubsection{Example}

An example of the dynamical state of planetesimals is provided in Fig.~\ref{fig:plan_disc_str}. The initial conditions and parameters are provided in Table~\ref{tab:params-ex}. This is the same initial disc as shown in Fig.~\ref{fig:gas_disc_str}, except than ten embryos were added to the disc, at the locations shown by the dashed vertical lines. In addition, both migration and \textit{N}-body interactions were artificially disabled so that the embryos remain at the same location throughout the simulation.

The different jumps in the initial surface density profile are due to the crossing of the different ice lines; the most consequential one at at about \SI{3}{\au} is due to the water-ice line. The surface density of planetesimals is equalised inside the feeding zone of each planet. It should be noted that besides this effect, we do not include planetesimals redistribution, as was found by, for instance, \citet{2010AJLevison}. In total, the planets accreted \SI{61}{\mearth} of planetesimals (\SI{47}{\mearth} of which by the giant planet) while \SI{89}{\mearth} were ejected (according to the prescription detailed in Sect.~\ref{sec:model-pla-ejec}; virtually all of them by the giant planet). The feeding zones are all nearly depleted by the planets due to accretion, except for the giant planets where \SI{65}{\percent} of planetesimals were ejected and \SI{35}{\percent} accreted.

The stirring by the protoplanets heats the planetesimals in the surrounding region. This effect is heavily dependent on the protoplanet's masses; the most massive one is the second outermost one (close to \SI{10}{\au}), which reaches a mass of about \SI{5.4}{\mj} at the end of the formation stage. That planet has a core mass of \SI{47}{\mearth}, which corresponds (for a pure H/He envelope) to a metallicity slightly lower than that of the star (\SI{2.8}{\percent} versus \SI{3.0}{\percent}). This is below the relationship found by \citet{2016ApJThorngren} for the planet's mass. This, however, is not unexpected for the idealised setup used here: first, planets that form in the in-situ case tend to have lower core masses than planets that migrated \citep[e.g.][]{2005ApJAlibertB}. Second, with \textit{N}-body interactions switched off here, giant impacts otherwise increasing the heavy element content are not possible. In the more realistic example in Sect. \ref{sect:highinitialsolidcontent}, where these effects are included, giant impacts strongly increase the solid content of the giant planets by a factor 2-3 relative to value at the moment gas runaway begins. The impacts are themselves triggered by the fast mass growth, destabilising neighbouring lower mass protoplanets.

As noted by \citet{2013A&AFortier}, the usual assumption that $\beta=\iplan/\eplan\approx1/2$ does not hold. We find that the stirring of eccentricities takes place over larger separation to the protoplanets than for the inclinations. This can be seen for instance in the region affected by the most massive planet.

Further, the effect of the planet is not only limited to the surrounding area because of the following effect: the massive planet is able to significantly reduce the inward gas flow such that the region inside its orbit becomes gas-poor. This greatly reduces the damping of the planetesimals dynamical state to a such point that their eccentricity becomes close to unity.

\section{Planet properties}
\label{sec:planet}

\subsection{Envelope structure}
\label{sec:struct}

In the \textit{Bern} model, the internal structure of the planets (and thus their gas accretion rate, radius, luminosity, and interior structure) are found at all stages (attached, detached, evolution) by directly solving the 1D structure equations. In contrast, many other global models use in contrast approximations and fits to find for example the gas accretion rate \citep[see][]{2019AAAlibertVenturini}. While the 1D hydrostatic picture is also not the final word for low-mass planets because of multidimensional hydrodynamic effects \citep[e.g.][]{2015MNRASOrmelB,2017AALambrechtsLega,2017MNRASCimerman,2021AAMoldenhauer}, the fits (except the deep neural networks) often fail grossly to reproduce the result of 1D structure equations that they should in principle recover \citep{2019AAAlibertVenturini}. Many fits also neglect the influence of the luminosity on the gas accretion rate \citep[e.g.][]{2004ApJIda1,2015AABitschB}. In reality, there is an important interplay between solid accretion which is dominant for the luminosity at early stages, and gas accretion. This leads to important feedbacks that can only be captured when solving the internal structure equations \citep{2014A&ADittkrist}.

Also, from the point of view of guiding and interpreting astronomical observations, it is crucial to solve the internal structure equations, as this gives self-consistently at each moment in time the planet’s radius and luminosity and associated magnitudes. These are the observable quantities for transit and direct imaging surveys. By predicting them self-consistently, the output of the \textit{Bern} model can be compared in population syntheses not just with methods measuring quantities depending on the planets’ mass (like RV, astrometry or microlensing), but also transit and direct imaging surveys.

The downside is that solving the internal structure for bodies ranging in mass from \SI{e-2}{\mearth} to beyond the deuterium limit requires an internal structure model that is very versatile and numerically stable in all stages of planetary formation and evolution. Solving the internal structure also comes with significant computational cost.

\subsubsection{Attached phase}

In the initial phase, known as the attached phase, the envelope is in equilibrium with the gas disc and the gas density smoothly transitions from the value in the protoplanetary envelope to the one in the background nebula. The planets do not yet have a well-defined outer radius. During this phase, the gas accretion rate is governed by the ability to radiate the gravitational energy liberated by the accretion of solids and gas, and the envelope's contraction. For the forming giant planets, this phase generally lasts until the planets reach a total mass in the range of \num{30} to \SI{100}{\mearth} where envelope contraction becomes fast, depending on the conditions. There is no fixed mass boundary; the transition occurs when the gas accretion rate obtained from solving the internal structure equations (that is the envelope's Kelvin-Helmholtz contraction) becomes larger than the disc-limited rate (Sect. \ref{sect:maximumgasaccertetion}. For low-mass planets which have very low gas accretion rates (very long Kelvin-Helmholtz timescales), the attached phase lasts (almost) until the gas disc dissipates.

Gas accretion is calculated by solving the classical 1D radially symmetric internal structure equations \citep{1986IcarusBodenheimerPollack},
\begin{eqnarray}
\partial M/\partial R & = & 4 \pi R^2 \rho \label{eq:intstr-mass} \\
\partial P/\partial R & = & -\ggrav M \rho / R^2 \label{eq:intstr-pres} \\
\partial T/\partial R & = & \partial P/\partial R \min{(\nabla_\mathrm{ad},\nabla_\mathrm{rad})},
\label{eq:intstr-temp}
\end{eqnarray}
with $M$ the mass enclosed in the radius $R$, $P$ the pressure, $T$ the temperature, $\rho=\rho(P,T)$ the density, computed using the \texttt{SCvH} equation of state \citep{1995ApJSSaumon}, and $\nabla_\mathrm{ad}$ and $\nabla_\mathrm{rad}$ the adiabatic and radiative gradients respectively. The minimum of the two indexes is the Schwarzschild criterion \citep[e.g.][]{1994BookKippenhahnWeigert}, and is used to ensure stability against convection. The adiabatic gradient comes from the equation of state, while the radiative gradient is given by
\begin{equation}
\nabla_\mathrm{rad}=\frac{3 \kappa L}{64 \pi \sigmaSB \ggrav M T^3},
\end{equation}
with $L$ being the luminosity.

The opacity in the envelope $\kappa$ is obtained in similar way as for the gas disc, but following \citet{2014AAMordasiniA}, the interstellar medium (ISM) grain opacity contribution in \citet{1994ApJBell} is multiplied by a factor $f_\mathrm{opa}=0.003$. This value was found in \citet{2014AAMordasiniA} to fit best the detailed simulations by \citet{2008IcarusMovshovitzPodolak} and \citet{2010IcarusMovshovitz} of the grain dynamics in protoplanetary atmospheres (growth, settling) and the resulting dust opacities. Using one global reduction factor of the ISM opacity can of course not reproduce the full complex behaviour of the grain opacity which depends on planetary properties like the core or envelope mass as found in grain dynamics models \citep{2008IcarusMovshovitzPodolak}. But as shown in \citet{2014AAMordasiniA}, it still provides a useful first approximation. The value is not increased when a planetary system with higher metallicity is simulated. The reason is that a higher dust input in the outermost layer (as possibly associated with a high metallicity system) does not lead to a strong increase of the opacity. This was found numerically in \citep{2008IcarusMovshovitzPodolak} and explained analytically in \citet{2014AAMordasiniB}: a higher dust input leads to a higher dust-to-gas mass ratio (which increases the opacity), but also larger grains (which decreases the opacity). These effects cancel each other out in the dominating growth regime of differential settling.

The boundary conditions for the integration are taken as follows: the outer radius is given by, following \citet{2009IcarusLissauer},
\begin{equation}
\frac{1}{\rtot} = \frac{1}{k_1\racc}+\frac{1}{k_2\rhill},
\label{eq:rtotattached}
\end{equation}
where
\begin{equation}
\racc=\frac{\ggrav\mtot}{c_s^2}
\label{eq:bondi}
\end{equation}
is the Bondi radius, $\rhill$ is the Hill's radius (Eq.~\ref{eq:rhill}), $k_1=1$ and $k_2=1/4$. The pressure and temperature are derived from the local properties of the disc with
\begin{eqnarray}
P(\rtot) & = & P_\mathrm{neb}(\aplanet)\ \ \mathrm{and} \\
T^4(\rtot) & = & T_\mathrm{neb}^4(\aplanet) + \frac{3\tau_\mathrm{out}L(\rtot)}{8\pi\sigmaSB\rtot^2},
\end{eqnarray}
and
\begin{equation}
\tau_\mathrm{out}=\max{\left(\kappa(\rho_\mathrm{neb},T_\mathrm{neb})\rho_\mathrm{neb}\rtot,\frac{2}{3}\right)}
\end{equation}
being the optical depth at the surface of the planet \citep{2012A&AMordasiniB}, using the reduced opacities for the grains. The more complex parts come from the luminosity and the mass. The calculation of the outer luminosity $L(\rtot)$ is described in Section~\ref{sec:luminosity}. In the case of the mass, what is known is the core mass, that is $M(\rcore)=\mcore$, while $M(\rtot)=\mtot$ is the quantity that is being searched for. We thus use an iterative method by guessing $\mtot$, which is then used to integrate the internal structure equations until the boundary condition at the inner boundary is fulfilled, that is $M(\rcore)=\mcore$. Once $\mtot$ is found, the envelope mass can be retrieved by $\menv=\mtot-\mcore$, and the gas accretion rate by taking the difference of the envelope mass between two successive steps of the envelope structure calculation $\mdotenv=(\menv(t)-\menv(t-\Delta t))/\Delta t$.

\subsubsection{Maximum gas accretion rate}
\label{sect:maximumgasaccertetion}

In the initial stages, the gas accretion is limited by the planet's ability to radiate away the potential energy provided of the accretion material, that is the Kelvin-Helmholtz process. The rate at which gas can be accreted is set by the Kelvin-Helmholtz time scale,
\begin{equation}
\tKH = \frac{\ggrav\mtot^2}{\rtot\ltot}.
\end{equation}
However, as the planet's core reaches a mass of about $\SI{10}{\mearth}$, the value of $\tKH$ becomes so low that the planet undergo runaway gas accretion. In this phase, the amount of gas that the planet can accrete is constrained by the supply from the gas disc. Therefore, we compute the quantity $\dot{M}_\mathrm{env,max}$, which is used to limit the value of $\mdotenv$ found by solving the internal structure equations.

Our approach to compute the maximum rate is similar to \citet{2012A&AMordasiniB} but using only the `local reservoir' component. This a major difference from the previous versions of the \textit{Bern} model, where gas accretion was constrained from the radial flow of the gas. Following \citet{2008ApJDAngeloLubow} and \citet{2007ApJZhouLin}, we adopt a Bond- or Hill-like accretion in a region of size $\rgc$ around the planet. For simplicity, we compute $\rgc$ according to Eq.~(\ref{eq:rtotattached}). Depending on the value of $\rgc$ with respect to $H$, the local disc's scale height, two different regimes occur. In the case where $\rgc<H$, the planet will not accrete from the full vertical extent of the disc, and so the gas flow through the gas capture cross section $\sigma_\mathrm{cross}=\pi\rgc^2$ is given by
\begin{equation}
\mdotenvmaxv = \rho \sigma_\mathrm{cross} v_\mathrm{rel}
\end{equation}
with $\rho\approx\Sigma/H$ the approximate density of the gas and $v_\mathrm{rel}=\max{\left(\Omega\rtot,c_\mathrm{s}\right)}$ the relative velocity between the gas and the planet.

On the other hand, in the case $\rgc>H$, the planet will accrete from the whole gas column and the approximation of constant gas density breaks down. In this situation, the gas flow through the planet's capture radius is provided only by the radial extension of the gas capture area, hence we have
\begin{equation}
\mdotenvmaxr = 2 \rgc \Sigma v_\mathrm{rel}.
\end{equation}

To distinguish between the two regimes, we use the lower rate of the two, that is
\begin{equation}
\mdotenvmax = \min{\left(\mdotenvmaxr,\mdotenvmaxv\right)}
\end{equation}

Finally, to ensure that no more gas than available in the feeding zone $\mfeed$ is accreted during one time step, we further constrain $\mdotenvmax<\mfeed/\dt$. We consider the limiting case to be that gap formation does not reduce the planetary gas accretion rate. Such a situation arises if the eccentric instability \citep{2001AAPapaloizou,2006AAKleyDirksen} allows the planets to efficiently access disc material even after a gap has formed. For circular orbits, gap formation would in contrast strongly reduce the gas accretion rate \citep{1999ApJLubow,1999ApJBryden}, and limit planetary masses to $\sim5-\SI{10}{\mj}$.

\begin{figure}
	\centering
	\includegraphics{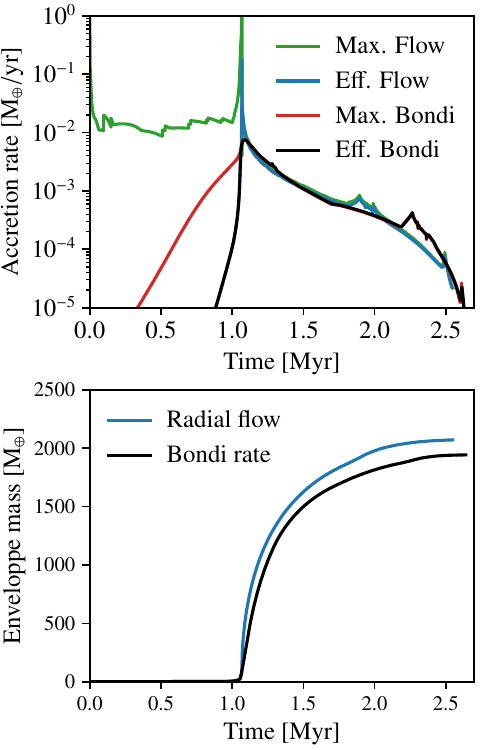}
	\caption{Comparison of two prescriptions for the maximum (i.e. disc limited) gas accretion rate, the one presented in this work (labelled `Bondi rate') with that of \citet{2012A&AMordasiniB} (labelled `Flow rate'). These are two different simulations (one for each prescription) whose initial conditions represent the second outermost planet in Fig.~\ref{fig:plan_disc_str}. \textit{Top panel:} Maximum value that can be supplied by the gas disc (labelled `Max.') and effective accretion rates (labelled `Eff.'), which is given by intrinsic cooling in the initial attached phase and the maximum rate in the detached phase. \textit{Bottom panel:} Corresponding enveloppe mass (i.e. total gas accreted).}
	\label{fig:mdotgas}
\end{figure}

The radial extent of the feeding zone is set by
\begin{eqnarray}
\rfeedinf & = & \aplanet\left(1-e\right)\left(1-f_\mathrm{feed}\sqrt{\frac{\mtot}{3\mstar}}\right) \\
\rfeedsup & = & \aplanet\left(1+e\right)\left(1+f_\mathrm{feed}\sqrt{\frac{\mtot}{3\mstar}}\right)
\end{eqnarray}
with $f_\mathrm{feed}=0.5$ so that the overall extent is a half a Hill radius larger than the radial excursion of the planet's orbit. This radial extent provides the location over which the disc properties ($\Sigma$, $H$, etc.) are averaged for the calculation of the maximum rate and the removal of the accreted gas, with
\begin{equation}
\sigmadotgplan=\frac{\mdotenv}{\pi\left(\rfeedsup^2-\rfeedinf^2\right)}.
\end{equation}
The planet's eccentricity consequently does not directly affect the maximum gas accretion rate, but only indirectly through the size of the feeding zone. The self-limitation of gas accretion by removal of local disc gas by the planet, which then needs to be replenished by the inflow from more distant disc regions (i.e. mass conservation) is fully taken into account in our scheme via the $\sigmadotgplan$ term entering the evolutionary equation of the disc gas surface density. We also take into account that for planets of any mass growing in multi-planet systems, the eccentricity can be increased via gravitational planet-planet interactions, which then affects the feeding zone width and thus indirectly the gas accretion rate. On the other hand, we currently do not take into account that the eccentric instability (i.e. the increase of a single giant's eccentricity because of gravitational interaction with the gas disc) in reality only acts for sufficiently massive planets \citep{2001AAPapaloizou,2006AAKleyDirksen}. This could lead to an overestimation of gas accretion at lower to intermediate giant planet masses. This could potentially explain why our current model of disc-limited gas accretion seems to too strongly reduce the stellar gas accretion rate \citep{2019AAManara,2020AABergez-Casalou}. Gap formation would reduce this effect, but could potentially lead to another issue: observationally, the giant planet mass function seems to extend smoothly to about \SI{30}{\mj} \citep{2011AASahlmann,2019GeoSciAdibekyan} (though \citealp{2017A&ASantosA} and \citealp{2018ApJSchlaufman} found a change in the metallicity dependency at about \num{4} to \SI{5}{\mj} and concluded that the planets above that threshold formed predominantly by gravitational instability). Reaching such high masses could be difficult given the expected reduction of gas accretion because of gap formation in the circular case.

The reduction of gas inflow into the inner disc because of an accreting giant planet can result in the clearing of the inner region of the protoplanetary disc by photoevaporation \citep{2013MNRASRosotti}. This effect is also automatically taken into account by our model.

To compare the prescription presented here with previous work, we provide in Fig.~\ref{fig:mdotgas} the comparison of the gas accretion rate for the second outermost planet from the case shown in Fig.~\ref{fig:plan_disc_str}. The previous methodology, using the radial gas flow and taking into account the geometry was described in \citet{2012A&AMordasiniB}, with a limit of 0.9 of the radial flow to allow some gas to flow through the gap \citep{2006ApJLubow}. The results show that using the Bondi rate, as we presented here, gives a somewhat stronger limitation of gas accretion by the forming planet, especially during the onset of the runaway gas accretion. As a result, the final planet's mass is a bit lower when using the Bondi rate.

\subsubsection{Detached phase}

Once the gas accretion rate exceeds the maximum that can be provided by the disc -- which includes the planet no longer being in a region where gas is present -- the accretion regimes changes to the detached phase \citep{2000IcarusBodenheimer}. In the detached phase, the solid and gas accretion rate are known (for the gas, it is given by the disc-limited rate), but not the planet's radius. The radius is determined following the approach of \citet{2012A&AMordasiniB,2012A&AMordasiniC}, that is by using the same internal structure equations as in the attached, but iterating on the radius until convergence is reached.

The pressure outer boundary conditions are modified to take into account that the disc and the envelope are no longer connected, and that the gas free-falls onto the surface of the planet
\begin{equation}
P(\rtot)=P_\mathrm{neb}(\aplanet)+P_\mathrm{edd}+P_\mathrm{ram}+P_\mathrm{rad}
\end{equation}
with $P_\mathrm{neb}(\aplanet)$ being the pressure at the midplane of the gas disc, $P_\mathrm{edd}=(2g)/(3\kappa)$ the Eddington expression for the photospheric pressure due to the material residing above the $\tau=2/3$ surface, $P_\mathrm{rad}=(2\sigmaSB T^4(\rtot))/(3c)$ the radiation pressure, $c$ being the speed of light in vaccum, and
\begin{eqnarray}
P_\mathrm{ram}=\frac{\mdotenv}{4\pi\rtot^2}\vff & ; & \vff^2 = 2\ggrav\mtot\left(\frac{1}{\rtot}-\frac{1}{\rhill}\right)
\end{eqnarray}
being the ram pressure due to the accretion shock and the free-fall velocity at the surface of the planet.

\subsubsection{Evolutionary phase}

For the evolutionary phase (after the dispersal of the gas disc), the outer boundary conditions are set to
\begin{eqnarray}
P(\rtot) & = & P_\mathrm{edd} + P_\mathrm{rad} \\
T^4(\rtot) & = & T^4_\mathrm{int} + (1-A)T^4_\mathrm{eq}
\end{eqnarray}
where $T_\mathrm{int}^4=L_\mathrm{tot}/(4\pi\sigmaSB\rtot^2)$ is the intrinsic temperature, $T_\mathrm{eq}=\tstar*\sqrt{\rstar/(2*\aplanet)}$, and $A=0.343$ is the albedo, which is taken be the same as Jupiter \citep{2005AREPSGuillot}. This value was selected for simplicity, although hot-Jupiter planets may have lower values \citep[e.g.][]{2019AAMallonn}.

We thus use an Eddington grey boundary condition taking the stellar irradiation into account, as described in \citet{2012A&AMordasiniB}. During evolution, we assume a solar-composition condensate-free gas for the opacities, using the opacity tables of \citet{2014ApJSFreedman}. Nebular grain opacity is neglected, at they are found to rain out quickly once gas accretion stops \citep{2008IcarusMovshovitzPodolak}. The identical envelope and atmospheric composition (pure H/He, solar composition opacities) in all planets means that for planets with identical bulk properties (orbital distance, core and envelope mass), the predicted radii will exhibit an artificially reduced spread. In reality, planets have different enrichment levels of heavy elements in the envelope \citep[e.g.][]{2013ApJForneyMordasini}. This affects the equation of state and opacity, resulting in particular in a larger spread of the radii \citep[e.g.][]{2011ApJBurrowsHeng,2020ApJMueller}.

\subsubsection{Example}

\begin{figure*}
	\centering
	\includegraphics{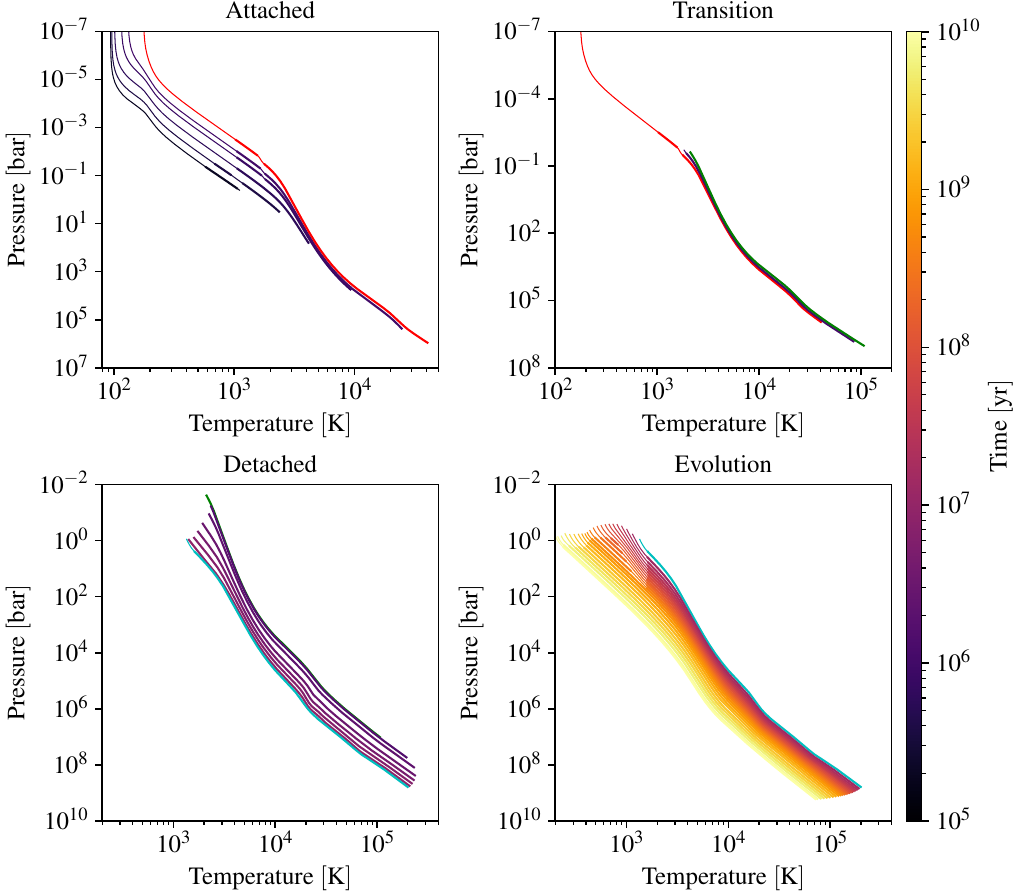}
	\caption{Snapshots of the internal structure of the second outermost planet of Fig.~\ref{fig:plan_disc_str}. The structures are split according to the phases, with attached (\textit{top left}), transition (the initial stage of the detached phase; \textit{top right}), detached (\textit{bottom left}) and evolutionary (\textit{bottom right}). The red line shows the first profile of the detached phase and is shown of both panels. The green and blue profiles lie at the transition between two stages and are shown of two panels each. In each profile, thin lines show the part where energy transport is radiative and thick lines for convective.}
	\label{fig:int_str}
\end{figure*}

\begin{figure*}
	\centering
	\includegraphics{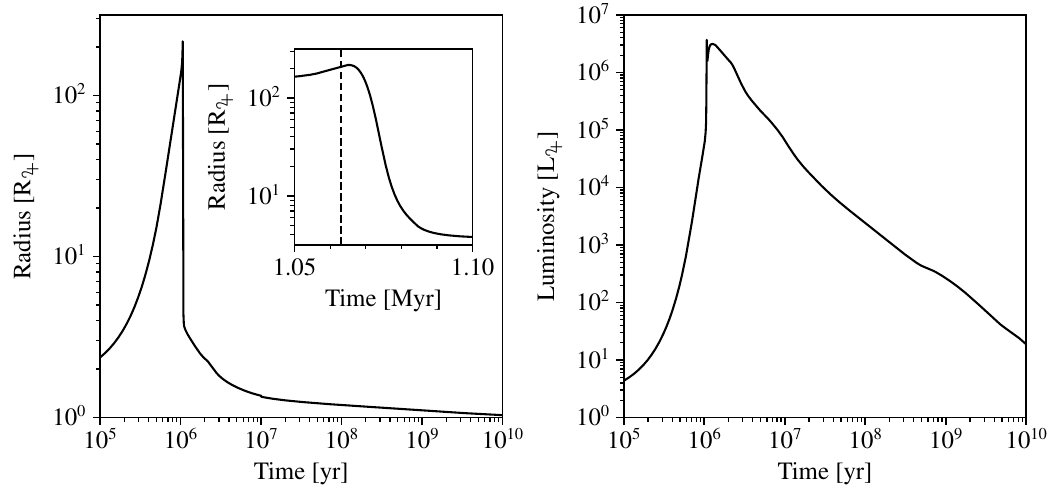}
	\caption{Time evolution of the planet's radius and luminosity of the second outermost planet of Fig.~\ref{fig:plan_disc_str}; the same as in Fig.~\ref{fig:int_str}. The insert on the left panel shows the contraction at the transition between attached and detached phase. The exact time where the model switches between the two phases is shown with the vertical dashed line.}
	\label{fig:tracks_int_str}
\end{figure*}

To illustrate the calculation of the internal structure, we provide snapshots of envelope structures in Fig.~\ref{fig:int_str} and the time evolution of the radius and luminosity in Fig.~\ref{fig:tracks_int_str}. These are taken from the second outermost planet of the system shown in Fig.~\ref{fig:plan_disc_str}, which is a giant planet whose final mass is \SI{6.4}{\mj}. Due to the different scales involved in the attached, detached, and evolutionary phases, they are shown in different panels. During the attached phase, the structure extends to the Bondi radius (Eq.~\ref{eq:bondi}), which is much larger than the core radius. Therefore, the structure spans a wider range of pressure. The upper part of the envelope is radiative while the lower part is convective, with several transitions in the mid region. The red profile shows the internal structure at the beginning of the transition from the attached to detached phase (the time marked with a dashed vertical line in the insert in Fig.~\ref{fig:tracks_int_str}).

Note that the planet is still accreting during the initial stages of the detached phase.

\subsection{Luminosity}
\label{sec:luminosity}

\subsubsection{Accretion and contraction}

The luminosity calculation suffers from the same problem as the total mass in the attached phase, or the outer radius in the detached phase; that is that the new structure needs to be known to retrieve its energy, hence the luminosity. This means that the total energy of the new structure needs to be estimated for a luminosity to be obtained.

The model uses the approach from \citet{2012A&AMordasiniB}. The total energy is given as
\begin{equation}
\etot = -\ggrav\int_0^{\mtot} \frac{M}{R} \mathrm{d}M + \int_{\mcore}^{\mtot} u \mathrm{d}M = -\kenerg\frac{\ggrav\mtot^2}{2\rtot}
\label{eq:etot}
\end{equation}
with $u$ being the specific internal energy of the gas, as obtained from the equation of state. The gravitational binding energy term includes the contribution from the core. For simplicity, we assume that it has a constant density, so its contribution is taken as $-3/5\ggrav\mcore^2/\rcore$. It should be noted that this is not strictly self-consistent with our model to determine its density or radius, which assumes differentiation \citep{2012A&AMordasiniC}; however, the difference remains small \citep{2019AALinder}. The parameter $\kenerg$ in Eq.~(\ref{eq:etot}) represents as in polytropic models the mass distribution and additionally the thermal energy content. It is retrieved from Eq.~(\ref{eq:etot}). The internal luminosity resulting from the accretion, cooling, and contraction $\lint$ can then be obtained as
\begin{equation}
\lint=\etotdot=\frac{\kenerg\ggrav\mtot}{\rtot}\mtotdot-\frac{\kenerg\ggrav\mtot^2}{2\rtot^2}\rtotdot+\frac{\ggrav\mtot^2}{2\rtot}\kenergdot,
\end{equation}
with $\mtotdot=\mdotcore+\mdotenv$ being the total accretion rate of the planet (solids and gas). The value $\mtotdot$ in the attached phase and of $\rtotdot$ in the detached phase are determined from the guess for the mass or radius during the iterations. The same is not true for $\kenergdot$. To circumvent this problem, we estimate the luminosity with
\begin{equation}
\lint\approx C\left(\frac{\kenerg\ggrav\mtot}{\rtot}\mtotdot-\frac{\kenerg\ggrav\mtot^2}{2\rtot^2}\rtotdot\right).
\end{equation}
The correction factor $C$ corrects for neglecting the $\kenergdot$ term. The value of $C$ can be calculated a posteriori by determining the actual total energy of the new planet, with $C=-(\etot(t)-\etot(t-\Delta t))/(\lint\cdot\Delta t)$. The value of $C$ is then used for the next time step.

\citet{2017ApJMarleau,2019ApJMarleau} conducted 1D radiation-hydrodynamic simulations of the planetary gas accretion shock, a feature that is seen in various 3D radiation-hydrodynamic simulations of accreting protoplanets of sufficiently high mass \citep[e.g.][]{2017MNRASSzulagyiMordasini,2020AASchulik}. High postshock entropies were found, suggesting that warm or hot gas accretion is more plausible than cold accretion \citep[see also][]{2017ApJBerardo,2017ApJBerardoCumming}. We therefore assume in our model that gas accretion in the detached phase is hot, which means that we do not subtract the accretion shock luminosity from $\lint$ \citep[see][]{2012A&AMordasiniB}.

In addition to the accretion and contraction luminosity, we include the luminosity from radioactive decay, bloating for close-in planets, and, in the case of brown-dwarfs, deuterium fusion. The radiogenic luminosity $\lradio$ includes contributions from the three most important long-lived radionucleides $^{40}$K, $^{238}$U and $^{232}$Th \citep{1964ScienceWasserburg}. To compute the luminosity contributions, we follow the procedure of \citet{2012A&AMordasiniC}: we assume the mantle of the protoplanets has a chrondritic composition and the energy production rate are retrieved from meteoritic values of \citet{2007BookLowrie}. The initial radiogenic contribution is $Q_0\approx \SI{5e-7}{\erg\per\gram\per\second}$ of mantle material (all elements besides iron).

\subsubsection{Bloating of close-in planets}
\label{sec:bloat}

Massive, close-in planets exhibit anomalously large radii \citep{2011ApJLaughlin}. To reproduce this effect, we include a bloating mechanism based on \citet{2021AASarkis}. For planets which are in the detached and evolutionary phase and directly irradiated by the host star, we include an additional luminosity contribution that is based on the best empirical fit formula obtained by \citet{2018AJThorngrenFortney}:
\begin{equation}
L_\mathrm{bloat} = \epsilon F_\star e^{-\tau_\mathrm{mid}}\pi\rtot^2
\label{eq:lumi-bloat}
\end{equation}
with
\begin{equation}
\epsilon = 2.37\exp{\left(-\frac{\left(\log{(F_\star/\SI{e9}{\erg\per\square\centi\meter\per\second})}-0.14\right)^2}{2\cdot0.37^2}\right)},
\end{equation}
$F_\star=L_\star/(4\pi\aplanet^2)$ the total stellar flux at the planet's location, and $\tau_\mathrm{mid}$ is the optical depth from the star to the planet location through the mid-plane of the disc, as in Eq.~(\ref{eq:tirr}).
We only apply bloating if the stellar flux $F_\star$ (in the evolutionary phase) or the stellar flux multiplied by the optical depth $ F_\star \exp{\left(-\tau_\mathrm{mid}\right)}$ (before the dispersal of the gas disc) is greater than \SI{2e8}{\erg\per\square\centi\meter\per\second} \citep{2011ApJSDemorySeager}.

\subsubsection{Deuterium-burning}

For the calculation of the luminosity due to deuterium fusion, we follow the procedure of \citet{2012A&AMolliere}. In this framework, the energy generation rate (per unit mass and time) is given by \citet{1994BookKippenhahnWeigert}, with the assumption that nuclei are fully ionised and non-degenerate. The energy released in each process is computed according to \citet{1967ARAAFowler}. The specific deuterium burning luminosity of a planet depends on the conditions in the planet's gaseous envelope, most notably the density, temperature, and the remaining deuterium nuclei. This implies that there is no universal mass at which deuterium burning starts, but as already found in \citet{2012A&AMolliere} (see also \citealp{2013ApJBodenheimer}), the mass where burning becomes important clusters around about \SI{13}{\mj}. The presence of a solid core does thus not significantly alter the mass where burning starts relative to (coreless) brown dwarfs \citep{2000ARAAChabrierBaraffe}. We use an initial deuterium number fraction [D/H]=\num{2e-5}, which is the primordial value.

Our model also includes the enhancing of the reaction rate by screening, that is the shielding of the positive charges by the surrounding electron. In turn, screening is affected by the electron degeneracy, as we are dealing with objects of high central densities. This procedure follows the work of \citet{1973ApJDewitt} and \citet{1973ApJGraboske}.

\subsubsection{Total luminosity}

The final luminosity is then given by
\begin{equation}
L(\rtot)=\ltot=\lint+\lradio+L_\mathrm{bloat}+\lburn.
\end{equation}
We assume that at a given time, the luminosity does not change within the envelope, that is $\partial L/\partial r=0$. This approximation is fine under most circumstances because energy transport is due to convection and the luminosity enters only in the radiative gradient. During rapid gas accretion in the detached phase, under the effect of hot accretion, the interior may become radiative \citep{2017ApJBerardo,2017ApJBerardoCumming} and we do not account for the decrease of the luminosity with depth. This will be addressed in future work.

\subsection{Accretion of solids}
\label{sec:model-core}

The growth of the astrophysical core of the planets can occur via three channels: 1) the accretion of planetesimals \citep[e.g.][]{1992IcarusGreenzweigLissauer,2003IcarusThommes}, 2) the accretion of pebbles \citep[e.g.][]{2010AAOrmelKlahr,2010MNRASJohansenLacerda,2012AALambrechtsJohansen}, and 3) by the collision with other embryos (which we call giant impacts). In the Generation III model, we consider accretion by planetesimals and giant impacts; the inclusion of pebble accretion is subject of ongoing work \citep{2020AAVoelkel}.

For planetesimals accretion, core growth is given by the probability of collisions with planetesimals in the oligarchic regime \citep{1993IcarusIdaMakino}, as described in \citet{2013A&AFortier}. This is a major difference to the first generation of the \textit{Bern} model which followed \citet{1996IcarusPollack} for the planetesimal accretion rate. According to \citet{2006IcarusChambers}, the core growth can be computed assuming a particle-in-a-box approximation is
\begin{equation}
\mdotcore = \Omega\sigmamean\rhill^2\pcoll,
\end{equation}
with $\sigmamean$ the mean surface density of planetesimals in the planet's feeding zone and $\pcoll$ the collision probability with planetesimals. As \citet{2008ApJIdaLinA}, we use the same prescription to calculate the planetesimal accretion rate independently of a protoplanet's orbital migration rate. In addition, we address the possible impact that orbital migration could have in the context of the shepherd/predator regimes proposed by \citet{1999IcarusTanakaIda}: in the idealised situation studied by \citet{1999IcarusTanakaIda} (single protoplanet per disc, no local reservoir of planetesimals, no growth via collisions with other protoplanets), shepherding was found to significantly reduce the planetesimal accretion rate for protoplanets migrating sufficiently slowly. However, in the more realistic N-body simulations by \citet{2006IcarusDaisaka} where multiple protoplanets (oligarchs) form and grow concurrently as expected in the oligarchic regime \citep{1998IcarusKokubo}, the trapping of planetesimals by the protoplanets is only tentative and does not significantly reduce their accretion rates. We similarly find that in the more realistic situation we consider here with many embryos per disc, the existence of a local reservoir of planetesimals in a protoplanet's initial feeding zone accessible without migration and a time sequence of a solid accretion dominated initially by planetesimals and later on collisions with other protoplanets (Sect. \ref{sec:twoexmples}), shepherding should only be of limited importance. We discuss these points further in Appendix \ref{appendixpredator}.

\subsubsection{Capture probability}

We distinguish three different accretion regimes depending on the random velocities: low-, mid- and high-velocity. The distinction is based on the reduced planetesimals' eccentricity $\eplanh=r\eplan/\rhill$ and inclination $\iplanh=r\iplan/\rhill$ ($r$ is the heliocentric distance): the high-velocity regime for $\eplanh,\,\iplanh\gtrsim2$, mid-velocity for $2\gtrsim\eplanh,\,\iplanh\gtrsim0.2$ and low-velocity for $0.2\gtrsim\eplanh,\,\iplanh$. According to \citet{2001IcarusInaba}, each regime has a different expression for the collision probability,
\begin{eqnarray}
p_\mathrm{high} & = & \frac{(\rcap+\rplan)^2}{2\pi\rhill}\left(I_\mathrm{F}(\beta)+\frac{6\rhill I_\mathrm{G}(\beta)}{(\rcap+\rplan)\eplanh^2}\right) \\
p_\mathrm{mid} & = & \frac{(\rcap+\rplan)^2}{4\pi\rhill\iplanh}\left(17.3+\frac{232\rhill}{\rcap+\rplan}\right) \\
p_\mathrm{low} & = & 11.3\left(\frac{\rcap+\rplan}{\rhill}\right),
\end{eqnarray}
where $\rcap$ is the planetesimal capture radius of the planet, $\beta=\iplan/\eplan$ and the $I_\mathrm{F}$ and $I_\mathrm{G}$ functions can be approximated as, following \citet{2006IcarusChambers}:
\begin{eqnarray}
I_\mathrm{F}(\beta) & \simeq & \frac{1+0.95925\beta+0.77251\beta^2}{\beta\left(0.13142+0.12295\beta\right)}\\
I_\mathrm{G}(\beta) & \simeq & \frac{1+0.39960\beta}{\beta\left(0.0369+0.048333\beta+0.006874\beta^2\right)}.
\end{eqnarray}
The final collision is then given by
\begin{equation}
\pcoll=\min{\left(p_\mathrm{med},\left(p_\mathrm{high}^{-2}+p_\mathrm{low}^{-2}\right)^{-1/2}\right)}.
\end{equation}

\begin{figure*}
    \centering
    \includegraphics{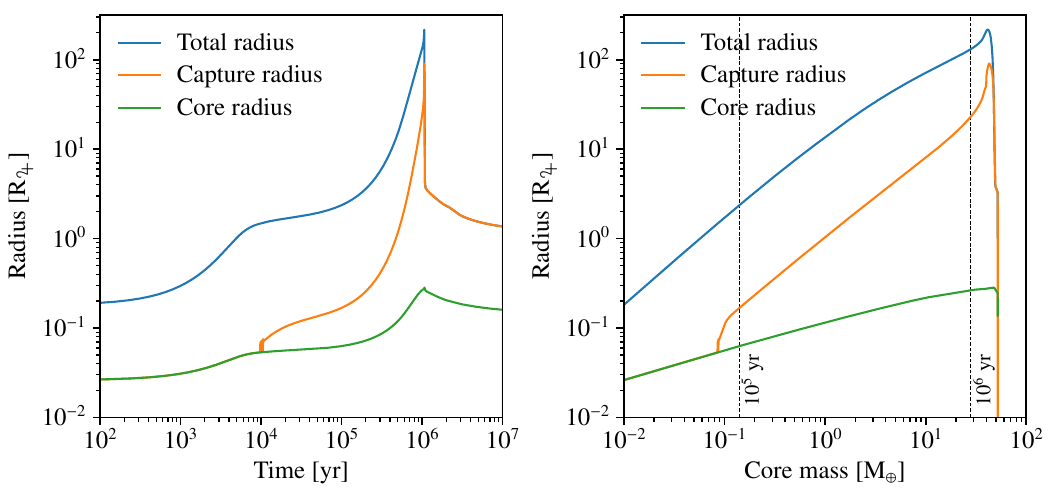}
    \caption{Evolution of the planet's core, planetesimals capture, and total radii as function of time (\textit{left panel}) and core mass (\textit{right panel}) for the second outermost planet of Fig.~\ref{fig:plan_disc_str}. In the right panel, we also include two time indicators at \SI{e5}{\year} and \SI{e6}{\year} with dashed vertical lines. The upper-right portion of the plot for the total radius in the left panel is also shown on the left panel of Fig.~\ref{fig:tracks_int_str}.}
    \label{fig:radius}
\end{figure*}

In the initial stage, the capture radius $\rcap$ is the physical radius of the core $\rcore$. Once the planet has sufficiently massive H/He envelope, it will enhance the capture cross-section of planetesimals. As in \citet{2013A&AFortier}, the capture radius is obtained following \citet{2003A&AInaba} by solving the implicit equation
\begin{equation}
\rplan = \frac{3}{2}\frac{\rho(\rcap)\rcap}{\rhoplan}\left(\frac{\vrel^2+2\ggrav M(\rcap)/\rcap}{\vrel^2+2\ggrav M(\rcap)/\rhill}\right).
\end{equation}

The enhancement of the capture radius over the physical radius is very important for increasing the planet's planetesimals accretion rate \citep{1988IcarusPodolak,2020AAVenturiniHelled}. We highlight this in Fig.~\ref{fig:radius}, which compares the planetesimals capture radius to that of the core for the same planet we highlighted in Fig.~\ref{fig:tracks_int_str}. The calculation of the envelope structure begins at about \SI{e4}{\year}, before that, the capture radius is equal to that of the core. At that moment, the core mass is \SI{9e-2}{\mearth}. By the time the core reaches \SI{1}{\mearth} at \SI{4.8e5}{\year}, the capture radius is 9 times the core radius. Therefore, for small roughly km-sized planetesimals as in our case, the enhancement of the capture radius is already important for low-mass bodies (starting about \SI{e-1}{\mearth}), and the calculation of gaseous envelopes cannot be omitted at any stage. Besides the factor that the eccentricity and inclination damping by nebular gas drag is more efficient for smaller planetesimals which leads to a larger gravitational cross section (a larger Safronov factor), the larger envelope drag enhancing the planet capture radius further is the second effect making the accretion of small planetesimal more efficient. This reflects that the accretion of km-sized planetesimals is not a pure gravitational process.

\subsubsection{Ejection of planetesimals}
\label{sec:model-pla-ejec}

Planets not only accrete material; they also induce gravitational perturbations on the planetesimals that come close-by but are not accreted. These planetesimals, if they receive a sufficient velocity kick from a close approach by a planet, can be ejected from the system. To estimate this effect, we follow a procedure similar to \citet{2004ApJIda1}. The planetesimals that receive a velocity kick greater than the escape velocity from the primary, $v_\mathrm{esc}=\sqrt{2\ggrav\mstar/\aplanet}$, will likely be ejected from the system. Thus, we have that the fraction of accreted-to-ejected planetesimals is \citep{2004ApJIda1}
\begin{equation}
\frac{\mdotcore}{\dot{M}_\mathrm{ejec}} = \left(\frac{v_\mathrm{esc}}{v_\mathrm{surf}}\right)^4
\end{equation}
with $v_\mathrm{surf}=\sqrt{\ggrav\mtot/\rcap}$ the characteristic surface velocity. The rate at which planetesimals are removed from the disc is then
\begin{equation}
\dot{M}_\mathrm{plan}=\mdotcore+\dot{M}_\mathrm{ejec}.
\end{equation}

It should be noted that our model does not include the redistribution of planetesimals by scattering \citep{2017IcarusRaymondIzidoro}.

\subsubsection{Feeding zone}
\label{sec:fz}

\begin{figure}
    \centering
    \includegraphics{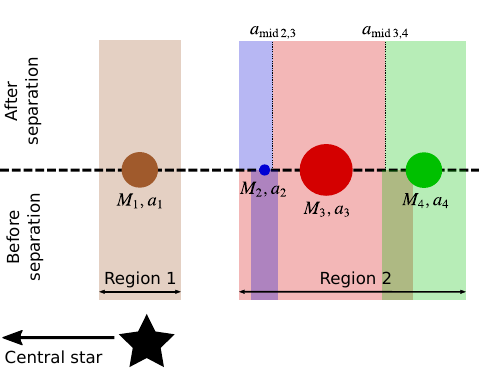}
    \caption{Illustration of the procedure to separate planetesimals' feeding zones when zones would otherwise overlap. The horizontal axis represents the separation to the central star and four planets are shown. The light colour areas below the horizontal line show the initial feeding zones while the ones above show the final zones. $a_\mathrm{mid2,3}$ and $a_\mathrm{mid3,4}$ are the edges of the new feeding zone.}
    \label{fig:separation}
\end{figure}

To obtain the mean surface density of planetesimals in the feeding zone, we must determine its extent. The half-width of the feeding zone (centred at the planet's location) is usually given in terms of the Hill radius with
\begin{equation}
R_\mathrm{feed}=b\rhill.
\end{equation}
For a planet on a circular orbit, conservation of the Jacobi energy implies that $b=\sqrt{12+4/3(\eplanh^2+\iplanh^2)}$ \citep[e.g.][]{1977PASJHayashi}. So, in a quiescent disc with $\eplanh,\iplanh\ll1$, $b=2\sqrt{3}\approx3.5$. For numerical stability reasons, however, we assume $b=5$, as in \citet{2013A&AFortier}.

In the general case, to account for a non-circular orbit of the planet, we take the feeding zone to span from $r_\mathrm{peri}-R_\mathrm{feed}$ to $r_\mathrm{apo}+R_\mathrm{feed}$, with $r_\mathrm{peri}$ and $r_\mathrm{apo}$ being the peri- and apocentre of the planet's orbit respectively.

When multiple planets are present in the same disc, their feeding zones may overlap. To avoid problems with two planets accreting from the same location, such as mass-conservation issues, we separate the feeding zones so that there is at most one planet accreting at any location the disc. A graphical representation of the following procedure is provided in Fig.~\ref{fig:separation}. First, we compute regions in the disc from where planets accrete. In the case a region contains a single planet, then the feeding zone is the same as in the single planet case (as in Region~1 on Fig.~\ref{fig:separation}). If there are multiple planets in one region (as in Region 2 on that figure), the inner edge of the innermost planet and the outer edge of the outermost planet are set to the edges of the region. For the other edges, we sort planets by distance, and for each pair, we compute the location of the limit between their feeding zones with
\begin{equation}
a_\mathrm{mid} = \frac{a_\mathrm{out}\sqrt{M_\mathrm{in}}+a_\mathrm{in}\sqrt{M_\mathrm{out}}}{\sqrt{M_\mathrm{in}}+\sqrt{M_\mathrm{out}}}
\end{equation}
where the subscripts indicate the inner (in) and outer (out) planets of the pair. We scale with the square root of the planet masses because the area of the feeding zone scales with the square of the distance. This scaling keeps the area of the feeding zones related to the planet masses. We tested alternative prescriptions, like using the cubic root of the mass (as in the Hill sphere) or the mid-point between the two planets and found that the prescription does not significantly affect the outcomes of the simulations.

\subsubsection{Core radius}

To obtain the radius of the core (and its density), we applied a methodology similar to \citet{2012A&AMordasiniC}. This model also accounts for the composition of the core and the pressure burden exerted by the envelope.

The principle is to solve similar structure equations as for the envelope, that is Eqs.~(\ref{eq:intstr-mass}) and~(\ref{eq:intstr-pres}), but with an equation of state that takes the form of a modified polytrope from \citet{2007ApJSeager}, which reads
\begin{equation}
\rho(P) = \rho_0 + cP^n.
\end{equation}
We include three different materials: iron, silicates (perovskite, MgSiO\textsubscript{3}) and ice, whose parameters $\rho_0$, $c$ and $n$ are taken from \citet{2007ApJSeager}. Because of the small thermal expansion coefficient of these materials compared to H/He, we neglect via the temperature-independent modified polytropic EOS a possible temperature dependency of the radius of the core. It should, in any case, be small \citep{2009ApJGrasset}.

For gas giant planets, where envelopes can reach masses of thousands of Earth masses, this can cause a significant compression of the core \citep{2008AABaraffe}. Thus, the pressure on the core's surface is taken as boundary condition of the calculation to include this effect. Core compression can be observed in Fig.~\ref{fig:radius}, where the core radius shrinks after the envelope contracts at \SI{1.06}{\mega\year}.

The core composition is retrieved from the accreted planetesimals described in Sect.~\ref{sec:disc-compo} and other embryos in case of giant impacts. The chemical composition is used to obtain the fraction of the different elements to compute the core radius. While in the chemistry model includes 32 \citep{2014AAThiabaud} refractory and 8 volatile \citep{2014AAMarboeufA} chemical species, the core radius calculation groups them into only three types: iron, silicates, and water ice. Thus, we map all ice species to water ice when calculating the core structure and all refractories except iron to the silicate mantle. The reason for this is that first, equations of state are only available for a limited number of species. Second, the differences between different types of, for instance, silicates is not very large \citep{2007ApJSeager}.

\subsection{Atmospheric escape}
\label{sec:model-escape}

During the evolutionary phase, that is after the dissipation of the gaseous disc, planets at small distances of their host star ($\sim\SI{0.1}{\au}$) receive intense XUV stellar irradiation, which will drive atmospheric escape. This effect is especially important for the low-mass planets, that can loose the whole of their gaseous envelope due to their low gravitational binding energy \citep[e.g.][]{2009AALammer,2012ApJLopez,2012MNRASOwenJackson,2014ApJJin,2018ApJJin}.

The stripping of the whole envelope has a significant effect on the planets radius. Due to the low density of gas, the presence of an envelope will result a significant increase of the planets' sizes even if the envelope mass is only on a percent level of the total planet mass. Bare cores are thus clearly separated from object that retain a gaseous envelope, and a gap is observed in the distribution of planetary radii \citep{2013ApJOwenWu,2013ApJLopezFortney,2014ApJJin,2016ApJChenRogers,2018AJFultonPetigura}.

The evaporation model is based on \citet{2014ApJJin}. It takes into account contributions from X-ray and extreme-ultraviolet (XUV) irradiation. At the early stages, the evaporation is typically X-ray driven. We describe this regime using the energy-limited rate from \citet{2012MNRASJackson} using the flux in the 1 to \SI{20}{\angstrom} range from \citet{2005ApJRibas} and assuming an efficiency factor $\epsilon=0.1$.

At later stages, the evaporation from EUV takes over. We also use the work of \citet{2005ApJRibas} to obtain the time-dependent EUV stellar luminosity for a Sun-like star. EUV evaporation can be divided into two sub-regimes \citep{2009ApJMurrayClay}. At low EUV fluxes, the same energy-limited approximation as for the X-ray flux is used. In this case, the escape flux is given by
\begin{equation}
\dot{M}_\mathrm{env,e}=\epsilon\frac{\pi F_\mathrm{EUV}R^3_\mathrm{base}}{\ggrav\mtot}
\end{equation}
where $F_\mathrm{EUV}$ is the EUV flux, $R_\mathrm{base}$ the radius of the photoionisation base, calculated as in \citet{2009ApJMurrayClay}, and $\epsilon=0.3$ is the heating efficiency, taken as in \citet{2009ApJMurrayClay}.

On the other hand, energy-limited evaporation is not suitable when the EUV flux is high ($>\SI{e4}{\erg\per\square\centi\meter\per\second}$), as a substantial part of the heating is lost in cooling radiation. In this regime, we adopt the radiation-recombination-limited approximation of \citet{2009ApJMurrayClay}. The mass loss rate is given by wind due to escape
\begin{equation}
\dot{M}_\mathrm{env,rr}\sim 4\pi \rho_\mathrm{s} c_\mathrm{s} R_\mathrm{s}^2
\end{equation}
at the sonic point $R_\mathrm{s}$, which is calculated the same way as $\racc$. Here $c_\mathrm{s}=\sqrt{\kB T/(m_\mathrm{H}/2)}$ is the isothermal sound speed of ionised gas with $T=\SI{e4}{\kelvin}$. The density can be related to the one at the ionisation base, where $\tau=1$, with
\begin{equation}
\rho_\mathrm{s}\sim\rho_\mathrm{base}\exp{\left[\frac{\ggrav\mtot}{\rtot c^2_\mathrm{s}}\left(\frac{\rtot}{R_\mathrm{s}}-1\right)\right]}.
\end{equation}
The photoionisation base is located where there is equilibrium between photoionisations and recombination:
\begin{equation}
\frac{F_\mathrm{UV}}{h\nu_0}\sigma_{\nu_0}n_\mathrm{0,base}\sim n^2_\mathrm{+,base}\alpha_\mathrm{rec}
\end{equation}
with $n_\mathrm{0,base}$ the density of neutrals at the base, $h\nu_0=\SI{20}{eV}$, $\sigma_{\nu_0}=\SI{6e-18}{\square\centi\meter}(h\nu_0/\SI{13.6}{eV})^{-3}$, $\alpha_\mathrm{rec}=\num{2.7e-13}$, and $\rho_\mathrm{base}=n_\mathrm{+,base}m_\mathrm{H}$.

The model also includes the effect of Roche lobe overflow. When solving the internal structure equations, there are sometimes solutions found in the detached and evolutionary phase where the radius is larger than the Hill sphere. This occurs in two situations: First, for close-in low-mass planets with a high envelope mass fraction. At the moment when the nebula dissipates (and thus the ambient pressure vanishes), and when the star starts to irradiate the planets directly (resulting in an increase of the temperature, see Fig.~\ref{fig:gas_disc_str}), these planets bloat. Second, giant planets that get very close to the star because of tidal spiral in (see Sect.~\ref{sec:dyn-tides}) can also overflow their Roche lobe. In this case, we remove at each time step the part of the H/He envelope that is outside of the Hill sphere.

\subsection{Initial conditions}

The simulation begin with a predetermined number of embryos whose initial mass is $\mstart=\SI{e-2}{\mearth}$ (approximately the mass of our Moon). They are randomly placed with an uniform probability in $\log{a}$, where $a$ is the semi-major axis, between $\rin$ and \SI{40}{\au}. The starting location zone is slightly more extended that in the previous studies, where the upper boundary was set to \SI{20}{\au}. Also, two embryos cannot be placed within 10 Hill radii from each other. It should be noted that for the simulations with largest initial number of embryos, 100, this represents an average spacing of 28 Hill radii.

The presence of a number of embryos right at the beginning of the simulations is a strong assumption we made because the model does not track the formation of the embryos themselves. This shortcoming of the model will be addressed in future evolutions of the model \citep{2020AAVoelkel}, where the evolution of the dust, pebbles and planetesimal and embryo formation is followed.

\section{Dynamical evolution: Orbital migration, N-body interaction, and tides}
\label{sec:dyn}

As the planet mass increases, it will generate a stronger perturbation in the density of the gas around the planet. This perturbation will cause the nebula to no longer be axis-symmetric, and as a consequence produces a torque back on the planet, leading to planetary migration.
At the same time, convergent migration can result in capture in mean-motion resonances or orbital destabilisation. Hence migration and dynamical evolution must be performed together to capture all the effects.

\subsection{Planetary migration}

We include two types of migration, Type~I for low mass planets embedded in the gas disc and Type~II for planets massive enough to open a gap in the disc.

\subsubsection{Type~I migration}

For Type~I migration, our model follows the approach of \citet{2014MNRASColemanNelson}. This includes the torques formulation from \citet{2011MNRASPaardekooper}, modified to consider that orbital eccentricity and inclinations attenuate the co-rotation torques \citep{2010AABitschKley,2011AABitschKleyA}.

The total Type~I torque on a planet, following Eqs.~(50) to (53) of \citet{2011MNRASPaardekooper} and (15) of \citet{2014MNRASColemanNelson}, is given by
\begin{equation}
\Gamma_1=F_\mathrm{L}\Gamma_\mathrm{L}+F_\mathrm{e}F_\mathrm{i}\left(\Gamma_\mathrm{c,baro}+\Gamma_\mathrm{c,ent}\right),
\label{eq:torque-t1}
\end{equation}
with
\begin{eqnarray}
\Gamma_\mathrm{c,baro} & = & \Gamma_\mathrm{hs,baro}F(p_\nu)G(p_\nu) + \Gamma_\mathrm{c,lin,baro}(1-K(p_\nu)) \\
\Gamma_\mathrm{c,ent} & = & \Gamma_\mathrm{hs,ent}F(p_\nu)F(p_\chi)\sqrt{G(p_\nu)G(p_\chi)} + \nonumber\\
& & \Gamma_\mathrm{c,lin,baro}\sqrt{(1-K(p_\nu))(1-K(p_\chi))},
\end{eqnarray}
where $\Gamma_\mathrm{L}$, $\Gamma_\mathrm{hs,baro}$, $\Gamma_\mathrm{hs,ent}$, $\Gamma_\mathrm{c,lin,baro}$ and $\Gamma_\mathrm{c,lin,baro}$ are the Linblad torque, barotropic and entropy part of the horseshoe drag and linear corotation torque respectively. They are given by Eqs.~(3) to~(7) of \citet{2011MNRASPaardekooper}. The function $F$ governs saturation, while $G$ and $K$ provide the cutoff at high viscosity, and are given by Eqs.~(22), (30) and (31) of \citet{2011MNRASPaardekooper}.

The other factors in Eq.~(\ref{eq:torque-t1}) account for the shape of the orbit. $F_\mathrm{L}$ provides the reduction of the Lindblad torque for eccentric or inclined orbits following \citet{2008AACresswellNelson}, with
\begin{equation}
F_\mathrm{L}^{-1}=P_\mathrm{e}+\left(\frac{P_\mathrm{e}}{|P_\mathrm{e}|}\right)\left(0.07\hat{\imath}+0.085\hat{\imath}^4-0.08\hat{e}\hat{\imath}^2\right)
\label{eq:flind}
\end{equation}
and
\begin{equation}
P_\mathrm{e}=\frac{1+\left(\frac{\hat{e}}{2.25}\right)^{1/2}+\left(\frac{\hat{e}}{2.84}\right)^6}{1-\left(\frac{\hat{e}}{2.02}\right)^4}.
\label{eq:pecc}
\end{equation}
Here, $\hat{e}=e/h=e/(H/r)$ and $\hat{\imath}=i/h=i/(H/r)$ are the planet's orbital eccentricity and inclination scaled by the disc's aspect ratio $h=H/r$. $F_\mathrm{e}$ and $F_\mathrm{i}$ provide the reduction of the corotation torques due to eccentricity and inclination \citep{2010AABitschKley}. We use
\begin{equation}
F_\mathrm{e}=\exp{\left(-\frac{e}{h/2+0.01}\right)}
\end{equation}
as suggested by \citet{2014MNRASFendykeNelson} for the reduction due to eccentricity and
\begin{equation}
F_\mathrm{i}=1-\tanh{(\hat{\imath})}
\end{equation}
for the reduction due to inclination \citep{2014MNRASColemanNelson}.

Eccentricity and inclination damping time scales follow \citet{2008AACresswellNelson}, with
\begin{equation}
\tau_\mathrm{e} = \frac{t_\mathrm{wave}}{0.78}\left(1-0.14\hat{e}^2+0.06\hat{e}^3+0.18\hat{e}\hat{\imath}^2\right)
\end{equation}
and
\begin{equation}
\tau_\mathrm{i} = \frac{t_\mathrm{wave}}{0.544}\left(1-0.3\hat{\imath}^2+0.24\hat{\imath}^3+0.14\hat{e}^2\hat{\imath}\right),
\end{equation}
where
\begin{equation}
t_\mathrm{wave}=\left(\frac{M_\star}{\mplanet}\right)\left(\frac{M_\star}{\Sigma\aplanet^2}\right)h^4\Omega^{-1}
\end{equation}
is the characteristic time of evolution of density waves \citep{2004ApJTanakaWard}.

\subsubsection{Type~II migration}

The criterion to detect gap opening and switch migration to Type~II is from \citet{2006IcarusCrida},
\begin{equation}
\frac{3H}{4\rhill}+\frac{50\nu\mstar}{\mplanet\aplanet^2\Omega}\leq 1,
\label{eq:gap}
\end{equation}
with $\nu$ is the viscosity from Eq.~(\ref{eq:nu}).

Type~II orbital migration follows the non-equilibrium approach from \citet{2014A&ADittkrist}. Here, the planet follows the radial velocity of the gas,
\begin{equation}
\vrad=\frac{1}{\sigmag\sqrt{r}}\frac{\partial}{\partial r}\left(\nu\sigmag\sqrt{r}\right)
\label{eq:vrad}
\end{equation}
\citep{1981ARAAPringle}, but is limited if the planet's mass is much larger than the local disc mass (the fully suppressed case, see \citealp{2009ApJAlexanderArmitage}). The radial velocity of the planet $\vplanet$ is given by
\begin{equation}
\frac{\vplanet}{\vrad} = \min{\left(1,\frac{2\sigmag\aplanet^2}{\mplanet}\right)}.
\label{eq:type2}
\end{equation}
For the larger planet masses, when the migration rate is constrained by the disc-to-planet mass ratio, this expression result in a similar behaviour as the formula obtained by \citet{2018ApJKanagawa}, although it does not take into account the aspect ratio of the disc $h$.

For our migration scheme, we convert the radial velocity into a torque according to
\begin{equation}
\Gamma_2=\frac{1}{2}\mplanet\Omega \aplanet\vplanet.
\end{equation}
This prescription allows in principle planets in Type~II to migrate outwards if the disc is decreting \citep{2004MNRASVerasArmitage}. However, in practice this mechanism is limited by the restriction to planets that are already at large distances or during the final moments of the disc, and limited by the small surface density \citep{2014A&ADittkrist}.

During type~II migration, the eccentricity and inclination damping time scales are set to
\begin{equation}
\tau_e=\tau_i=\frac{1}{10}|\tau_\mathrm{a}|=\frac{1}{10}\frac{\aplanet}{|\vplanet|}.
\end{equation}
This relationship was selected because hydrodynamical simulations of migrating planets in this regime have shown that eccentricity and inclination damping act on time scales that are shorter than migration \citep{2004AAKley,2019SAASKley}.

\subsubsection{Migration map}
\label{sec:dyn-mig-map}

\begin{figure}
    \centering
    \includegraphics{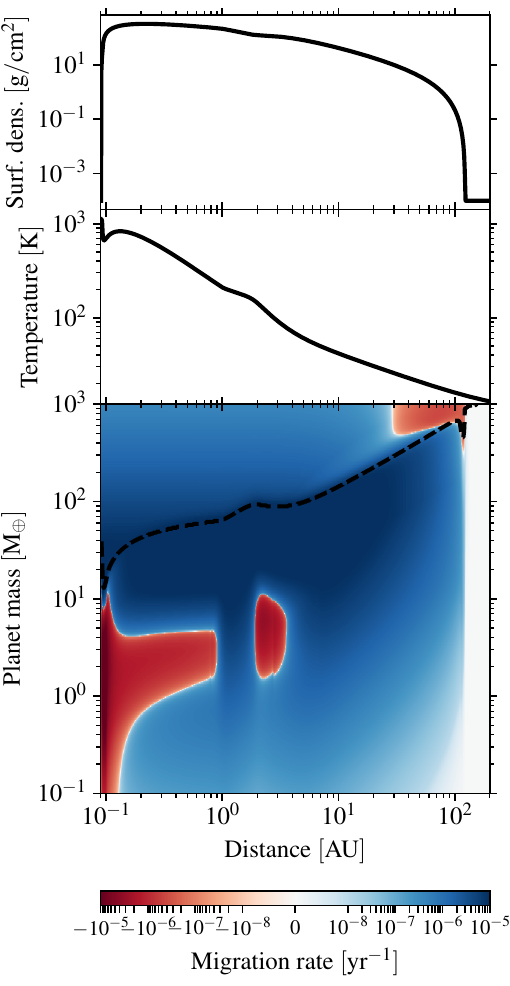}
    \caption{Radial surface density profile (top panel), temperature profile (middle panel), and migration map (bottom) as function of the planet mass (assuming zero eccentricity and inclination), for the same disc presented in Fig.~\ref{fig:gas_disc_str} at $t=\SI{1}{\mega\year}$. The value plotted in the bottom panel is the relative migration rate $1/\tau_\mathrm{a}=-\vplanet/\aplanet$; blue regions indicate inward migration, red regions outward migration. For both directions, the locations in bright colours are where migration is inefficient while dark tones indicate efficient migration. The dashed black line shows the boundary between type I (below) and type II (above) migration regimes.}
    \label{fig:mig-map}
\end{figure}

An example of the outcome of the whole migration scheme for one disc profile is provided in Fig.~\ref{fig:mig-map}. The disc is the same as the example shown in Fig.~\ref{fig:gas_disc_str} at \SI{1}{\mega\year}; at this time the disc mass is \SI{1.46e-02}{\msun}. Its outer radius is \SI{123}{\au}, so we cut the figure at \SI{200}{\au} since there is no migration outside this distance.

Migration is most efficient for intermediate mass planets, above about \SI{10}{\mearth} up to the transition to Type~II migration (shown with the dashed black line on the migration map). The outward migration at large separation for the type~II migration regime is due to the outward spreading of the gas disc. We also note two convergence zones for low- to mid-mass planets. These are due to opacity transitions \citep{2010ApJLyra} or structures in the gas disc \citep{2012ApJKretke} such as the increase of the surface density close to the inner edge of the disc \citep{2006ApJMassetA}. These are the locations where, for a given planet mass, outward migration happens on the inner side and inward migration on the outer side. Hence, at this moment of evolution, planets with masses less than $\approx\SI{8}{\mearth}$ cannot reach the inner edge of the disc by migration only. However, as time goes and gas becomes scarcer, the zones of outward migration (hence the convergence zones) shift towards lower planetary masses. Thus, by the end of the gas disc, planet with masses down to $\approx\SI{2}{\mearth}$ could reach the inner edge of the disc.

\subsection{N-body integration}

Gravitational interactions between the protoplanets are now modelled with the \texttt{mercury} N-body code \citep{1999MNRASChambers} using the hybrid method. Unlike the direct resolution of the equation of motion \citepalias[as performed in][]{2013A&AAlibert}, this use a symplectic integration scheme \citep[see e.g.][for a review]{1992ActaNumSanzSerna}. The basic principle is to use the solution of Hamilton's equations,
\begin{equation}
\dot{x_i}=\frac{\partial\mathcal{H}}{\partial p_i},\hspace{1em}\dot{p_i}=-\frac{\partial\mathcal{H}}{\partial x_i}
\end{equation}
where $x$ denotes the position coordinates, $p$ the momentum coordinates, and
\begin{equation}
\mathcal{H}=\sum_{i=0}^{N}\frac{p_i^2}{2M_i}-\ggrav\sum_{i=0}^{N}M_i\sum_{j=i+1}^{N}\frac{M_j}{\Delta x_{ij}}
\end{equation}
is the Hamiltonian of the system, with $\Delta x_{ij}=|x_i-x_j|$. Here, the index $i=0$ refers to the central star and $M_0=\mstar$ while the subsequent are the planet with $M_i=M_\mathrm{planet,i}$ so that $N$ is the number of planets in the system.

However, while $\mathcal{H}$ has no analytical solution for $N>1$, it is possible to split the Hamiltonian into several pieces, solving the simpler problems to finally combine them back so that a solution close to that of the original system. The Hamiltonian is divided into three components, so that $\mathcal{H}=H_\mathrm{K}+H_\mathrm{S}+H_\mathrm{I}$, and
\begin{eqnarray}
H_\mathrm{K} & = & \sum_{i=1}^N\left(\frac{p_i^2}{2M_i}-\ggrav\frac{\mstar M_i}{\Delta x_{i0}}\right) \\
H_\mathrm{S} & = & \frac{1}{2\mstar}\left(\sum_{i=1}^N\mathbf{p}_i\right)^2 \\
H_\mathrm{I} & = & -\ggrav \sum_{i=1}^N\sum_{j=i+1}^N\frac{M_i M_j}{\Delta x_{ij}}.
\end{eqnarray}

Here, $H_\mathrm{K}$ represents the unperturbed Keplerian orbits of the planets about the central star, $H_\mathrm{S}$ the kinetic energy of the star and $H_\mathrm{I}$ the interactions between the planets. The separation into three different Hamiltonians (rather than two) is required because the scheme uses mixed-centre coordinates (also called `democratic heliocentric'): heliocentric positions and barycentric velocities. These coordinates are chosen so that $H_\mathrm{K}\gg H_\mathrm{S},H_\mathrm{I}$, unless two planets come close together.

The evolution of such a system by splitting is done using a second-order method,
\begin{equation}
H_\mathrm{I}\left(\frac{\tau}{2}\right)H_\mathrm{S}\left(\frac{\tau}{2}\right)H_\mathrm{K}\left(\tau\right)H_\mathrm{S}\left(\frac{\tau}{2}\right)H_\mathrm{I}\left(\frac{\tau}{2}\right),
\end{equation}
where the notation $H_\mathrm{...}(\tau)$ is used to represent the evolution under the given Hamiltonian for a step $\tau$. For $H_\mathrm{I}$, this means that the planets receive a kick in velocity due to the interactions with the other bodies (except the central star). In our case, $H_\mathrm{I}$ is extended to include additional forces representing the effect of the gas disc, see Section~\ref{sec:add-forces}. The evolution under $H_\mathrm{S}$ results in a shift $\tau/(2\mstar)\sum \mathbf{p}_i$ while the evolution under $H_\mathrm{K}$ is a Keplerian motion around the central star for a period $\tau$.

As we noted, the assumption that $H_\mathrm{I}$ is small compared to $H_\mathrm{K}$ is no longer valid when two bodies become close together. In that situation, the idea is to bring the interaction between the two close-by bodies into $H_\mathrm{K}$ so that the interaction Hamiltonian remains small. This implies that $H_\mathrm{K}$ is no longer analytically integrable during that period, but only for the orbits of the involved bodies. In practice, the orbits of the two close-by bodies are integrated with a conventional Bulirsch-Stoer method \citep{1980BookStoerBulirsch} for the duration of the encounter. That algorithm is described in detail in \citet{1999MNRASChambers}.

The symplectic integration scheme has a huge advantage in terms of computational requirements compared to a standard Bulirsch-Stoer method, as the interaction between the planets, the one part that is $\mathcal{O}(N^2)$, is only computed once per step.

We do not use the N-body when there is only one protoplanet in a system as the solution is analytical. This happens either for populations with one embryo per system or in the unlikely case that only one planet survives in a planetary system with initially multiple embryos per system.

\subsubsection{Additional forces}
\label{sec:add-forces}

Migration and damping are included as additional forces in the N-body. The contributions from migration and eccentricity damping apply in the orbital plane and are split into tangential ($\theta$) and radial ($r$) components, while the inclination damping acts on the vertical component ($z$), resulting in
\begin{eqnarray}
a_\mathrm{\theta} & = & -\frac{v_\mathrm{\theta}-\vkep}{2\tau_\mathrm{e}} + \frac{\Gamma}{\mplanet} \\
a_\mathrm{r} & = & -\frac{v_\mathrm{r}}{\tau_\mathrm{e}} \\
a_\mathrm{z} & = & -\frac{2v_\mathrm{z}}{\tau_\mathrm{i}}
\end{eqnarray}
with $a$ denoting the additional accelerations, $v$ the planet's velocity along each direction. Here, $\vkep=\Omega r$ is the Kelperian velocity.

\subsubsection{Collision detection}

Collisions are detected when two planets come closer than a predetermined distance, which is the sum of their radii. When the closet approach is found inside to be during one of the substeps of the N-body, the minimum distance is retrieved by fitting a third-degree polynomial equation whose condition are set by the relative separation and their radial velocity at the beginning and end of the substep (similar to \citetalias{2013A&AAlibert}).

For planets with a significant and extended envelope (like during the attached phase), the assumption that planets have a unique radius which decides whether a collisions occurs or not is no very accurate, as the outcome is determined by gas dynamics inside the merging envelopes. As we do not have the full envelope structure in the N-body, we nevertheless remaining with a unique radius approach. In the attached phase, the envelope transitions smoothly to surrounding nebula. The outer radius, as provided by Eq.~(\ref{eq:rtotattached}), is unsuitable for the detection of collisions, as it corresponds to very low gas densities. Thus, the radius used to detect collisions is computed assuming that the whole planet mass has the same density as its core. This is an approximation, but reflects that the gas density in the envelope is much higher close to the (solid) core surface. In the detached phase, we use the planetesimals' capture radius $\rcap$; this is normally an overestimation of the effective collision radius, larger bodies needing to penetrate deeper down in the envelope to be captured. However, in this phase the envelope scale height is small compared to the radius except for the very short time directly after detachment, so the actual error is small.

\subsubsection{Collision treatment}
\label{sec:collisiontreatment}

\begin{figure}
	\includegraphics{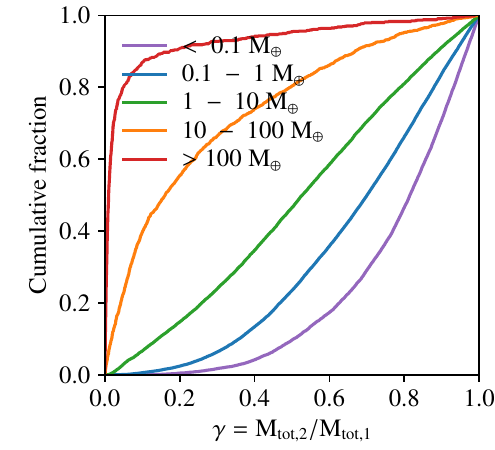}
	\caption{Cumulative distribution of the impactor-to-target mass ratio $\gamma=M_\mathrm{tot,2}/M_\mathrm{tot,1}$ for different target mass ranges (as provided in the legend). Data come from the 100-embryos population presented in \papertwo.}
	\label{fig:coll-mratio}
\end{figure}

When a collision is detected, the following procedure is applied: the cores merge, the eventual envelope of the less massive body is deemed to be ejected, and the impact energy is added as a additional contribution to the luminosity for the structure calculation of the new body.

The merger of the cores will make that a part of the impact energy will already be taken into account consistently with the luminosity calculation described in Section~\ref{sec:luminosity}; so the additional energy is calculated using
\begin{equation}
E_\mathrm{impact}=\max{\left(\frac{1}{2}\mu v_\mathrm{imp}^2 - E_\mathrm{acc,core},0\right)}
\end{equation}
where $\mu=M_\mathrm{tot,1}M_\mathrm{tot,2}/(M_\mathrm{tot,1}+M_\mathrm{tot,2})$ is the reduced mass, and the indexes 1 and 2 refer to the quantities of the larger and smaller body respectively. $v_\mathrm{imp}$ is the relative velocity at time of contact. Here
\begin{equation}
E_\mathrm{acc,core} = \ggrav\frac{M_\mathrm{tot,1}M_\mathrm{core,2}}{R_\mathrm{core,1}+R_\mathrm{core,2}}
\end{equation}
is the centre-of-mass impact energy of two bodies with the total mass of the target and the core mass of the impactor colliding at their mutual escape velocity. Also, we restrict the supplementary energy to positive values. Negative value can arise if the bodies are colliding at below the mutual escape velocity, which is possible due to the drag by the gas disc or in the case of specific configuration, such as co-orbitals. However, the impact velocity is never quite lower than the mutual escape velocity, so that the error remains small.

The addition of the core mass and luminosity is performed via
\begin{eqnarray}
\dot{M}_\mathrm{core,supp} & = & \frac{M_\mathrm{core,2}}{\tau_\mathrm{impact}\sqrt{2\pi}} \exp{\left(-\frac{1}{2}\left(\frac{t-t_\mathrm{impact}}{\tau_\mathrm{impact}}-3\right)^2\right)} \\
L_\mathrm{core,supp} & = & \frac{E_\mathrm{impact}}{\tau_\mathrm{impact}\sqrt{2\pi}} \exp{\left(-\frac{1}{2}\left(\frac{t-t_\mathrm{impact}}{\tau_\mathrm{impact}}-3\right)^2\right)}
\end{eqnarray}
where $t_\mathrm{impact}$ is the time of the impact, $\tau_\mathrm{impact}=\SI{e4}{yr}$ is the time scale of release taken as in \citet{2012A&ABroeg}. These two terms are added to the core accretion rate due to planetesimal accretion, and to the luminosity (Sect.~\ref{sec:luminosity}) used in the internal structure calculation, respectively.

This impact model was tailored for the most common collisions that we find in our simulations. We highlight this by showing cumulative distrubutions of impactor-to-target mass ratio $\gamma$ for different ranges of target masses in Fig.~\ref{fig:coll-mratio}. At low masses, most target/impactor pairs are of similar masses, thus this source of growth cannot be neglected. In contrast, most collisions involving giant planets are with much smaller impactors (the red curve in Fig.~\ref{fig:coll-mratio}). Our models neglect the envelope of the impactor, but there are only few collisions where this could provide significant source of mass.

\subsection{Tidal evolution}
\label{sec:dyn-tides}

During the evolution phase we include the inward migration of planets due to tides they raise onto the central star. In addition to planets that are pushed inwards due to capture in mean-motion resonances, this gives another channel to obtain planets that are within the inner boundary of the gas disc. For the tidal migration rate, we compute the rate according to
\begin{equation}
\frac{\partial \aplanet}{\partial t}=-\frac{9}{2}\sqrt{\frac{\ggrav}{\mstar}}\frac{\rstar^5\mplanet}{\qstar}\aplanet^{-11/2}
\label{eq:tidal-mig}
\end{equation}
\citep{2008CeMDAFerrazMello,2009ApJJackson,2011AABenitezLlambay}, where $\qstar=10^6$ is the stellar dissipation parameter. It is clear that this model for the tidal spiralling-in is strongly simplified. It will be improved in future work along the lines of, for example, \citet{2016CeMDABolmont}.

\section{Terrestrial planet formation}
\label{sec:terrestrial}

We begin by studying whether the new generation of the \textit{Bern} model with a higher initial number of embryos, but which still includes a statistical description of planetesimals, is capable of reproducing models of terrestrial planets that use purely \textit{N}-body \citep[e.g.][]{2001IcarusChambers}, that is where the planetesimals are represented as individual (test) particles. This test is crucial to assess whether we can reach our goal of having a formation model which is able to simulate the growth of planets with a very large mass range from about that of Mars, to brown dwarfs. This is in contrast with earlier generations of the Bern Model, where mainly more massive planets were at the focus (or more specifically, planets for which the giant impact phase after disc dissipation is not very important).

The formation of terrestrial planets does not have the same time constraint as for gas giants. In the case of planets with a significant H/He envelope, a sufficiently massive core must be formed before the dispersal of the gas disc, but this does not apply to terrestrial planets. Indeed, in the case of the Earth, cosmochemical evidences point to a formation time between a few tens \si{\mega\year} \citep{2002NatureYin,2002NaturKleine} to roughly \SI{100}{\mega\year} \citep{2007NatureTouboul,2008EPSLAllegre,2009GeCoAKleine}. This is longer than the expected lifetime of the solar system's nebula of \SI{4}{\mega\year} \citep{2017ScienceWang} by about an order of magnitude or more. Hence the modelling of formation of planetary systems with terrestrial planets needs to span a longer time period for dynamical effects (i.e. the `late stage') than for gas-dominated planets.

\subsection{Setup}

For this test cases, we performed a few modifications to our main model to mimic earlier work like \citet{2001IcarusChambers} and \citet{2005ApJRaymond}. Orbital migration has been disabled; as for the envelope structure calculation and the evolution phase, here, all planets are treated as purely rocky. We adopt an initial surface density profile close the minimum-mass solar nebula \citep[MMSN;][]{1977ApSSWeidenschilling,1981PThPSHayashi}, with a reference surface density of $\Sigma_\mathrm{0,s}=\SI{7.1}{\gram\per\square\centi\meter}$ at $r_0=\SI{1}{\au}$, but truncated at \SI{2}{\au}, as we are primarily interested in the inner planets. This also helps to determine more precisely which fraction of the planetesimal disc has been accreted by the terrestrial planets during their formation. This gives a solids mass of \SI{3.67}{\mearth}. The initial number of embryos is selected to have a similar spacing as the two populations presented \citet[][\papertwo]{NGPPS2} with most embryos per system, which means that we have initially 23 (correspond to 50 in \papertwo) and 46 (corresponding to 100) lunar-mass ($\SI{0.01}{\mearth}$) embryos. In addition to that, we perform one run with 9 embryos initially in Sect.~\ref{sec:terrestrial-numemb}, which corresponds to 20 embryos in \papertwo.

It should be noted that the model lacks the `dynamical friction' obtained in \textit{N}-body simulations with a large number of small bodies \citep{2006IcarusOBrien,2006IcarusRaymond} because we do not include the effect of the damping of eccentricities and inclinations of the embryos by the planetesimals. However, after all material has been accreted onto the planets, the remainder of the formation process is similar to pure \textit{N}-body simulations of terrestrial planet accretion, as all the mass is now contained in bodies that are directly followed by the \textit{N}-body.

For some simulations, we include Jupiter and Saturn to determine the effects they have on the formation of the inner planets. In that case, Jupiter and Saturn are on their present-day orbits, but they are rotated so that their invariant plane coincides with that of the disc \citep[as in][]{2001IcarusChambers,2013IcarusChambers,2020ApJEmsenhuberA}. We do not model the formation of these planets, because they form over a period that is much shorter than the terrestrial planets.

To obtain a better overview of the influence of the parameters we are studying, and to reduce (and better understand) the stochastic effects of \textit{N}-body interactions, we perform 10 simulations for each combination of parameters (initial number of embryos and presence of the outer planets). The only differences between the 10 simulations are the initial position of the terrestrial planet embryos. For the 10 simulations, we consider the average outcomes as being representative (e.g. Fig.~\ref{fig:avg}).

The simulations starts with a gas disc, which lives for roughly \SI{4.4}{\mega\year}. Its only effect however is to damp the eccentricities and inclinations of the planetesimals. Planetesimals accretion continues after the dispersal of the gas disc. As the planets do not have envelopes, we perform only the formation stage of the calculation. However, the duration of that stage has been extended to \SI{400}{\mega\year} to account for the much longer time needed for the solar system's terrestrial planets to converge.

\subsection{Gravitational interactions}

\begin{figure}
	\includegraphics{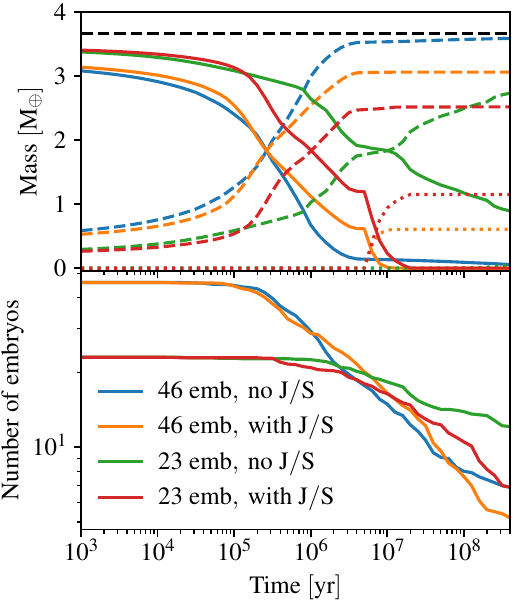}
	\caption{Average over the 10 simulations for each set of parameters. The top panels show the masses of solids (excepting the outer giant planets in the relevant cases) in the protoplanetary disc, that is still in planetesimals (solid lines), accreted by the embryos (dashed lines) and ejected (dotted lines). The dashed black line denotes the total mass of solids in each simulation. The bottom panel shows the number of embryos that remain. The `J/S' in the legend refers to Jupiter and Saturn.}
	\label{fig:avg}
\end{figure}

If the embryos remain at their initial locations during the whole formation process, then they grow to their isolation mass \citep{1987IcarusLissauer}. In our model, we obtain this behaviour if we artificially remove the \textit{N}-body interactions, unless the feeding zones of two adjacent embryos overlap at some point, in which case the masses become slightly lower. When using this mode, the runs starting with 46 embryos have accreted roughly half of the disc's mass onto the embryos by about \SI{4}{\mega\year} (the time at which the gas disc disperses) and accrete very slowly thereafter. For the runs starting with 23 embryos, only a quarter of the mass ends in the embryos by \SI{4}{\mega\year}.

For the other parameter sets (all with gravitational interactions), Fig.~\ref{fig:avg} provides the averaged results over 10 simulations, for the masses of solids and the number of embryos. The story is quite different when \textit{N}-body interactions are included. We see for instance that in the case with 46 embryos and no outer giant planets, nearly all the planetesimals have been accreted onto the embryos. For the case with 23 embryos initially and no outer giant planets, more than half of the planetesimals end up accreted.

There are two aspects we point out here. First, in the figure, the planetesimal mass accreted by embryos that have been later ejected is accounted as accreted. Second, our planetesimal model does not include redistribution of material by interactions with the embryos. For instance, in their less realistic setup where embryos only populate a limited orbital distance range in the disc, \citet{2010AJLevison} found that planetesimals can be redistributed to locations outside of the embryos' feeding zone rather than be accreted. However, when they add the mechanism that embryos can reside in all parts of the disc (which is more realistic, \citealt{2010AJLevison}) no gap in the planetesimal disc opens, as embryos mutually scatter planetesimal into their vicinity and accrete them eventually. This leads to an efficient formation of massive planets. Thus, feeding zones overlap in these simulations, therefore the effect of planetesimal redistribution should play little role in our case as there are very few locations in the disc where planetesimals would not be accreted by the local embryo and/or scattered back into the feeding zone of other embryos.

\subsection{Interactions lead to more massive planets}
\label{sec:terrestrial-numemb}

\begin{figure}
	\includegraphics{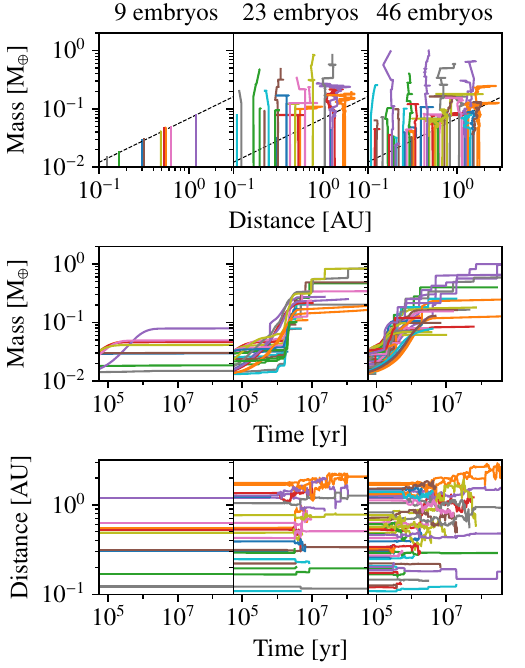}
	\caption{Comparison of formation tracks for a system with a MMSN-like surface density of planetesimals, and with 9 embryos (\textit{left}), 23 (\textit{centre}), and 46 (\textit{right}), and no outer giant planets. Each line represents one embryo. \textit{Top panels}: mass versus semi-major axis; embryos start at the bottom and move upwards as they grow. The final positions of the remaining planets are shown by dots. The dashed black line denotes the isolation mass \citep{1987IcarusLissauer}. \textit{Middle panels}: mass versus time; sudden increases in mass are due to embryo-embryo collisions. \textit{Bottom panels}: semi-major axis versus time.}
	\label{fig:tracks_o}
\end{figure}

To understand how the embryo-embryo interactions lead to a quasi-complete accretion of the planetesimals disc, we show the formation tracks for one particular system with a varying number of embryos in Fig.~\ref{fig:tracks_o}.

We can easily observe that the larger the number of embryos, the more and the sooner they start to move around. In the system with only 9 embryos, they basically remain where they started and grow slightly above their isolation mass. For the other two simulations, however, the local isolation mass is sufficient to trigger significant embryo-embryo interactions that will change their positions in the disc. This in turn enables them to accrete from regions that would otherwise inaccessible, which creates a positive feedback since more massive planets will result in yet more interactions. This feedback only ends when nearly all planetesimals have been accreted onto the embryos.

Thus, closer packed embryos lead to enhanced stirring of their eccentricities, which has two consequences: the increase of the feeding zone size because of radial excursion for eccentric orbits, and collisions between embryos. Embryos having a greater eccentricity can sample a broader region of the disc, thus grow to a larger mass before depleting the disc. Collisions with other embryos are capable to bring material from more distant regions of the disc that would otherwise not be accessible to one embryo. At the end, we arrive at a result that is maybe counter-intuitive at first: the larger the number of embryos, the less planets remain. We observe this for instance in the bottom panel of Fig.~\ref{fig:avg}.

\subsection{Time needed for formation}

We find a similar pattern for the timing at which interactions start in the two simulations with the higher number of embryos of Fig.~\ref{fig:tracks_o} (23 and 46 embryos). In the early phase (a few \SI{e5}{\year}), no dynamical interactions occur, because the embryos need to reach a certain mass before the eccentricities can be significantly excited. Then, the first embryos to show an increased eccentricity are located at $\sim\SI{0.3}{\au}$, and then this propagates both inwards and outwards. In the inner part of the system, collisions happen rather rapidly so that the system has essentially obtained its final configuration by several \si{\mega\year}.

On the other hand, in the outer region we observe that embryos remain on eccentric orbits for a certain amount of time before suffering from collisions. It takes more than \SI{10}{\mega\year} for the planets located at about \SI{1}{\au} to reach their final mass. In the even more distant regions, it takes even longer, and we see the phase with several embryos on eccentric orbits remaining for more than \SI{100}{\mega\year}. Such a growth wave travelling from the inside to the outside is expected, as the growth process scales with the local Keplerian frequency.

Therefore, our choice of the integration time dictates the location where and how accurately the model can follow the formation of the terrestrial planets. With our choice of an integration time limited to \SI{20}{\mega\year} for the formation phase, the model can only track most of the giant impact stage inside of roughly \SI{1}{\au} for systems that have a MMSN-like surface density of solids. Even within \SI{1}{\au}, the giant impact stage is not entirely finished within our set time frame, as it can be see in the innermost planet by about \SI{300}{\mega\year} in the bottom right panel of Fig.~\ref{fig:tracks_o}. Nevertheless, these events remain rare. Locations further away or systems with a lower amount of solids (as formation is slower for less massive systems, \citealp{2006ApJKokubo,2015MNRASDawson}) will, however, not have reached a final state by end of the formation stage at \SI{20}{\mega\year}.

\subsection{With outer giant planets}

\begin{figure}
	\includegraphics{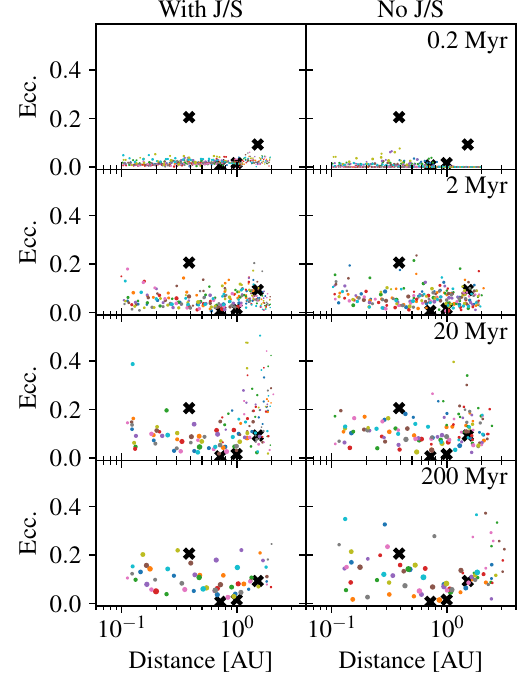}
	\caption{Stacked eccentricity versus distance snapshots of 10 simulations with each 46 embryos initially. The left column shows the runs with outer giant planets whereas the right column has no outer giant planets. In each column, the 10 systems are represented with a different colour for each one. The bodies are shown by points whose sizes are proportional to their physical ones. Black crosses show the solar system planets.}
	\label{fig:dyn}
\end{figure}

\begin{figure}
	\includegraphics{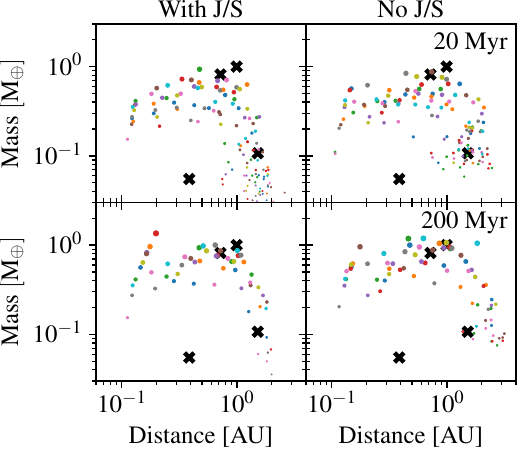}
	\caption{Same as Fig.~\ref{fig:dyn}, but showing mass versus distance stacked diagram of 10 simulations with each 46 embryos initially. As before, the black crosses show the solar system planets.}
	\label{fig:massdist}
\end{figure}

As the final stage of terrestrial planet formation (the giant impact stage) takes longer than formation of the giant planets, we also want to consider the effects of their presence on terrestrial planet formation. Here we perform the same simulations again, each time with the addition of two outer giant planets that represent Jupiter and Saturn.

To provide a better comparison point between the two cases, we provide in Fig.~\ref{fig:dyn} several snapshots of the simulations. One general consequence at earlier times is that there is slower growth for the embryos beyond \SI{1.5}{\au}. We see in the two top rows than the outermost embryos remain smaller in the runs with outer giant planets. Also, their eccentricities have already increased in the first snapshot, while this is not the case at all for the runs without giant planets. The underlying cause is stirring of planetesimal's eccentricity and inclination by the giant planets; this heavily reduces the collision probability with the low-mass protoplanets \citep{2001IcarusInaba} and hence the accretion rate.

A consequence of the longer timescales of accretion in the outer part of the disc is the state at the moment of the dispersal of the gas disc. In the runs with giant planets, a larger percentage of the planetesimals remains unaccreted at the moment the gas disperses. In addition, after that point, there is no longer gas present to counterbalance the effects of the stirring by the giant planets. This means that after a short moment, the planetesimals will reach eccentricities of the order of unity and will be ejected. This can be observed in Fig.~\ref{fig:avg}, where we see that up to a quarter of the original initial mass is ejected from the planetesimals disc. The final eccentricities of the terrestrial bodies are similar in both cases (Fig.~\ref{fig:dyn}), as the inner region is subject to the self-stirring while in the outer region, excitation by the outer planets makes up from a weaker self-stirring as the masses are lower.

Thus, the outer giant planets will limit and delay growth of the terrestrial planets in the outer region. The number of objects is a bit higher than the one obtained by pure \textit{N}-body simulations of terrestrial planet formation, but we are using a somewhat smaller initial surface density profile compared to, for example, \citet{2006IcarusRaymond}, which prevents the accretion into a lower number of higher mass bodies \citep{2006ApJKokubo}.

\subsection{Summary for terrestrial planets}

To summarise, we have just seen that as long as the separation between the embryos is sufficiently small that dynamical interactions are triggered before the embryos reach their local isolation mass, the model is capable of reproducing the main features of the formation of terrestrial planets in good agreement with pure \textit{N}-body models. This is due to embryo-embryo interactions being able to increase the eccentricities, so that the embryos can move out of their original locations, and almost entirely depletes the planetesimals.

An integration period (for the formation stage) longer than the lifetime of the protoplanetary disc is necessary to follow the giant impact phase. The time required for the bodies to obtain their final characteristics increase with distance (as shown here) and with decreasing initial amount of solids \citep[e.g.][]{2006ApJKokubo,2015MNRASDawson}. The limitation of the formation stage to \SI{20}{\mega\year} (Section~\ref{sec:model-desc}) permits to capture all the accretion of planetesimals (provided there are enough embryos initially) and most of the dynamical interactions of Earth-mass and larger planets forming via giant impacts up to roughly \SI{1}{\au}, and sub-Earths planets in the first few tenths of an \si{\au} (corresponding to periods of roughly 100 days). For the population syntheses in \papertwo, we estimate from tracking major changes of the planets' orbits, that for orbital distances of $\lesssim\SI{1}{\au}$ around \SI{90}{\percent} of the major instabilities should have been captured when integrating the systems for \SI{20}{\mega\year}.

The integration time needed to capture most instabilities within a given orbital distance range is a function of the architecture of the planetary systems that results from the previous growth stages. If the growth and migration during the presence of the gas disc leads for example to very closely packed systems of massive planets, instabilities will often occur shortly after disc dispersal. On the other hand, if at gas disc dispersal only low-mass, widely-space planets are present, they will first have to grow further via accretion of remaining planetesimals and embryos - which can take a very long time - to eventually (or also never) become unstable. For larger planet masses, gravitational interactions can extend further: Even on distant orbits, massive planets can destabilise the system as noted by \citet{2020AABitsch} and \citet{2021AAMatsumura}. This could explain why \citet{2021AAIzidoro} find that by \SI{20}{\mega\year} only a fraction of the instabilities between the planets have happened in their setup, whereas \citet{2020ApJMulders} on the contrary find that increasing the integration time from \SI{10}{\mega\year} to \SI{100}{\mega\year} only leads to minor further evolution in their simulations. For systems lacking outer planets, \citet{2021AAIzidoro} also found a convergence after \SI{\sim30}{\mega\year} (their Model III). This is in better agreement to the analysis for planets with $a<\SI{1}{\au}$ done in \citet{2020ApJMulders}.

The purpose of the model here is to obtain planetary systems that can be compared with observations at the population level. For this, it is important to see that the region where long-term growth will be most important (distant low-mass planets) represents at the same time the parameter space currently not accessible to most detection techniques of extrasolar planets (radial velocity, transits, and direct imaging). This should minimise the impact of this limitation. We acknowledge, however, that generally speaking, not all dynamical interactions will have taken place by the end of the integration time of \SI{20}{\mega\year} in the model. While the later evolution should not be substantial enough to strongly affect the statistical results in the inner systems, this limitations must be critically kept in mind when comparing for example to microlensing surveys \citep[e.g.][]{2018ApJSuzuki} that probe more affected regions.

Nevertheless, we conclude that the new generation of syntheses can be used to describe in a much more comprehensive way planetary sub-populations ranging from sub-Earths to super-Jupiters.

\section{Giant planets}
\label{sec:giants}

\begin{figure*}
	\centering
	\includegraphics{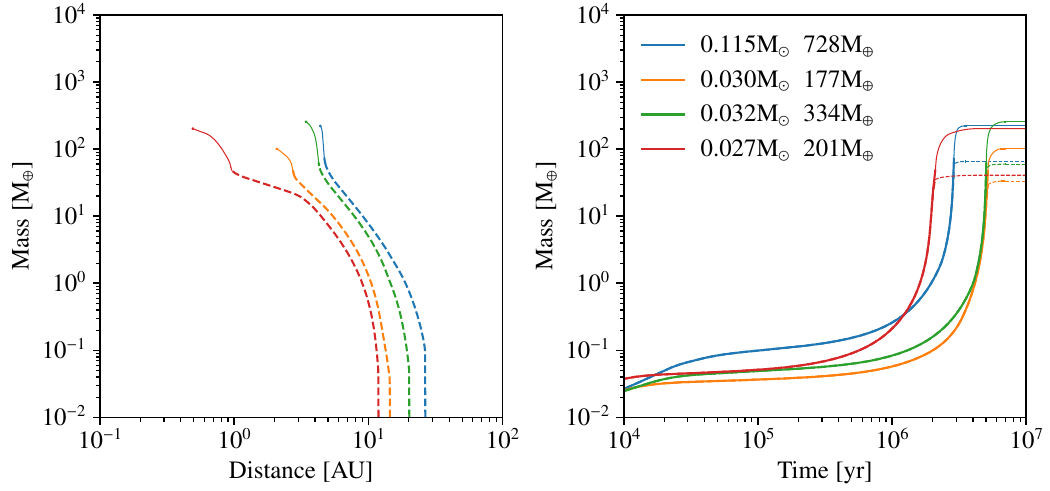}
	\includegraphics{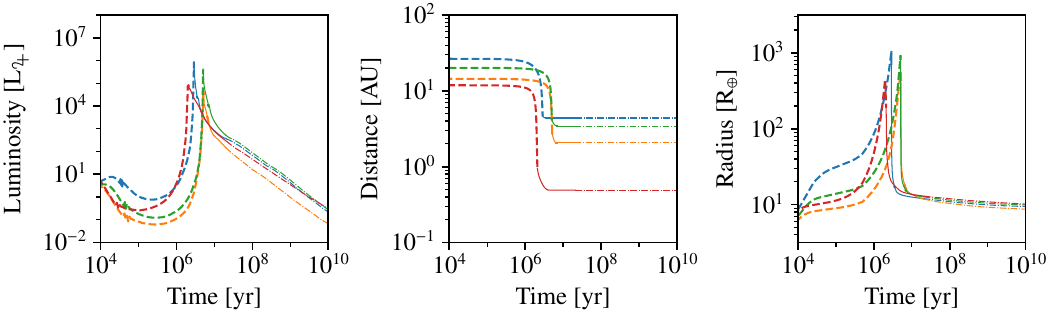}
	\caption{Formation and evolution tracks of four giant planets with final masses in the $1/3$ to \SI{2}{\mj} range in discs with a single embryo. The top panels present the formation tracks with total mass $\mtot$ versus distance (time goes towards the top) and total mass $\mtot$ (solid lines) and core mass $\mcore$ (dashed lines) versus time. The three panels on the bottom row show the time dependence of the outer luminosity $\ltot$ (bottom left), the distance (bottom centre) and the total radius $\rtot$ (bottom right). For all panels except for the mass versus time (top right), the line styles denote the phase: dashed lines for the attached phase, solid line for the detached phase during formation and dash-dotted lines for the evolution stage. Line widths denote the migration regime, with tick lines for Type~I and think lines for Type~II. The legend in the top right panel shows the gas (in Solar masses) and planetesimals (in Earth mass) disc masses.}
	\label{fig:subgi-single}
\end{figure*}

\begin{figure*}
	\centering
	\includegraphics{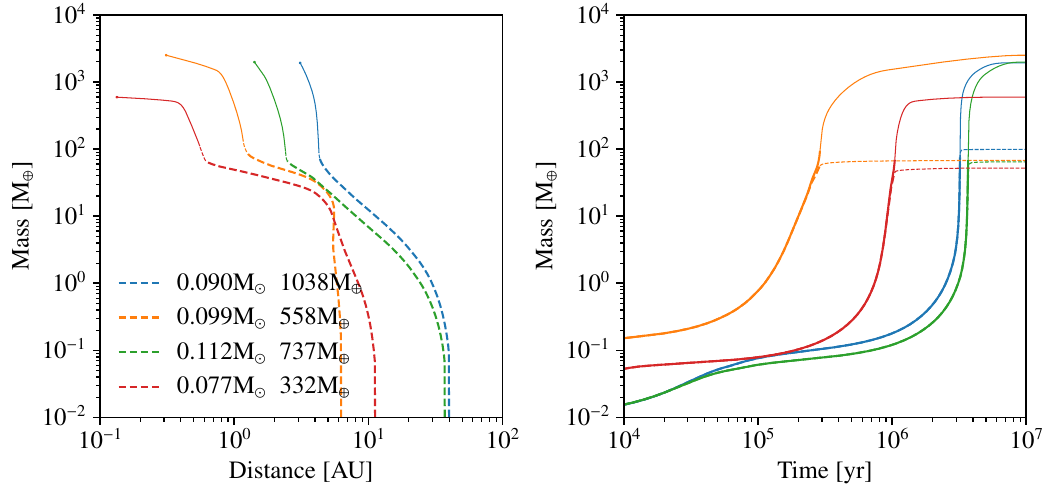}
	\includegraphics{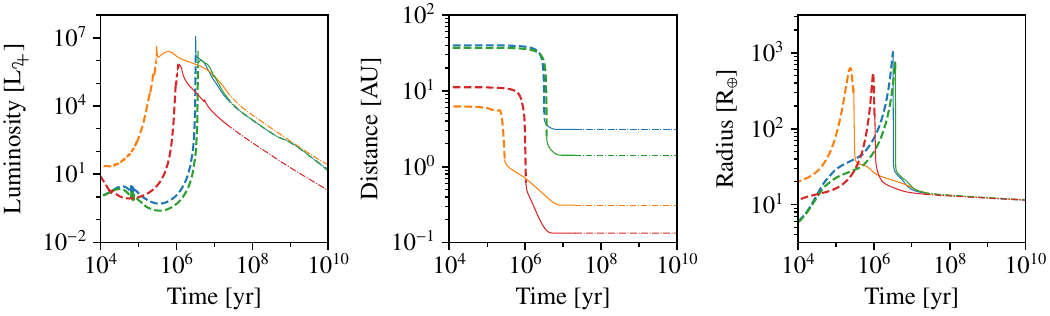}
	\caption{Formation and evolution tracks of two groups of four giant planets with final masses between \num{2} and \SI{10}{\mj} in discs with a single embryo. Panel and line descriptions are the same as Fig.~\ref{fig:subgi-single}.}
	\label{fig:giant-single}
\end{figure*}

The formation of giant planets is quite different. Cores must form before the dispersal of the gas disc so that they can undergo runaway gas accretion, and since we have massive cores in a gas disc, migration is efficient. To gain an understanding of the interplay of accretion and migration, we here show some illustrative cases with a single embryo per disc. For this case, we use the model without modifications, but the \textit{N}-body is not used. The following examples are taken from the single-embryo population of \papertwo. Simulations parameters are the same as provided in Table~\ref{tab:params-ex}, except for disc masses and surface density (both gas and planetesimals), inner edge, characteristic radius, and external photoevaporation rate of the gas disc. In the following simulations, the inner radius has negligible effect on the final outcome, as we do not study close-in planets, and so we do not mention it. The characteristic radius $\rcutg$ of the gas disc is set as
\begin{equation}
\frac{\mgas}{\SI{2e-3}{\msun}}=\left(\frac{\rcutg}{\SI{10}{\au}}\right)^{1.6}.
\end{equation}
(see \papertwo{} for the motivation). We provide the remaining two parameters, the initial masses of the gas and planetesimals discs in the following.

\subsection{Formation and evolution of Jupiter-mass planets}

We show in Fig.~\ref{fig:subgi-single} the formation tracks of a few synthetic giant planets whose masses are in the \num{100} to \SI{500}{\mearth} range and have a wide range of final positions. Due the inclusion of migration in the model, we observe that the final position of these planets is closer-in that the initial location of the embryo: all the embryos start beyond \SI{10}{\au}, with one close to \SI{30}{\au}, while all the planets end up inside \SI{10}{\au}.

During the initial stage, both accretion and migration are slow, but accretion is still faster. As the planets grow, migration becomes more efficient; we observe that most of the migration occurs while the planets are close to the transition to gas giants, with masses between \num{20} and \SI{50}{\mearth}. The innermost planet shows a strong inward migration at this stage, but this is due to limited accretion while migration remains at the same rate. Once the planets undergo the runaway accretion of gas and switch to type~II migration, accretion is strong, and they experience limited migration. This leads again to near-vertical tracks. The two changes (from an attached to a detached envelope and from type~I to type~II migration) happen in the same period, not always in the same order. In one case (the inner most planet shown in red), the change of the migration regime occurs first, while in the three other cases it is the reverse. Once the migration regime changes to type~II, the rate slows down (bottom centre panel of Fig.~\ref{fig:giant-single}) but the accretion remains mostly constant. Thus, accretion dominates at the onset of this stage, but this reverses at the end. In contrast, \citet{2009A&AMordasinia} used the equilibrium values of the radial gas flow for both gas accretion and migration. Thus, the slope of detached planet migrating with the planet-dominated case of the Type~II regime exhibited a common slope in the mass-distance diagram. It should also be noted that for planets inside roughly \SI{1}{\au}, it happens that the criterion limiting the gas accretion rate changes to the mass in the feeding zone which leads to a reduction of the rate at the end of the formation. It can also be noted that our model allows for the growth of embryos at large separation (up to about \SI{30}{\au}, unlike the work of \citet{2019AAJohansenBitsch}. The difference is mainly related to the planetesimals size. Smaller planetesimals have lower eccentricities and inclinations because of more efficient damping by the disc gas, and in addition, a larger capture probability by the planets for a given surface density because of the more strongly drag-enhanced capture radius for small planetesimals. This results in a larger accretion rate of solids, which enables planets to sufficiently grow to undergo runaway gas accretion before the dispersal of the gas disc also in the outer parts of the disc.

The formation of Jupiter-like planets with migration and planetesimals accretion follows a different pattern in the one-planet-per-disc approximation studied here than what was found by some other models using the in situ (and one-embryo-per-disc) approximation. For the latter, the favoured scenario is that a core between \num{10} and \SI{20}{\mearth} forms early (less than $\SI{e5}{\year}$) and undergoes runaway gas accretion only close to the dispersal of the gas disc \citep{1996IcarusPollack,2018NatAsAlibert}. The slow accretion of planetesimals, resulting in a steady luminosity, is able to prevent runaway gas accretion during the intermediate stage. This intermediate stage is the problematic part when migration is included; the reason being that migration is most efficient for planets that are between \num{10} and \SI{50}{\mearth} (see Sect.~\ref{sec:dyn-mig-map} and Fig.~\ref{fig:mig-map}). Hybrid pebble-planetesimals \citep{2018NatAsAlibert} or pure pebble \citep{2019AABitschA} accretion models can account for the migration during the intermediate phase, as the cores are able to form at larger separation, provided that far out, bodies emerge early and massive enough to be able to efficiently capture pebbles.

The simulations presented here show a situation with fast migration where no intermediate stage is possible, because the planets would otherwise end up at the inner edge of the gas without the opportunity to undergo runaway gas accretion. This means that the cores must form just at the time to undergo runaway gas accretion. The usual picture of the formation of Jupiter-mass planets in our model is then more similar that what was found by \citet{2005A&AAlibert}, with an almost nonexistent intermediate phase (left panel on the second row of Fig.~\ref{fig:giant-single}). As the accretion time scale are longer at large separation, the embryos will accrete their mass over a longer period. At the same time, the inward migration experienced by the protoplanets means that their feeding zone is not depleted as in the in situ formation scenario.

In contrast, for the multi-embryos simulations that we present in \papertwo{} (see also Sect. \ref{sec:twoexmples}), migration can be significantly altered by mean-motion resonances chains. In that case, the torque acting on one planet must be spread over all the bodies, meaning that the planet with the largest specific torque will migrate slower than it would were it not in a resonance chain. This provides a way to obtain an intermediate stage and less overall efficiency of migration, as we show in that work. This effect leaves open the possibility to have an intermediate stage for the formation of giant planets, as obtained in \citet{2018NatAsAlibert}. Thus, once multiplicity is included, the simulations here become more similar to Jupiter formation models like \citet{2018NatAsAlibert}.

\begin{figure*}
	\centering
	\includegraphics{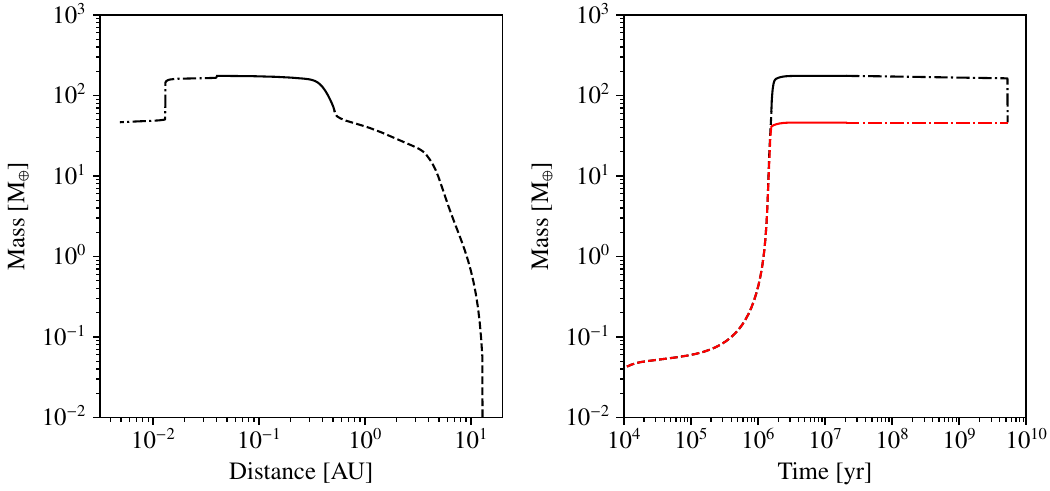}
	\includegraphics{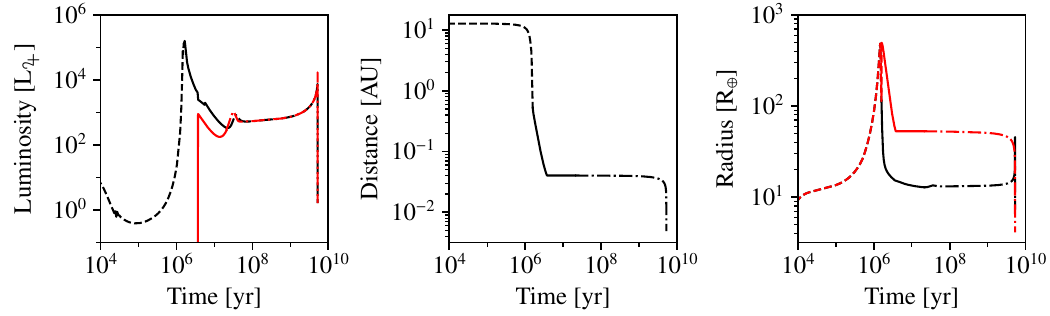}
	\caption{Formation and evolution tracks of one giant planet that ends up being accreted by the star during the evolution stage due to tidal migration. \textit{Top left}: total mass $\mtot$ versus distance; \textit{Top right}: total mass $\mtot$ (black) and core mass $\mcore$ (red) versus time; \textit{Bottom left}: total luminosity $\ltot$ (black) and the bloating contribution $L_\mathrm{bloat}$ (red) versus time; \textit{Bottom centre}: distance versus time; \textit{Bottom right}: total radius $\rtot$ (black) and outer (attached phase) or Hill (detached phase) radius (red) versus time.}
	\label{fig:giant-tidal}
\end{figure*}

\subsection{More massive planets}

Figure~\ref{fig:giant-single} shows the formation tracks of planets that are in the \num{2} to \SI{10}{\mj} range. Compared to the planets previously discussed, these ones show a greater range of initial locations (from \num{6} to \SI{40}{\au}) and overall effect of migration. The planet shown in orange is the quickest to accrete a massive core and undergo runaway gas accretion, due to both the more massive disc and the inner location. The latter is made possible due to the disc's mass. This is also the one to migrate the least before reaching \SI{10}{\mearth} because 1) the fast formation limits the effect of migration and 2) enters a convergence zone (see Fig.~\ref{fig:mig-map} and the discussion in Sect.~\ref{sec:dyn-mig-map}). As the boundary of convergence zone moves inward \citep{2010ApJLyra,2014A&ADittkrist} and to lower planetary masses over time, the planet shown in red will encounter the convergence zone at a different location, which will not affect the planet as much.

Unlike the Jupiter-mass planets, all the ones of this group first switch to type~II migration before going to the detached phase. This is seen on the top right panel of Fig.~\ref{fig:giant-single}, where the tracks become dashed and thin during a brief section. The slope break that was discussed in for the Jupiter-mass is stronger for the two innermost planets. Comparing the time evolution of the two, it can be noted that the migration rate remains mostly constant while in the type~II regime, while the accretion rate decreases. Concerning the radius and luminosity, we observe that all the planets show a similar behaviour even with the difference in the final location.

\subsection{Giant planets ending in the star by tidal migration}

As an illustration how close-in planets are affected by the newly added physical processes during evolution, we finally discuss the formation and evolution of a close-in giant planet. These will raise tides onto the star, which will result in tidal migration. The consequence is that the planet can be accreted by the star at some point during its evolution. We show such a case in Fig.~\ref{fig:giant-tidal}. The formation stage looks quite similar to the previous example, with the difference that the planet ends at a close-in location, \SI{0.04}{\au}. The radius shrinks already before the planet goes to the detached phase, because it experiences a strong inward migration at the same time (as it can be seen in the lower right panel of Fig.~\ref{fig:giant-tidal}). As the planet migrates inward, the Hill radius shrinks. Once the detached phase begins, the Hill radius continues th shrink as further inward migration continues.

As this planet is close to the star (\SI{0.04}{\au}), the evolution stage is different from the case shown previously. The luminosity increases over time time, the envelope gradually expands and looses mass due to atmospheric escape and the planet migrates further inward due to the tides raised onto the star. The migration rate increases over time due to its strong dependence on the distance between the planet and the star (see Eq.~(\ref{eq:tidal-mig})). To determine the reason for the luminosity increase, we print alongside the total value, the contribution from bloating (Eq.~(\ref{eq:lumi-bloat})). We see that from late in the formation stage until the end, this contributes to nearly all the planet's luminosity. And as it goes with the stellar flux, it increases at late times due to tidal migration. The luminosity increase in turns leads to an expansion of the envelope, which increases the loss rate by atmospheric escape. But rather than this being the main cause of gas loss, we see that the bulk of the envelope is removed because it overflows the Hill sphere. This occurs suddenly at the end of the planet's life, once the outer radius gets larger than the Hill sphere. Only a bare core remains, which get accreted by the star shortly thereafter.

\subsection{Summary for giant planets}

The formation and evolution of giant planets involves multiple concurrent processes. Migration being most efficient during the onset of the gas runaway accretion, this phase must occur in a relatively short time for the planets to not end up at the inner edge of the disc, in the absence of another planet to prevent migration. This also means that the cores must form late (i.e. shortly before the dispersal of the disc) to prevent a massive envelope from being accreted. Close-in planets will experience addition effects during their evolution, such as atmospheric escape and inward tidal migration that can lead to accretion by the star. In the latter case, it is possible for Hill sphere overflow to cause the loss of most of the envelope.

\section{Individual systems}
\label{sec:results}

After discussing formation pathways of terrestrial under idealised conditions, and of single giant planets, we finally show results obtained with the full model. Using many embryos per system, the model is able to produce a very large variety of planetary systems. These range from terrestrial planets (as we saw in the previous section) to giant planets. We first provide two examples of the temporal emergence of planetary systems and then show the variety of the final architecture of 23 systems.

\subsection{Two examples of the temporal evolution of the emergence of planetary systems}\label{sec:twoexmples}
Figures \ref{fig:aMsystem30} and \ref{fig:tasystem30} show the formation of a planetary system in a protoplanetary disc with a low initial content of solids (System 30 in \texttt{NG76}, see \papertwo). The initial disc gas mass is \SI{0.023}{\msun} while [Fe/H] is -0.13, corresponding to a dust to gas ratio of 0.011. This results in a low initial solid content of \SI{65.1}{\mearth} in the disc of planetesimals. The disc is seeded with 100 lunar mass embryos at $t=0$, distributed uniformly in the logarithm of the semi-major axis inside of \SI{40}{\au} (see \papertwo{} for more details on the initial conditions).

\subsubsection{Low initial solid content}

\begin{figure*}
	\centering
	\includegraphics[width=\textwidth]{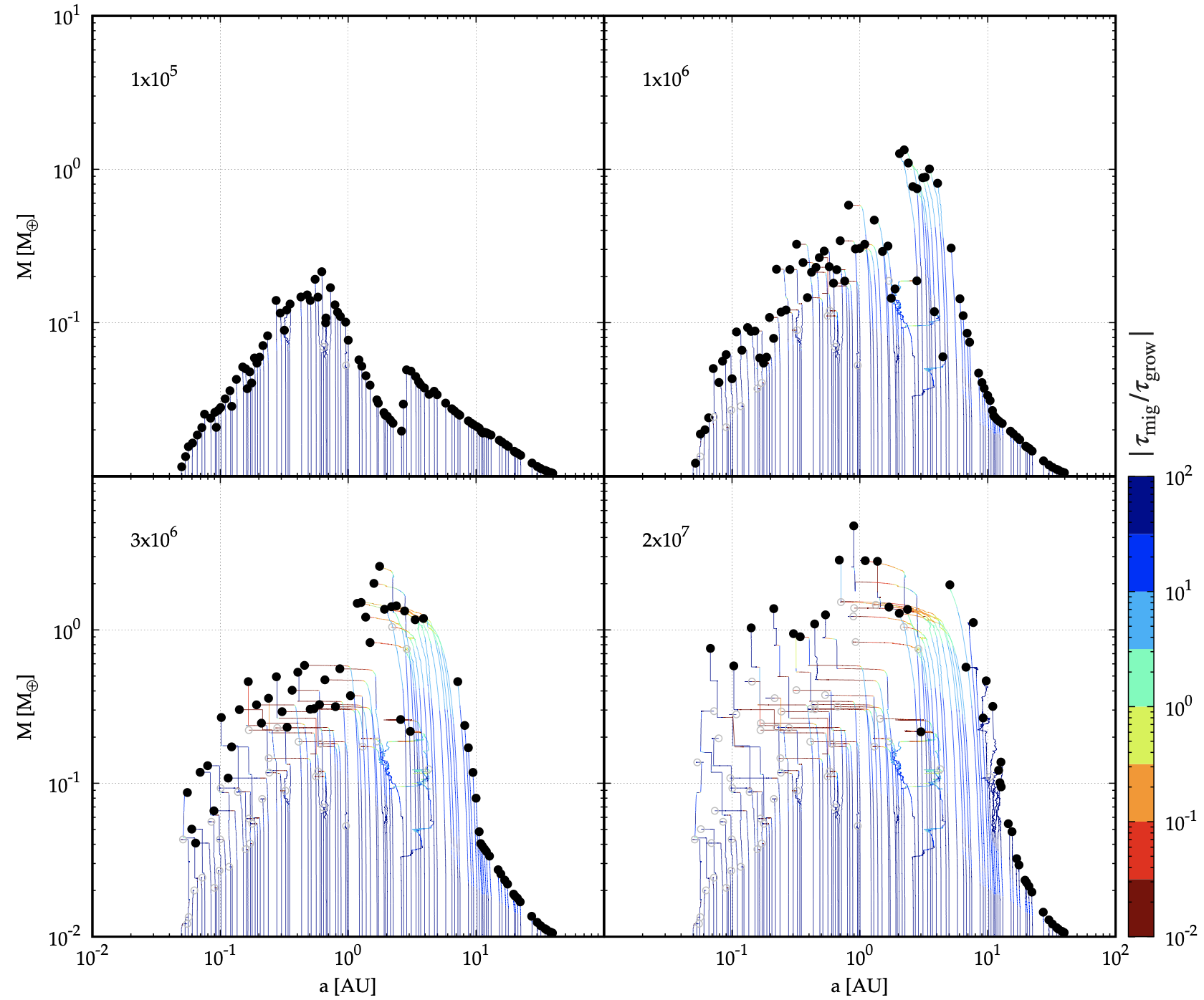}
	\caption{Example of the formation of a planetary system from initially 100 lunar-mass embryos in low gas mass initial mass \SI{0.023}{\msun}), low metallicity ([Fe/H]=-0.13) disc. The initial mass of planetesimals is \SI{65.1}{\mearth}. Four moments in time (in years) are shown. Lines shows the growth tracks in the semi-major axis-mass plane. Black points show (proto)planets existing at a given epoch. Grey open circles show the last position of protoplanets that were accreted by another more massive body in a giant impact. The colours of the lines are $|\tau_{\rm mig}/\tau_{\rm grow}|=|d \ln{m}/d \ln{a}|$.}
	\label{fig:aMsystem30}
\end{figure*}

\begin{figure*}
\includegraphics[width=\textwidth]{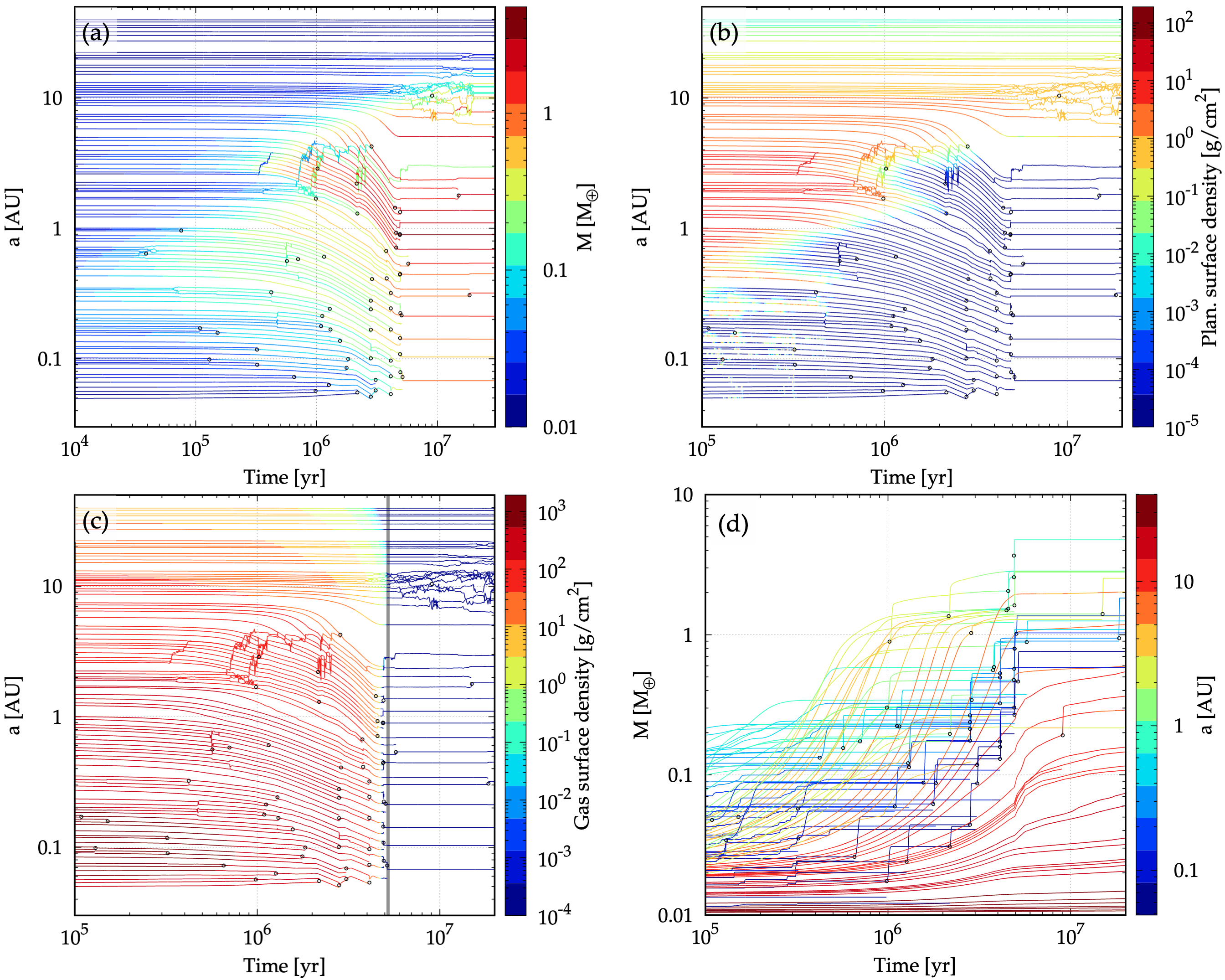}
\caption{Same system as in Fig. \ref{fig:aMsystem30}, but now showing the semi-major axes $a$ of the planets as a function of time, colour coding in panel (a) the planets' mass, in (b) the planetesimal surface density in the planets' feeding zone, and in (c) the local gas surface density. Here, the vertical line indicates the moment of gas disc dissipation. Panel (d) shows mass as a function of time, colour coding the semi-major axis. Small black circles indicate giant impacts, by showing the position or mass of the target (the more massive collision partner) at the moment of the impact.}
\label{fig:tasystem30}
\end{figure*}

Many aspects of the emergence of the planetary systems can be understood with the comparison of the timescales of growth and migration, and the consequences of (large-scale) dynamical instabilities caused by the gravitational interactions of protoplanets. Therefore we colour code in Fig. \ref{fig:aMsystem30} showing the temporal evolution of the system in the $a-M$ plane the tracks of the planets by the ratio $|\tau_{\rm mig}/\tau_{\rm grow}|=|d \ln{m}/d \ln{a}|$. Regarding the timescales, it is of fundamental importance that the oligarchic planetesimal accretion timescale increases with increasing planet mass \citep[e.g.][]{2003IcarusThommes}, whereas the orbital migration timescale in the Type I regime decreases with planet mass \citep[e.g.][]{1989ApJWard}.

At the beginning (\SI{e5}{\year}, top left panel of Fig.~\ref{fig:aMsystem30}), the quasi in-situ accretion of planetesimals present in the initial)feeding zone of the embryos is the dominating process. Migration occurs at these very low masses on a much longer timescale, leading to nearly vertical upward tracks. We note that the model does not include any artificial reduction factors of Type~I migration. The specific distance dependency of the mass to which the protoplanets have grown by \SI{e5}{\year} is given by the following interplay of growth timescale as a function of orbital distance and the local availability of solids: from the innermost embryo at about \SI{0.03}{\au} to the one at about \SI{0.6}{\au}, the protoplanets have already grown to the local planetesimal isolation mass \citep{1987IcarusLissauer}. Given the planetesimal surface density scaling with $r^{-3/2}$, the isolation mass increases with orbital distance. As can be seen in Panel b of Fig. \ref{fig:tasystem30} which shows the mean planetesimal surface density in the feeding zone of the planets, at \SI{e5}{\year}, the surface density is already strongly depleted in the inner parts of the disc. Between the local maximum at \SI{0.7}{\au} and the water iceline at \SI{2.7}{\au}, the mass is in contrast decreasing with distance because protoplanets further out grow slower. The next feature is a sharp increase of the protoplanets' mass by about a factor 2 across the water iceline because of the increase of the solid surface density. One protoplanets grows in the transition zone, giving it an intermediate mass. Outside of the iceline, the masses decrease again with distance because of the longer growth timescales. For the protoplanets in the inner part that have already reached the isolation mass, the growth is temporally stalled. Because of the very low (isolation) masses of these protoplanets, orbital migration is nevertheless negligible.

At the very beginning, all protoplanets grow as if they were the only bodies in the disc, not feeling the influence of the other protoplanets. With increasing mass, the interaction (either directly via N-body interactions) or indirectly via resonant migration, become important. By \SI{e5}{\year}, the first dynamical interactions have started among some of the more massive protoplanets, which is visible as a `jitter' in some tracks, and two collisions, which are shown by two open grey circles.

At \SI{1}{\mega\year} (top right panel in Fig. \ref{fig:aMsystem30}), inside of the iceline, the character of growth has changed from planetesimal-dominated, to some first growth via giant impacts (embryo-embryo collisions) for some protoplanets or stalled growth for others. As can be seen in Panel a of Fig.~\ref{fig:tasystem30} which shows the semi-major axis of the (proto)planets as a function of time colour coding the mass, about 10 further giant impacts have occurred. This has allowed the protoplanets in the inner disc to grow beyond the local isolation mass. As visible in Panel b of Fig. \ref{fig:aMsystem30}, at \SI{1}{\mega\year}, the planetesimal disc is now depleted out to about \SI{1.3}{\au}, and as time proceeds, the depletion moves even further out. We thus see a growth wave moving outwards \citep{2003IcarusThommes}. All solid mass has been transferred into the embryos in this part, and their mutual interaction (giant impacts) governs the further mass growth. This implies that the accretion of planetesimals is only important at the early phases when the planets grow mostly in-situ. In the outer disc beyond the iceline, growth in contrast still proceeds mainly via planetesimal accretion, as there is a larger mass reservoir available. Between 2 to \SI{4}{\au}, a group of about 10 protoplantes with a mass of about \SI{1}{\mearth} has formed, meaning that the most massive planets are now found further out than before. These protoplanets originate from (just) beyond the iceline. The colours of the lines in Fig.~\ref{fig:aMsystem30} show that migration is still much slower than accretion for these planets at \SI{1}{\mega\year}, but some slight inward migration is now occurring, causing the tracks to bend inwards. This applies also to the inner disc, where horizontal tracks are visible. They result from the depletion of the planetesimal disc, and the fact that the cores are of such a low mass that virtually no gas accretion is possible.

At \SI{3}{\mega\year} (bottom left panel of Fig. \ref{fig:aMsystem30}), in the inner disc, the dominant effect is further growth via giant impacts. About 25 protoplanets with masses between the one of Mars and Earth are now present. In the outer disc, beyond the iceline, the aforementioned group of the about 10 most massive protoplanets has grown further now reaching a maximum mass of \SI{3}{\mearth}, and has also migrated further inward. As these planets migrate into zones that have been previously depleted by inner planets (in particular inside of the iceline), planetesimal accretion is quickly stalled. This means that planetesimal accretion for migrating planets is usually limited in low-mass multiple systems like the one present here. This means that a possible shepherding effect \citep{1999IcarusTanakaIda} that we do not include in the model should not affect the outcome very much, except for a transition phase where $\tau_{\rm mig}\approx\tau_{\rm acc}$ for some planets. This phase can be seen for the outer group from the cyan line colours. As can be seen in Panel a of Fig.~\ref{fig:tasystem30}, the planets capture each other in very large resonant convoys and migrate together \citep[e.g.][]{2008AACresswellNelson,2013A&AAlibert}. In this configuration, outer more massive planets push inner smaller planets.

As visible in Fig. \ref{fig:tasystem30} by the small black circles, many giant impacts seem to occur in groups (i.e. at similar moments in time in fast sequence): a first group occurs at about \SI{3}{\mega\year}, a next one at \SI{4}{\mega\year}, and again one at the moment when the disc inside of about \SI{2}{\au} becomes free of gas. This is visible in Panel c of Fig.~\ref{fig:tasystem30}, which colour codes the gas surface density at the planets' position. This moment corresponds to the opening of the inner hole in the gas disc because of internal photoevaporation (cf. Fig.~\ref{fig:gas_disc_str}). At this moment, the damping effect of the gas vanishes, allowing orbit crossings and collisions \citep[e.g.][]{2010ApJIdaLin}. The outer gas disc dissipates a bit later, at \SI{5.1}{\mega\year}, shown by the vertical line in Panel c of Fig.~\ref{fig:tasystem30}. After the dissipation of the disc, only 3 more giant impacts occur in this system to \SI{20}{\mega\year}.

\begin{figure}
\includegraphics[width=9cm]{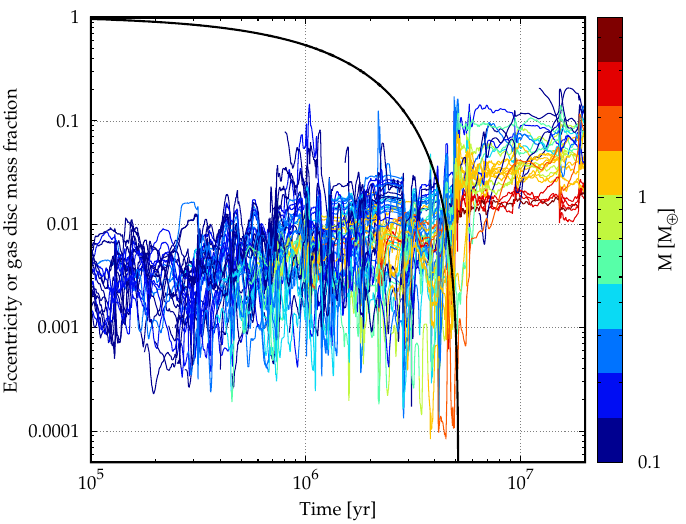}
\caption{Temporal evolution of the eccentricities of the planets of the system emerging in the low-mass disc shown in Fig.~\ref{fig:aMsystem30}. Colours indicate the planet mass. For better visibility, only planets more massive than \SI{0.1}{\mearth} are shown. The curves are running averages such that one sees more clearly the mean values instead of rapid variations of the eccentricities. The thick black line is the mass of the gas disc relative to the value at 10$^5$ years, which is in turn very similar to the initial value. The increase of the eccentricities at around \SI{5}{\mega\year} when the gas disc dissipates is visible.}
\label{fig:tesystem30}
\end{figure}

The temporal evolution of the eccentricities is shown in Figure \ref{fig:tesystem30}. The colours show the planet mass. For clarity, only planets with a mass of at least \SI{0.1}{\mearth} were included. One can clearly see the increase of the typical values of the eccentricities near the time the gas disc dissipates at around \SI{5}{\mega\year}. Before, typical values of the eccentricities are of the order of $10^{-3}$ to a few $10^{-2}$. After disc dissipation, they increase to values between about 0.02 to 0.2. Such values are expected from the increase of the velocity dispersion of the orbits until they are comparable to the escape velocity from their surfaces resulting from close encounters, once the damping by the gas is gone \citep{2004ApJGoldreich}. One also sees that more massive bodies tend to be less eccentric, likely a consequence of energy equipartition.

In our model, dynamical friction by residual planetesimals is neglected. This would reduce the eccentricities and inclination of the protoplanets. This implies that our model tends to overestimate the eccentricities and inclinations of lower mass planets for which dynamical friction by planetesimals would play a role.

The general sequence of solid growth that is first dominated by the near in-situ accretion of planetesimals followed by the second phase of growth via giant impacts is well visible in Panel d of Fig. \ref{fig:tasystem30}. It shows the mass of the protoplanets as a function of time. The line colours show the semi-major axis. We note how the transition between the two regimes occurs the later the more distant a planet is. At the largest orbital distances where embryos were inserted into the disc (maximum starting distance is \SI{40}{\au}), nearly no growth at all has occurred during the simulated period. As described in Sect. \ref{sec:collisiontreatment}, numerically speaking, we add the mass of the impactor in a giant impact over a timescale of \SI{e4}{\year} to the target. This is the reason why the vertical steps in the curves corresponding to giant impacts (indicated with the black circles) are not strictly vertical. This is visible particularly at the early ages.

The bottom right panel of Fig. \ref{fig:aMsystem30} shows the system at \SI{20}{\mega\year}, which corresponds to the time where we stop the N-body integration and planetesimal accretion. Between 3 and \SI{20}{\mega\year}, numerous giant impacts have reduced the number of planets and destroyed the mean motion resonances (see also Fig. \ref{fig:tasystem30}). The inner system now contains 8 roughly Earth-mass planets, exhibiting a certain inter-system similarity of the mass scale \citep{2017ApJMillholland} with an increase towards the exterior \citep{2018AJWeiss}. At \SI{0.7}{\au}, there is a sudden increase in the typical mass, corresponding to the transition from volatile-poor planets that have formed inside of the iceline, to very volatile-rich planets originating from beyond the iceline. Compared to the original location of the iceline at \SI{2.7}{\au}, there was thus an inward shift in this transition by about \SI{2}{\au} because of orbital migration.

In the end, the planetesimal disc is depleted out to about \SI{5}{\au}. Outside, about \SI{35}{\mearth} remain in the form of planetesimals. This corresponds to a fraction of about \SI{46}{\percent} of the initial planetesimal mass that was converted into planets. Since we follow the accretion for only \SI{20}{\mega\year}, this remaining planetesimal mass must be considered an upper limit for the actual mass of remaining planetesimals, as over longer timescales, the distant protoplanets would continue to accrete. However, since the accretion timescales at several \SI{10}{\au} in the absence of eccentricity damping (corresponding to orderly growth) become extremely long \citep{2004ApJIda1}, at least some part of these planetesimals could remain to eventually form a debris disc, in analogy to the Kuiper belt beyond the orbit of Neptune in the Solar System.

\subsubsection{High initial solid content}
\label{sect:highinitialsolidcontent}

\begin{figure*}
	\centering
	\includegraphics[width=\textwidth]{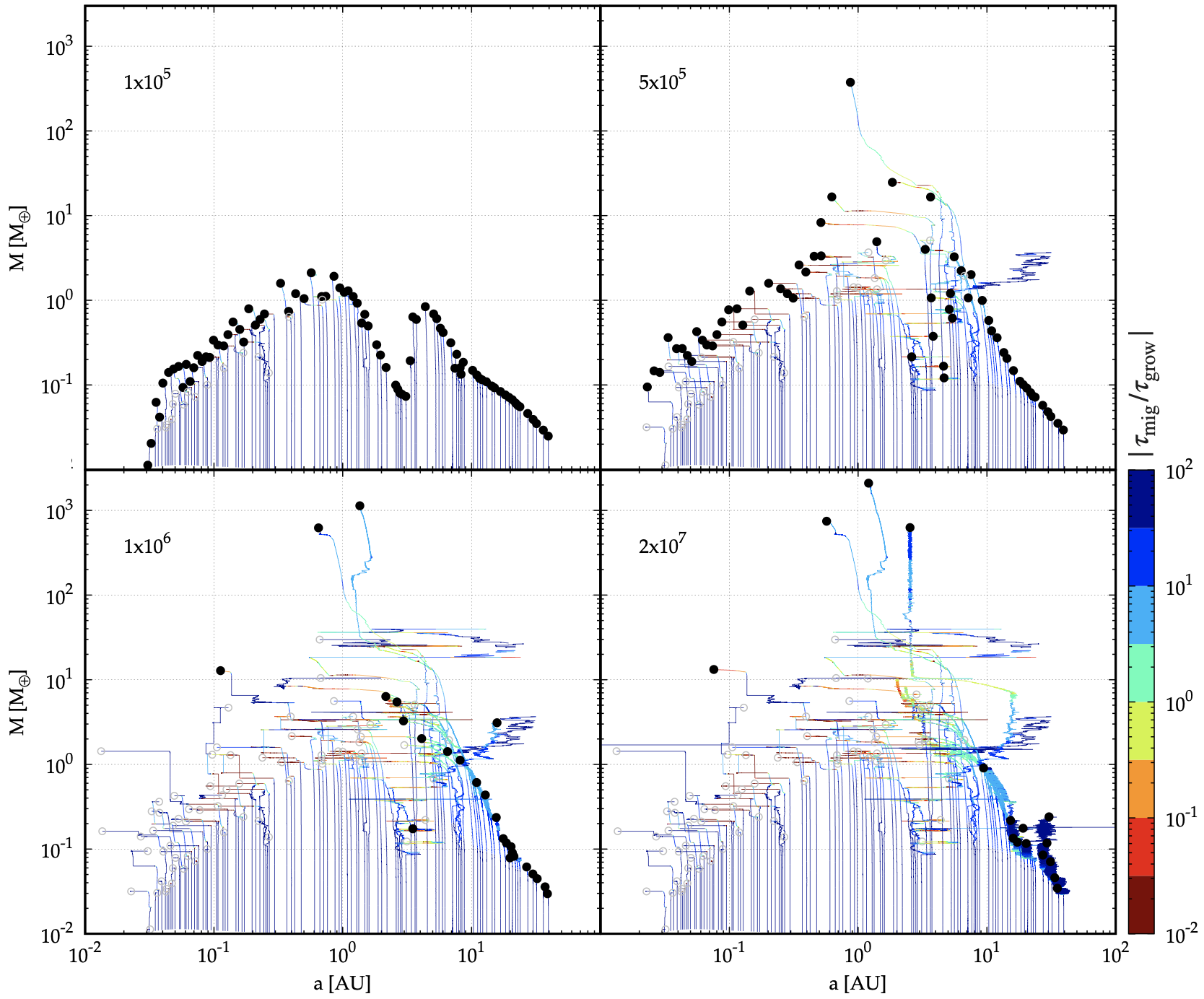}
	\caption{Example of the formation of a planetary system from initially 100 lunar-mass embryos in a high gas mass (initial mass \SI{0.066}{\msun}), high metallicity ([Fe/H]=0.23) disc. The initial mass of planetesimals is \SI{432}{\mearth}. The plot is analogous to Fig. \ref{fig:aMsystem30}, but the y-axis now extends to much higher masses, and the moments in time that are shown are different. At the end of the simulation at \SI{20}{\mega\year}, this system contains one close-in sub-Neptunian planet, three giant planets, and a group of outer very low-mass planets.}
	\label{fig:aMsystem852}
\end{figure*}

\begin{figure*}
\includegraphics[width=\textwidth]{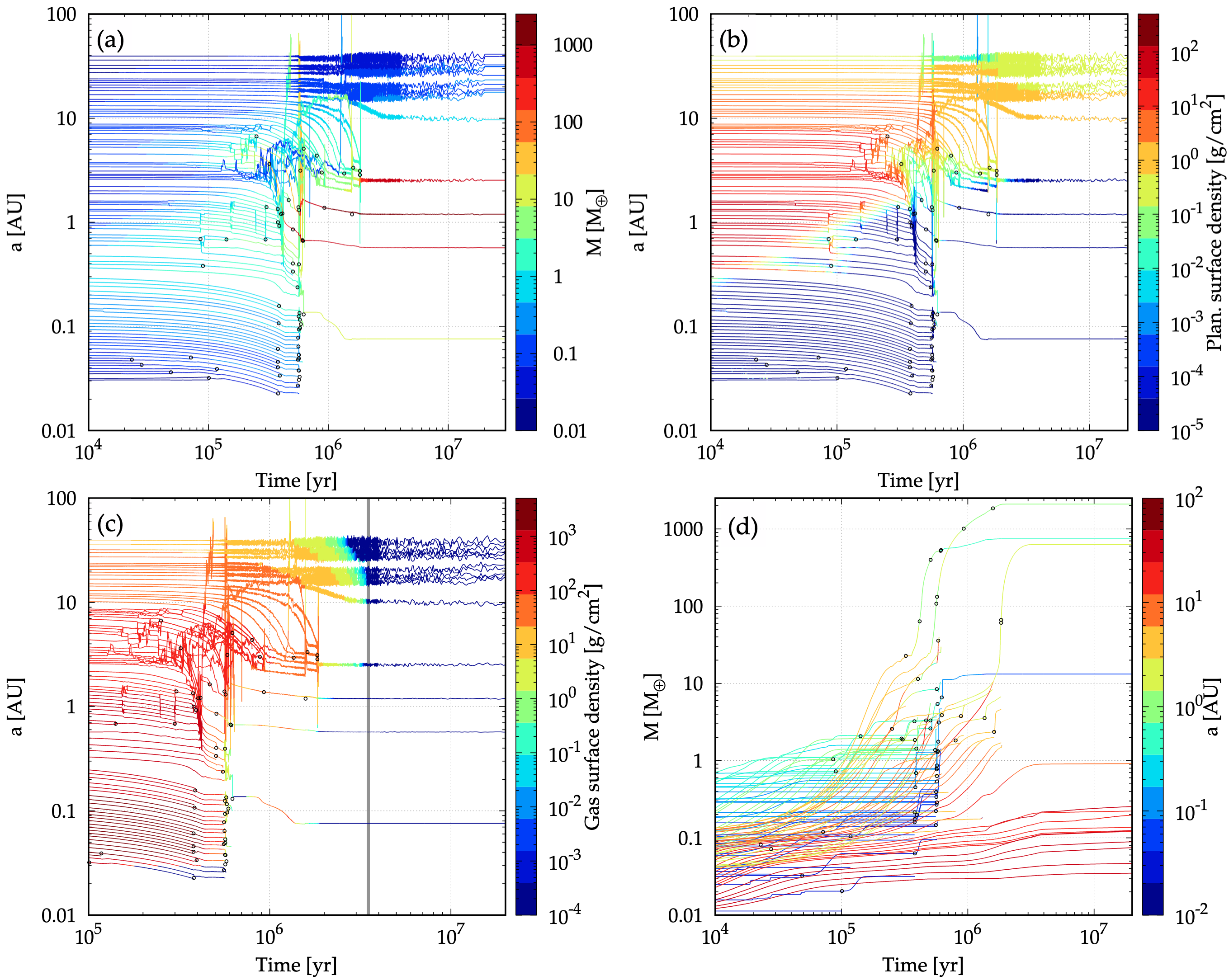}
\caption{Temporal evolution of the system shown in Fig. \ref{fig:aMsystem852}. The four panels have the same meaning as in Figure \ref{fig:tasystem30}.}
\label{fig:tasystem852}
\end{figure*}

The second system we consider is System 852 in \texttt{NG76} (\papertwo). The initial conditions are here a disc mass of \SI{0.066}{\msun} and a metallicity of [Fe/H]=0.23. This leads to an initial planetesimal mass of \SI{432}{\mearth}, 6.6 times as much as in the first example. As in the previous case, 100 lunar mass embryos are put into the disc at the beginning, uniform in log of the semi-major axis out to an orbital distance of \SI{40}{\au}. The evolution in the $a-M$ plane is shown in Fig.~\ref{fig:aMsystem852}. The semi-major axis and mass as function of time is shown in Fig.~\ref{fig:tasystem852}.

In the top left panel of Fig. \ref{fig:aMsystem852} we see that at \SI{e5}{\year}, the basic picture regarding the (relative) mass of the protoplanets as a function of orbital distances is analogous to the one in the low-mass disc at the same time. In absolute terms, the planet masses are, however, about one order of magnitude larger. As can be seen in Panel b of Fig. \ref{fig:tasystem852}, the planetesimal disc is already strongly depleted out to about \SI{1}{\au}. Some giant impacts have also already occurred in the inner disc. This fast development away from the initial conditions is a sign that the early phase of solid growth (from dust to embryos) should be treated more explicitly \citep[e.g.][]{2021AAVoelkel}.

The situation at \SI{0.5}{\mega\year} is already quite different, as a first core has undergone runaway gas accretion, at about \SI{0.35}{\mega\year}. By \SI{0.5}{\mega\year}, its mass has already grown to about \SI{350}{\mearth}. In the end, it will have a mass close to \SI{750}{\mearth} and be the innermost giant planet. The starting position of this embryo was \SI{4.5}{\au}. The water iceline in this system is for comparison found at about \SI{3.4}{\au}. The formation of this first giant planet does not yet strongly affect the rest of the system, at least at this moment. In the inner system, we in particular see a similar development as in the low-mass disc: the formation of very large resonant convoys and some giant impacts. However, shortly after \SI{0.5}{\mega\year}, a second core, located about \SI{0.5}{\au} outside of the first giant, also starts runaway gas accretion. The embryo of this planets started at about \SI{5.3}{\au}, and was for some time in a resonant configuration with the first giant-to-be. It will eventually become the most massive giant planet in the system (about \SI{2100}{\mearth}) at \SI{1.2}{\au}.

The growth of this second giant planet has important system-wide consequences, as can be seen in the panel at \SI{1}{\mega\year}. It not only destabilises several Neptunian planets in the vicinity of the forming giants, but it also sends a protoplanet of about \SI{3}{\mearth} from about \SI{0.9}{\au} into the inner system (close to \SI{0.1}{\au}). The orbit of this planet is eccentric, and triggers numerous giant impacts among the protoplanets in the inner system (see Panel d of Fig. \ref{fig:tasystem852}). These orbit crossings and impacts are facilitated because the runaway gas accretion by the two forming giants strongly reduces the gas surface density in the inner disc temporarily, reducing eccentric damping (Panel c of Fig. \ref{fig:tasystem852}). In the end, only the intruder from the exterior remains, the mass of which has increased to about \SI{13}{\mearth} by accreting the local protoplanets. The formation of the second giant also scatters an initially low-mass protoplanet (\SI{0.7}{\mearth}) from about \SI{2}{\au} onto a very eccentric orbit with a semi-major axis of about \SI{15}{\au}. This protoplanet grows then out there (potentially in a monarchical growth mode, \citealp{2005SSRvWeidenschilling}), reaching a mass of about \SI{3}{\mearth} by \SI{1}{\mega\year}.

By about \SI{1.4}{\mega\year}, its mass has increased to \SI{8}{\mearth}, and a phase of rapid inward migration sets in. It then runs from outside into a group of 7 protoplanets at about 2 to \SI{4}{\au} that are captured in MMRs with the giant planet that had formed second (see Panel a of Fig. \ref{fig:tasystem852}. A series of giant impacts occur, and at \SI{1.8}{\mega\year}, the protoplanet coming from the outside starts runaway gas accretion. It eventually becomes the third giant planet in the system with a mass of about \SI{630}{\mearth} at \SI{2.5}{\au}. Interestingly enough, this implies that giant planets in a system need not to be strictly coeval, which could be of importance for example for direct imaging observations. Here, the outermost giant is nearly \SI{1.5}{\mega\year} younger, and starts runaway accretion only when the inner two planets have already reached nearly their final mass. Actually, the fact that this third outer planet forms strongly reduces the gas accretion rate of the middle giant, by reducing the gas surface density in the inner system (see Panel c of Fig. \ref{fig:tasystem852}). So, more precisely speaking, the formation of this third giant actually sets the final mass of the giant planet inside of it. A comparable, transient depletion of the inner gas disc is already also seen when the inner two giants form, as mentioned. It should be noted that the degree of depletion of the inner disc because of gas accreting giant planets might be overestimated in our model \citep{2019AAManara,2019MNRASNayakshin,2020AABergez-Casalou}. Then, this indirect interactions via the disc would be reduced. The lifetime of the gas disc is in this example about \SI{3.4}{\mega\year}. This is less than the lifetime of the low mass system studied in the previous section, despite the higher starting mass. The difference is mainly a consequence of the higher external photoevaporation by nearly a factor 5 (it is an independent initial condition, see \papertwo). The gas accretion of the giant planets also contributes to the dispersal by them containing in the end about \SI{0.01}{\msun} of gas (out of the initial disc mass of \SI{0.066}{\msun}).

The temporal evolution of this system shows how the growth of multiple giant planets strongly affects the overall system architecture. This also has important consequences for the giant planets themselves (see Panel d of Fig. \ref{fig:tasystem852}): while they accrete their gas envelopes, they get hit by several lower-mass protoplanets that they destabilise. This increases the core mass of the three giants from about 24, 14 and \SI{10}{\mearth} at the onset of gas runaway accretion to clearly higher finales values of 64, 26, and \SI{21}{\mearth}, respectively. Such giants impacts thus strongly influence the final heavy element content \citep{2018AJThorngrenFortney}, and could potentially lead to the existence of diluted cores as found in Jupiter \citep{2019NatureLiuHori}.

At the end of the simulation at \SI{20}{\mega\year}, the system contains four planets more massive than one Earth mass. During the emergence of the system, eight protoplanets have collided with the host star and four were ejected. About \SI{244}{\mearth} of planetesimals remain out of the starting value of \SI{432}{\mearth}, corresponding to a difference of \SI{188}{\mearth}. However, the planets actually existing at the end contain only \SI{123}{\mearth}, meaning that about \SI{65}{\mearth} of planetesimals were `lost' because they were either directly ejected or contained in planets that were themselves ejected or fell into the star. This correspond to a solid conversion efficiency of planetesimals into planets of about \SI{28}{\percent}. Over gigayear timescales, atmospheric escape reduces the mass of the close-in planet at \SI{0.08}{\au} from 13.2 to \SI{11.6}{\mearth}, but it does retain a remaining H/He envelope. Under the effect of bloating, the planet therefore has a relatively large radius of \SI{5.3}{R_\oplus} at \SI{5}{\giga\year}. It is an example of an inner sub-Neptunian planet in a system with outer giant planets (see \paperthree).

Finally, it is worth mentioning that systems with three giant planets are statistically a very rare outcome in the population synthesis (\papertwo): There are only five such systems among the 1000 synthesised in the nominal population \texttt{NG76}. Systems with one or two giants are in comparison much more common (each about 100 systems). In the system at hand, orbital stability is provided by the giant planets residing in the 3:1 mean motion resonance for both pairs of planets. This allows them to remain stable \citep{2016CeMDAAlves} despite their relative proximity to each other, corresponding for both pairs to about 6-7 mutual Hill radii, and their significant eccentricities (about 0.08, 0.18, and 0.40 for the inner, middle, and outer planet). We have further tested the stability of this system by extending the orbital integration (including all bodies in the system) from 20 to \SI{100}{\mega\year}. At least on this timescale, the system remained stable without secular growth of the eccentricities.

\subsection{Overview of the diversity of system architectures}

\begin{figure*}
	\centering
	\includegraphics{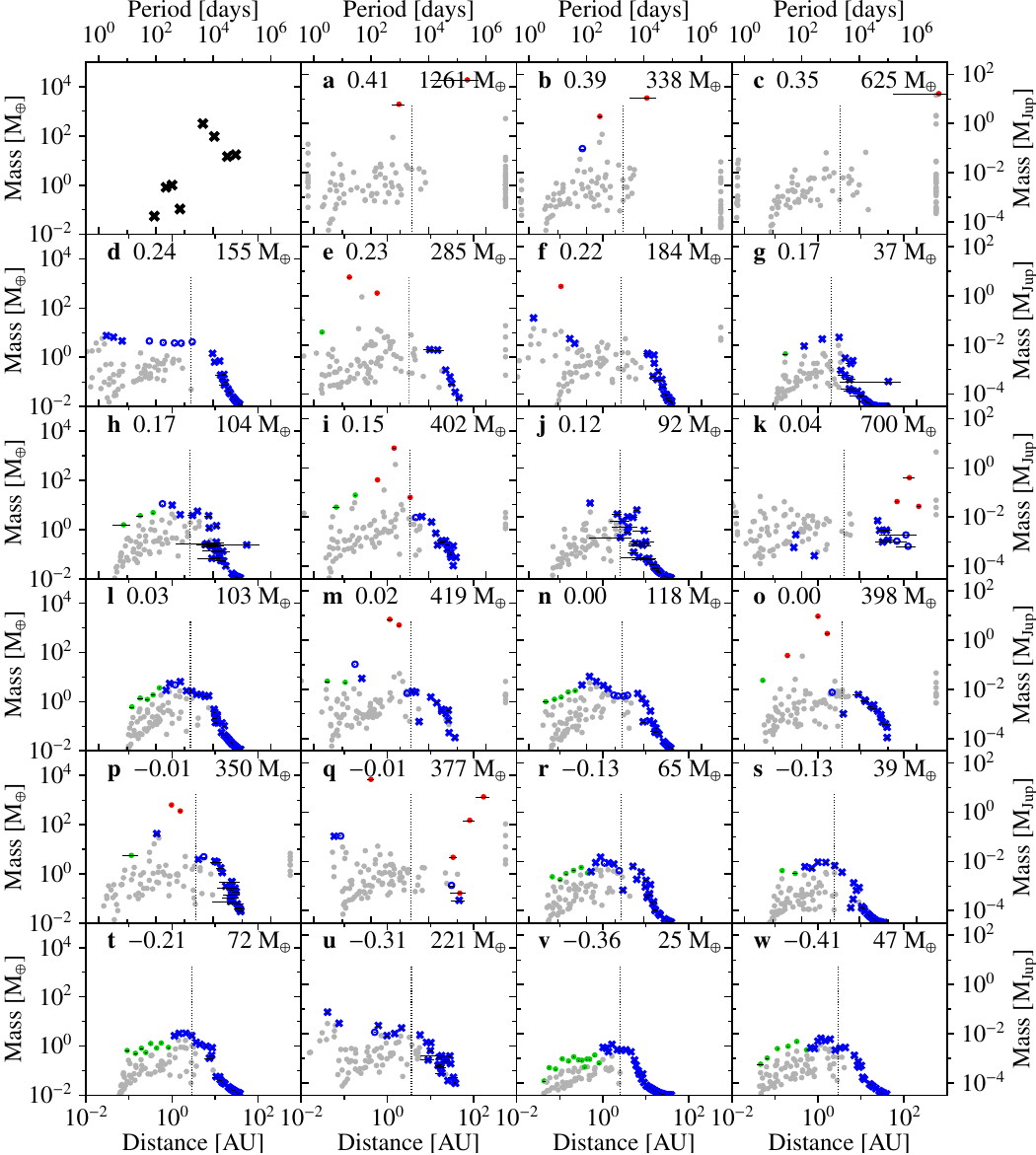}
	\caption{Mass-distance diagrams of specific systems with 100 embryos initially (panels a to w), which are taken from the nominal population predicted for a \SI{1}{\msun} star (\texttt{NG76}). Symbols are as follows: red points show gas-rich planets where $\menv/\mcore>1$. Blue symbols are planets that have accreted some volatile material (ices) outside of the ice line(s). Green symbols are planets that have only accreted refractory solids. Open green and blue circles have 0.1$\leq\menv/\mcore\leq1$ while filled green points and blue crosses have $\menv/\mcore\leq0.1$. For all these bodies, the grey horizontal bars go from $a-e$ to $a+e$. The top left panel with black crosses shows the solar system. Bodies lost because of collisions or ejections are shown in light grey. Planets accreted by the central star are show in the very left of each panel, the ejected ones on the very right and planets that collided with another (more massive) planet are shown at their last position on the diagram. The dotted vertical line in each system shows the location of the ice line. The number after each panel name is the metallicity [M/H] of the system expressed in dex, while the value on the top right is the initial mass of the planetesimals disc.}
	\label{fig:examples}
\end{figure*}

\begin{figure*}
	\centering
	\includegraphics{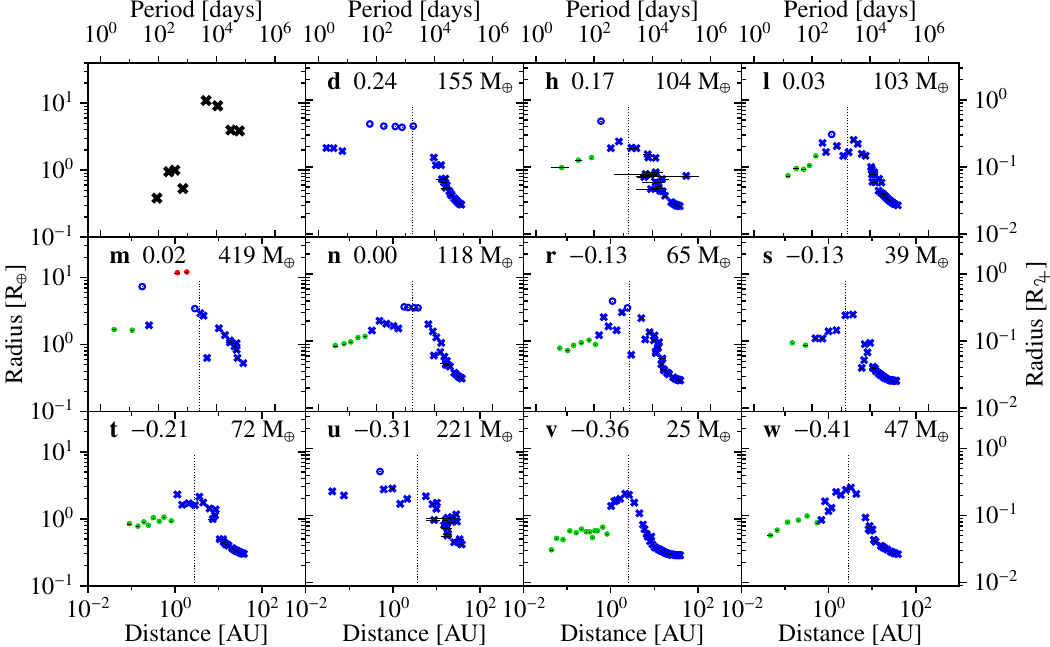}
	\caption{Radius-distance diagrams for a subset of the systems shown in Fig.~\ref{fig:examples}. Here, we focus on systems with multiple low-mass planets; panel letters correspond to the same system in Fig.~\ref{fig:examples}. In contrast to Fig.~\ref{fig:examples}, lost planets are not shown for clarity.}
	\label{fig:examples-radius}
\end{figure*}

Figures Fig.~\ref{fig:examples} and \ref{fig:examples-radius} show the mass-distance and radius-distance of 23 synthetic systems. The solar system is shown in the top-left panel for comparison. All these systems are again taken from the nominal synthetic population \texttt{NG76} for \SI{1}{\msun} star that will be presented in \papertwo. However, here we study these as individual systems without taking into account the likelihood of such systems in populations.
Hereafter we give an overview of some major correlations that we find. For quantitative results, we refer to the next papers of the series.

It is important to point out a potential complication concerning the formation of the outer two giant planets in the system shown in the panel \textit{q} of Fig.~\ref{fig:examples}. These planets accreted their cores and started undergoing runaway gas accretion in the inner region of the system.
They were subsequently moved to the outer region of the disc where they continue to accrete gas. However, the reason for their final distant locations are not planet-planet scatterings. The presence of a inner massive giant planet (in this case, the one at \SI{0.2}{\au} with nearly \SI{30}{\mj}, which corresponds to \SI{3}{\percent} of the mass of the central star, for an initial disc mass of \SI{9}{\percent} of the stellar mass) results here in the outer planets obtaining large eccentricities. This, in turn, causes the prescription for the modulation of the torque (Eqs.~\ref{eq:flind} and \ref{eq:pecc}) to reverse its sign. Via the additional forces added to the \textit{N}-body integrator (Sect. \ref{sec:add-forces}), this if found to lead to outward migration in the present case.

Generally speaking, a positive torque means that the angular momentum of a planet has to grow. For an eccentric planet, this can occur via two ways \citep{2007AACresswell}: by eccentricity reduction (circularisation) or outward migration (increase of the semi-major axis). The different approaches how to translate the positive torque found in hydrodynamical simulations into the additional N-body forces have been inconsistent with one another in the literature in the past in this regard \citep{2020MNRASIda}. A re-assessment was recently made in \citet{2020MNRASIda}, but is not yet included in the simulations shown here.

The problem we encounter in the special setup here (the presence of an inner very massive giant planet) is likely that the eccentricity state towards which eccentricity damping is acting is becoming ill-defined. The setups used to derive the eccentricity and inclination damping expressions and their translation into additional N-body forces \citep[e.g.][]{2000MNRASPapaloizou,2008AACresswellNelson,2013AABitsch,2020MNRASIda} assume that the disc orbits on a nearly circular orbit centred on the star. However, in the case here, the planet and outer disc will tend to orbit the barycenter of the star-inner giant pair, which means that the eccentricity can likely not be stabilised near zero. Figuring out the consequences for the orbital evolution of the different involved planets would likely require dedicated hydrodynamical simulations. This shows a limitation of our N-body approach with additional forces instead of direct hydrodynamical simulations. This implies that the model results for distant giant planets with an inner massive planet must be taken with caution.

\subsubsection{Mass and final number of planets}

The number of planets that remain past the formation stage is anti-correlated to the mass of the formed planets. Systems forming giant planets loose more embryos than the ones forming low-mass planets only. We obtained some systems where only one giant planets remains, for instance in panels \textit{a} and \textit{c} (including a single one in the latter case), where all the other embryos were removed during the formation stage. When this occurs, at least one of the final planets remains on a wide orbit, as it needs to clear the outer embryos. If this is not the case, then we observe that some embryos with low masses remain in the outer region (e.g. panels \textit{e} and \textit{i}).

Systems that still form giant planets, but of lower masses, are able to retain more bodies. We have a few examples that have an architecture in the fashion the solar system, with terrestrial planets inside of giants, such as in panels \textit{e}, \textit{i}, \textit{m} and \textit{p}. However, those are not comparable to the solar system for several reasons. First, the giant planets are quite more massive than in the solar system; it is not uncommon to find masses of the order of \num{5} to \SI{10}{\mj}. Likewise, the terrestrial planets are many Earth masses. Further, the location of the giants is much closer in that Jupiter, with distances that are around \SI{1}{\au}. These findings indicate that 1) the gas accretion rate in the disc-limited regime could be too high and 2) the simple Type~II migration model we employ in this work leads to too much inward migration.

Finally, systems that form low-mass planets only remain with the largest number of bodies. This is seen for instance in panels \textit{l}, \textit{n}, \textit{r}, \textit{t}, \textit{v} and \textit{w}, where many ice-free bodies (shown in green circles) are present at the end.

\subsubsection{Similarity in the low-mass systems}

Systems where only terrestrial planets are present have planets with similar properties. It can be seen in panels \textit{d}, \textit{g}, \textit{h}, \textit{l}, \textit{n}, \textit{r}, \textit{s}, \textit{t}, \textit{v}, and \textit{w}. This is result consistent with observational results about masses and spacing \citep{2017ApJMillholland}. To provide a comparison point with the similarity of planet radii \citep{2018AJWeiss}, we provide a radius-distance diagram in Fig.~\ref{fig:examples-radius}. For the rocky planets, both masses and radii show the same similarity. The transition from rocky to icy planets affects the radii only slightly. More important is the presence of (remaining) H/He envelopes that were not removed by photoevaporation.

We observe a general slight increase of mass with distance, at least in the inner region. This is most likely linked to the surface density profile of solids. The isolation mass $\miso\propto r^{(1.5(2-\betas))}$ \citep{1987IcarusLissauer}, and so since we have $\betas=1.5$, the value increases with distance. This increase stops at locations usually slightly outside of \SI{1}{\au}, which could be due to our limited integration time, as we discussed in the previous section.

\subsubsection{Composition of the close-in planets}

We find that close-in terrestrial planets are likely to be rocky, which is in agreement with inferences from observations \citep{2018ApJJin}. This is especially the case for systems where no planets grow to more than a few Earth masses. We observe in all systems that icy planets are found inside the location of the ice line (the dotted vertical line). Nevertheless, the innermost planets only accrete from the inner region of the disc where the planetesimals are rocky. This indicates that these planets neither migrate from beyond the ice line to their current position, nor get moved to other locations by mean of dynamical instability. It should be noted that in our simulations, planetesimals composition is set from the initial temperature and pressure profile of the gas disc (Sect.~\ref{sec:disc-compo}).

Nevertheless, there are systems without giant planet that consist of only ice-bearing bodies; these are shown in panels \textit{d}, \textit{g} and \textit{u}. These systems form planets that are more massive than the previous ones, with most of them having at least one planet above \SI{10}{\mearth}. The Type~I migration timescale decreases with increasing mass; therefore these more massive planets can migrate from outside of the water ice line to their current position, increasing the compositional diversity of the systems \citep{2018MNRASRaymond}.

Systems with giant planets exhibit different behaviours. Some have only ice-bearing planets (panels \textit{b}, \textit{f} and \textit{q}) while others have also terrestrial planets. In the latter case, the giant planets do not necessarily separate rocky bodies from icy ones. Panels \textit{e} and \textit{o} show systems where rocky and icy planets are separated by giants, while in panels \textit{i}, \textit{m} and \textit{p} icy planets are present both inside and outside of the gas giants. This points at a high diversity of the composition of planets in systems containing both giant and low-mass planets. Correlations between the occurrences of giant planets and others in planetary systems will be investigated in more details in \papertwo. \citet[hereafter \paperthree]{NGPPS3} will look thoroughly at correlations between close-in Super-Earth planets and long-period giants.

\section{Summary and conclusions}

In this work, we presented the Generation III version of the \textit{Bern} global model of planetary formation and evolution. In this generation, the following two main aspects were improved. First is the ability to simulate planets with a mass range from Mars to deuterium-burning planets. Older generations of the \textit{Bern} model could not address terrestrial planets, as they we lacking the giant-impact stage. To reach this goal, we improved the \textit{N}-body integrator so that per disc, hundreds of concurrently forming embryos can now be included. This is crucial for the formation of low-mass planets in general and the Solar System. We also added several new physical processes to take into account the consequences of stellar proximity, allowing us to simulate with the new model planets that cover the widest range of orbital separations, from star-grazing to distant and even rogue planets. Second, the ability to predict self-consistently for multi-planet systems as many directly observable quantities as possible: not only masses and orbital elements as in the past, but also other key observables like luminosities, magnitudes, transit radii, or evaporation rates. To achieve this, we coupled our planet formation model (to \SI{20}{\mega\year}) to our planet evolution model (\SI{20}{\mega\year} to \SI{10}{\giga\year}). Thanks to this, we can now self-consistently and statistically compare the same population to all important observational techniques, as will be done in the series of NGPPS papers. This is crucial, as different methods probe distinct planetary sub-populations. This combined comparison puts extremely compelling and powerful constraints on any theoretical model.

The formation and evolution model follows the envelope structure of the giant planets during they entire lifetime. This allows for example to study the luminosities at any time \citep{2017A&AMordasini}, and enables the comparison with directly-imaged exoplanets \citep[e.g.][]{2017AAVigan}.

The model now includes a multitude of physical processes (see Fig.~\ref{fig:modules}). The following are included during both the formation and evolution phase:
\begin{itemize}
\item A solution of 1D radially symmetric internal structure equations \citep{1986IcarusBodenheimerPollack} is used to calculate the internal structure of the H/He envelope and thus the gas accretion rate (during the attached phase), radius and luminosity, which includes Deuterium burning \citep{2012A&AMolliere} and bloating of close-in planets.
\item The solution of the 1D internal core structure is used to obtain the radius of the solid core with a modified polytropic EOS \citep{2007ApJSeager}.
\item An atmospheric model yields the outer boundary conditions during the attached, detached, and evolutionary phase. For the detached phase, we assume hot gas accretion. For the evolutionary phase, we use a simple grey atmosphere.
\item The host star properties are retrieved from tabulated stellar evolution tracks \citep{2015AABaraffe}.
\end{itemize}
During formation, the following processes are included:
\begin{itemize}
\item The radial structure of the protoplanetary gas disc is computed with a 1D radial (axis-symmetric) constant $\alpha$-disc model. The effects of internal and external photoevaporation are included.
\item The vertical structure of the disc is modelled by building on radiative equilibrium \citep{1994ApJNakamoto}, including viscous heating and stellar irradiation \citep{2012A&AFouchet}. Irradiation now also includes the direct irradiation in the disc midlplane important when the disc becomes optically thin.
\item Planetesimals are presented by a 1D radial (axis-symmetric) disc, with a surface density and a dynamical state (eccentricity, inclination). The temporal evolution of $e$ and $i$ are explicitly followed, including the dynamic excitation by protoplanets and planetesimals, and damping from gas drag \citep{2013A&AFortier}. The composition of the planetesimal and the position of ice lines is found from an equilibrium condensation model \citep{2015A&AThiabaud}.
\item The equation for the planetesimal accretion rate of the protoplanet is computed assuming the oligarchic regime \citep{2006IcarusChambers}. The enhancement of the planetesimal capture radius because of the planetary H/He envelope is included \citep{2003A&AInaba}.
\item A prescription based on Bondi- and Hill-type gas accretion in the 2D and 3D cases limits the planetary gas accretion rate in the disc-limited regime.
\item Gas-driven Type I and Type II orbital migration are computed including the effects of non-isothermality and of the planet’s eccentricity and inclination \citep{2011MNRASPaardekooper,2014MNRASColemanNelson,2014A&ADittkrist}.
\item Full N-body interaction between all the embryos forming concurrently in one disc are tracked using the \texttt{mercury} integrator \citep{1999MNRASChambers}. Orbital migration and the damping of eccentricity and inclination are input in the integrator via additional forces. In
case of a collision, the impact energy is added as an additionally luminosity term \citep{2012A&ABroeg} to the internal structure model. This can lead to the loss of the H/He envelope.
\end{itemize}
During the evolutionary phase we include:
\begin{itemize}
\item XUV-driven atmospheric photoevaporation in the energy and radiation-recombination-limited approximation \citep{2014ApJJin},
\item for close-in planets, the addition of a bloating luminosity modelled with the empirical relation of \citet{2018AJThorngrenFortney}, and
\item tidal spiral-in because of stellar tides \citep{2011AABenitezLlambay}, along with Roche-lobe overflow.
\end{itemize}

We show in Sect.~\ref{sec:terrestrial} where we study the formation of terrestrial planets that provided there are initially enough embryo in each disc, mutual gravitational interactions will stir their eccentricities. Due to the radial excursions, embryos will have access to more material until all the planetesimals are accreted. Afterwards, a phase of giant impacts sets in. Thus, despite the use a fluid-like description for the planetesimals, the model is able to reproduce the giant impact phase of terrestrial planet formation. Due to the limitation of the integration time (\SI{20}{\mega\year}), this is only completely modelled within a distance of roughly \SI{1}{\au}. Giant planets, in contrast, are not affected by the integration time limitation as they must anyway form before the dispersal of the gas disc. The model is then able to track the formation of all planets in the inner part of planetary systems.

After the description of the model, we study how the many different sub-models included in the Bern Generation III Model interact in the full end-to-end model by simulating the formation of two planetary systems. To understand the results, it is helpful to compare the timescales of growth and migration, revealing which process is dominant. It is also helpful to study the planetesimal surface density, revealing the solid accretion mode (planetesimal accretion versus growth by giant impacts). Other key processes occurring during the emergence of the planetary systems include the capture of many protoplanets into large resonant convoys, and the consequences of dynamical instabilities caused by the gravitational interactions between the protoplanets. This includes the destabilisation of other protoplanets at the moment a giant planet (especially a second one in the system) starts runaway gas accretion as well as series of giant impacts at the moment the gas disc dissipates.

We also give a short overview of the diversity of planetary systems that were obtained using the model. We find that systems containing giant planets can have a great diversity of configurations, while for systems forming only low-mass (Earth-like) planets exhibit arranged planets with similar masses.

This work is the first of a series. Here we present the outline of the series:
\begin{itemize}
\item \papertwo{} will introduce the methods to calculate population syntheses. Several populations for Solar-mass stars with different numbers of initial embryos per system are computed. The effects of this parameter at the population level will be investigated.
\item \paperthree{} will look for correlations between of the occurrence of inner low-mass and outer giant planets.
\item Paper IV \citep{NGPPS4} will extend the population synthesis to lower-mass stars (down to late M-dwarfs) and analyse the effects of the stellar mass.
\item Paper~V \citep{NGPPS5} will study the mapping of disc initial conditions to planet properties with machine learning.
\item Paper~VI \citep{NGPPS6} will look for the diversity between planets in each system compared to diversity of the overall population \citep{2018AJWeiss}.
\item There are then three papers on the quantitative comparison with various observational techniques: radial velocity with HARPS and CARMENES, and transits with Kepler.
\end{itemize}
The discrepancies uncovered in these comprehensive and multi-aspect comparisons with observations will be helpful to improve the understanding of planet formation and evolution.

\begin{acknowledgements}
The authors thank David Swoboda, Natacha Brügger, Martin Schlecker, and the members of the Theoretical Astrophysics and Planetary Science (TAPS) group at the University of Bern for fruitful discussions. We also thank the anonymous reviewer, whose remarks and suggestions helped improve the quality of this manuscript.
A.E. and E.A. acknowledge the support from The University of Arizona.
A.E. and C.M. acknowledge the support from the Swiss National Science Foundation under grant BSSGI0\_155816 `PlanetsInTime'. Parts of this work have been carried out within the frame of the National Center for Competence in Research PlanetS supported by the SNSF.
The plots shown in this work were generated using \textit{matplotlib} \citep{2007CSEHunter}.
\end{acknowledgements}

\bibliographystyle{aa}
\bibliography{manu,add}

\appendix

\section{Possible impact of migration on the accretion of planetesimals}
\label{appendixpredator}

\begin{figure*}[ht!]
	\centering
	\includegraphics[width=0.99\textwidth]{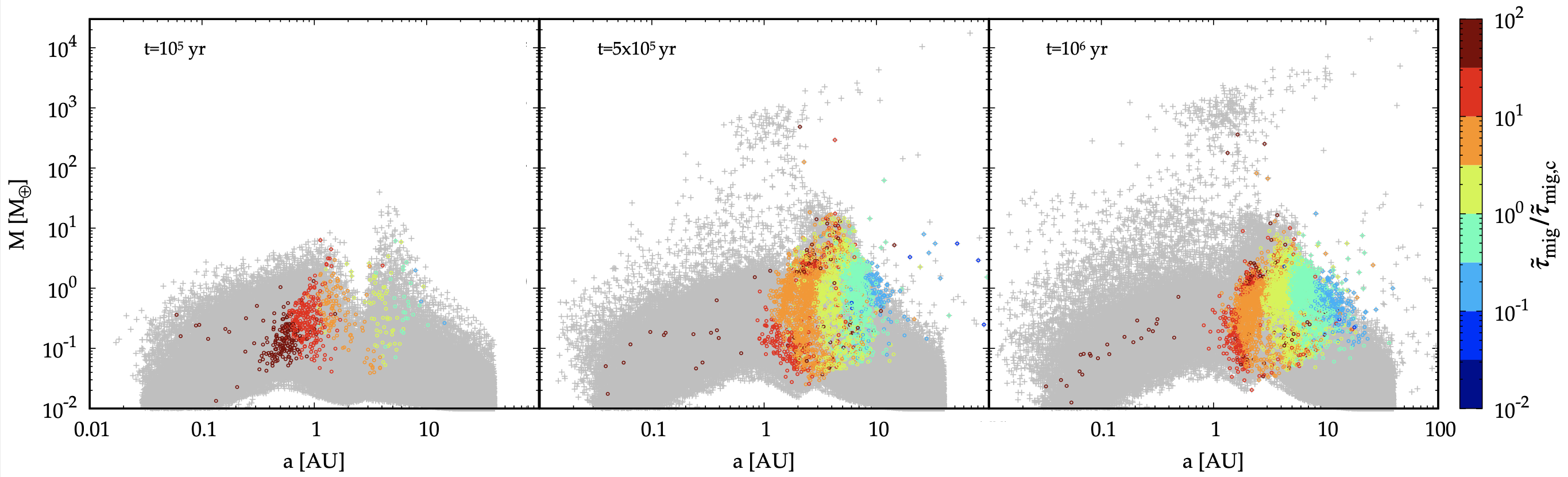}
	\caption{Mass-distance diagram of the nominal synthetic population NG76 of solar-like starts which stars with initially 100 moon-mass embryos per disc (see \papertwo). The epochs of 0.1, 0.5, and \SI{1}{\mega\year} are shown. Coloured points show protoplanets that can no longer accrete planetesimals of the initial local reservoir, which have a planetesimal accretion timescale of less than \SI{3}{\mega\year}, and which are still embedded in the parent gaseous disc. When $\tilde{\tau}_{\rm mig}\gtrsim \tilde{\tau}_{\rm mig,c}$, the planetesimal accretion of these planets could in principle be affected by shepherding if they would be the only protoplanets growing in the disc.}
	\label{fig:aMpredshep}
\end{figure*}

\subsection{General considerations}

\citet{1986IcarusWard} and \citet{1989ApJWard} were the first to suggested that the orbital migration of a protoplanet could strongly accelerate its planetesimal accretion if the planet is able to catch most planetesimals through which it is sweeping during its migration. In the terminology of Ward, the protoplanet would then act as a predator.

On the other hand, \citet{1999IcarusTanakaIda} found that in the case of a slow migration timescale $\tau_{\rm mig}$, the protoplanet rather carves a gap in the planetesimals disc around its orbit and shepherds the planetesimals, which stalls the accretion of the protoplanet. They derived a criterion in terms of a critical migration timescale $\tau_{\rm mig,crit}$ only below which the protoplanets can act as predator, $\tau_{\rm mig} \leq \tau_{\rm mig,crit}$. Otherwise, the protoplanets acts as shepherd.

\citet{2005A&AAlibert} studied this effect with the Generation I Bern Model and found that the migration rates were generally high enough for the protoplanets to be predators, and thus ignored the shepherding effect. Since the Generation 3 Model differs in numerous aspects (like the oligarchic growth mode, the disc model, the planetesimal size, the migration model, or the multiplicity of forming planets) from the model of 2005, we here re-assess this question. We do this based on existing simulations and published criteria. It is clear that for a more definitive assessment, direct N-body simulations of thousands of concurrently growing and migrating oligarchs accreting planetesimals and gas in the setup that we consider here would be needed. In these simulations, the planetesimals would have to be included as individual fully interacting bodies in the N-body. This differs from our current approach of representing planetesimals statistically as a surface density with a dynamic state (eccentricity and inclination, see Sect. \ref{sec:solids disc}).

\subsection{Applicability of \citet{1999IcarusTanakaIda}}

Coming to the general applicability of the work of \citet{1999IcarusTanakaIda}, one should note that they studied a highly special setup. In what follows, we discuss several important effects they neglected. When considering them, it becomes much less clear whether the shepherding effect can at all represent a major impediment to planetesimal accretion. We mention these points also in view of the specific results discussed in Sect. \ref{sect:predatorspec} where we try to identify planets for which shepherding could potentaially be relevant.

First, the predator-shepherd mechanism of \citet{1999IcarusTanakaIda} was found for a single core growing alone in a disc of planetesimals. This seems a highly unlikely setup given that many protoplanets (oligarchs) emerge concurrently from the pre-dating runaway planetesimal accretion phase and that subsequently grow further in lockstep \citep[e.g.][]{1998IcarusKokubo}. As \citet{1999IcarusTanakaIda} state actually themselves, in the case of such multiple protoplanets, `the protoplanets push planetesimals into the feeding zones of others and they can grow.'

Indeed, the more recent and realistic work of \citet{2006IcarusDaisaka} who run N-body simulation including type-I migration and tidal damping of eccentricity and inclination, starting from 7000–14'000 equal-mass self-gravitating planetesimals whose size is roughly \SI{1000}{\kilo\meter} showed a different picture than the single protoplanet simulations of \citet{1999IcarusTanakaIda}: namely that in the multiple protoplanet situation, the trapping of planetesimals by cores is only tentative and does not significantly reduce their accretion rates. This was already noted by \citet{2008ApJIdaLinA}, motivating them to not modify the core accretion rate for migrating protoplanets.

Second, according to the \citet{1999IcarusTanakaIda} timescale criterion one finds (at least formally) that whenever the migration timescale is longer than the critical value, no planetesimal accretion occurs. This includes in particular the case that a planet does not migrate at all. However, as shown by many works \citep{1998IcarusKokubo,2010AJLevison,2013A&AAlibert}, multiple protoplanets forming concurrently can grow in-situ by planetesimals that are already in their feeding zones. When a protoplanet grows, its feeding zone which is proportional to the Hill sphere and thus the mass of the protoplanet, also expands, bringing new planetesimals in reach of the protoplanet in a process that is unaffected by shepherding. The protoplanet's mass growth can occur via the accretion of solids, but also via the accretion of gas. In fact, the interplay of gas accretion leading to an extension of the solid feeding zone which in turn increases the core mass, and then again the gas accretion rate is the underlying process for Phase II in \citet{1996IcarusPollack}. Because of the absence of a gap in planetesimal disc around the planet if many protoplanets grow concurrently thanks to scattering \citep{2010AJLevison,2013A&AAlibert}, each protoplanet has a local reservoir of planetesimals from which it grows. Thus, when judging whether shepherding could affect growth, we do not need to consider planets that have not (yet) migrated outside of the planetesimal feeding zone around their starting location, since they still accrete their local reservoir.

Third, the temporal sequence of how solid accretion proceeds in forming multi-planet systems itself reduces the impact of shepherding. Here it is important to consider that solid growth of protoplanets occurs via two channels: the accretion of planetesimals, but also via protoplanet-protoplanet collisions (giant impacts). As can be seen in the growth tracks in Figures \ref{fig:aMsystem30} and \ref{fig:aMsystem852}, at the lower masses, when the planetesimal accretion timescale is less than the migration timescale, the protoplanets grow nearly in-situ by accreting planetesimals from their local feeding zone. As explained before, in this case shepherding is not important. In the subsequent phase, when the planets have grown to a mass where they start migrating, they also start accreting other protoplanets via giant impacts, which is also unaffected by shepherding. They also often migrate into parts of the disc which were previously depleted in planetesimals by another protoplanet that has formed further in. Here, shepherding would not have an effect. This insight also further justifies why using the in-situ prescription for the planetesimal accretion rate is appropriate. This also implies that when identifying planets that could be affected by shepherding, we do not need to consider planets that accrete in any case only very slowly planetesimals even when neglecting shepherding.

Finally, we also do not need to consider protoplanets where the gas disc has already dissipated. In this case, no gas disc-driven orbital migration occurs and thus no shepherding.

\subsection{Assessment of the population-wide importance of shepherding} \label{sect:predatorspec}

Except for the first point (the effect of protoplanet multiplicity) which questions the existence of shepherding in a fundamental way, the considerations from the previous section can - in an approximate way - be cast into a set of conditions where shepherding could be important. This allows us to identify these protoplanets which might at least in principle be affected by shepherding.

Figure \ref{fig:aMpredshep} shows the mass-distance diagram of the nominal synthetic population NG76 from \papertwo. Three moments in time are shown where planetesimal accretion is in general important. We colour code the absolute value of the ratio of the normalised migration timescale of a planet $\tilde{\tau}_{\rm mig}$ to the normalised critical migration timescale $\tilde{\tau}_{\rm mig,c}$ which both are calculated as in \citet{1999IcarusTanakaIda}. When this ratio is larger than approximately unity, shepherding would occur for a single protoplanet migrating alone through a disc of planetesimals. Only protoplanets which could in principle be affected by shepherding by fulfilling the following criteria are colour coded: First, the distance a planet has migrated away from its starting location is larger than five times the size of its Hill sphere. This means that it can no longer accrete from its initial local reservoir of planetesimals. Second, the planetesimal accretion timescale is less than three million years (the typical disc lifetime), meaning that planetesimal accretion (as opposed to growth via giant impacts) is still relevant. Third, the gas disc has not yet dissipated. Other protoplanets should in any case not be significantly affected by shepherding and are shown in grey.

The plot first shows that the large majority of protoplanets are grey, meaning that shepherding should not be important for them in any case. Then, more specifically, at \SI{0.1}{\mega\year}, there is a group of Mars- to Earth-mass protoplanets inside of the ice line where $\tilde{\tau}_{\rm mig} \gg \tilde{\tau}_{\rm mig,c}$. At \SI{0.5}{\mega\year}, there is a radial interval from about 1 to \SI{4}{\au} where $\tilde{\tau}_{\rm mig}$ is longer than $\tilde{\tau}_{\rm mig,c}$, however in most cases by less than one order of magnitude. These are regions where usually groups of tens of protoplanets form together (see Sect. \ref{sec:twoexmples}), so that it is not clear if shepherding would occur at all. Planets where the ratio is clearly larger, and thus where the effect could in principle be particularly strong, are rare. At \SI{1}{\mega\year}, a similar pattern is seen, but the potentially affected region is reduced.

It is clear that this simple a posteriori analysis cannot be seen as a final result - for this, simulations where planetesimal are included directly in the N-body would be necessary. Nevertheless, together with the finding of \citet{2006IcarusDaisaka} that shepherding is by principle not important when several protoplanets form concurrently, they indicate that shepherding can only affect a relatively limited part of all growing protoplanets.

\end{document}

%% file: defs.tex
\def\kB{k_\mathrm{B}}
\def\sigmaSB{\sigma_\mathrm{SB}}
\def\ggrav{\mathcal{G}}

\def\mearth{M_\oplus}

\def\msun{M_\odot}
\def\lsun{L_\odot}
\def\mj{M_{\textrm{\tiny \jupiter}}}

\def\mstar{M_\star}
\def\lstar{L_\star}
\def\rstar{R_\star}
\def\tstar{T_\star}
\def\qstar{Q_\star}

\def\mcore{M_{\rm core}}
\def\mdotcore{\dot{M}_{\rm core}}

\def\rcore{R_\mathrm{core}}
\def\menv{M_\mathrm{env}}
\def\mdotenv{\dot{M}_\mathrm{env}}
\def\mtot{M_\mathrm{tot}}
\def\mtotdot{\dot{M}_\mathrm{tot}}
\def\lint{L_\mathrm{int}}
\def\ltot{L_\mathrm{tot}}
\def\rtot{R_\mathrm{tot}}
\def\rtotdot{\dot{R}_\mathrm{tot}}
\def\etot{E_\mathrm{tot}}
\def\etotdot{\dot{E}_\mathrm{tot}}
\def\kenerg{\xi}
\def\kenergdot{\dot{\xi}_\mathrm{tot}}

\def\mplanet{M_\mathrm{planet}}
\def\aplanet{a_\mathrm{planet}}

\def\vplanet{v_\mathrm{planet}}

\def\mplanetj{M_\mathrm{planet,\jmath}}
\def\aplanetj{a_\mathrm{planet,\jmath}}

\def\rhill{R_\mathrm{H}}
\def\racc{R_\mathrm{acc}}
\def\miso{M_\mathrm{iso}}

\def\mstart{M_\mathrm{emb,0}}

\def\lradio{L_\mathrm{radio}}
\def\lburn{L_\mathrm{D-burn}}
\def\vkep{v_\mathrm{K}}

\def\vrad{v_\mathrm{rad}}

\def\sigmag{\Sigma_\mathrm{g}}

\def\sigmadotgplan{\dot{\Sigma}_\mathrm{g,planet}}
\def\sigmadotgphoto{\dot{\Sigma}_\mathrm{g,photo}}
\def\sigmadotgphotoint{\dot{\Sigma}_\mathrm{g,photo,int}}
\def\sigmadotgphotoext{\dot{\Sigma}_\mathrm{g,photo,ext}}
\def\betag{\beta_\mathrm{g}}
\def\rnorm{r_\mathrm{0}}
\def\sigmanorm{\Sigma_{g,0}}
\def\rin{r_\mathrm{in}}
\def\rcutg{r_\mathrm{cut,g}}
\def\rmax{r_\mathrm{max}}
\def\rhomid{\rho_\mathrm{mid}}
\def\tsurf{T_\mathrm{s}}
\def\tmid{T_\mathrm{mid}}
\def\tirr{T_\mathrm{irr}}
\def\tcd{T_\mathrm{cd}}
\def\mwind{\dot{M}_\mathrm{wind}}
\def\mgas{M_\mathrm{g}}

\def\sigmasol{\Sigma_\mathrm{s}}
\def\sigmas0{\Sigma_\mathrm{s,0}}
\def\rcuts{r_\mathrm{cut,s}}
\def\betas{\beta_\mathrm{s}}

\def\eplan{e_\mathrm{plan}}
\def\iplan{i_\mathrm{plan}}
\def\eplanh{\tilde{e}_\mathrm{plan}}
\def\iplanh{\tilde{\imath}_\mathrm{plan}}
\def\esqdotplan{\dot{e}_\mathrm{plan}^2}
\def\isqdotplan{\dot{i}_\mathrm{plan}^2}
\def\vrel{v_\mathrm{rel}}
\def\rplan{R_\mathrm{plan}}
\def\rhoplan{\rho_\mathrm{plan}}
\def\mplan{M_\mathrm{plan}}
\def\hplan{h_\mathrm{plan}}

\def\sigmamean{\bar{\Sigma}_\mathrm{s}}
\def\rcap{R_\mathrm{cap}}
\def\pcoll{p_\mathrm{coll}}

\def\rfeedinf{r_\mathrm{feed,inf}}
\def\rfeedsup{r_\mathrm{feed,sup}}
\def\rgc{R_\mathrm{gc}}
\def\mdotenvmax{\dot{M}_\mathrm{env,max}}
\def\mdotenvmaxr{\dot{M}_\mathrm{env,max,2D}}
\def\mdotenvmaxv{\dot{M}_\mathrm{env,max,3D}}
\def\mfeed{M_\mathrm{feed}}

\def\vff{v_\mathrm{ff}}

\def\dt{\Delta t}

\def\fopa{f_{\rm opa}}

\def\tKH{\tau_{\rm KH}}

